\documentclass[]{article}
\usepackage{fullpage}
\usepackage{booktabs} %
\usepackage[ruled]{algorithm2e} %
\usepackage{amssymb}

\SetAlFnt{\small}
\SetAlCapFnt{\small}
\SetAlCapNameFnt{\small}
\SetAlCapHSkip{0pt}
\IncMargin{-\parindent}

\usepackage[utf8]{inputenc} %
\usepackage[T1]{fontenc}    %
\usepackage{url}            %
\usepackage{booktabs}       %
\usepackage{amsfonts}       %
\usepackage{nicefrac}       %
\usepackage{microtype}      %
\usepackage{xcolor}         %

\usepackage[right, outer, inline, final]{showlabels} %
\usepackage{nicematrix}

\usepackage{tikz}
\usepackage{float}
\usetikzlibrary{shapes,backgrounds,patterns,calc,arrows,arrows.meta}
\usepackage{chemfig}

\catcode`\_=11
\definearrow5{-x>}{%
    \CF_arrowshiftnodes{#3}%
    \CF_expafter{\draw[}\CF_arrowcurrentstyle](\CF_arrowstartnode)--(\CF_arrowendnode)%
        coordinate[midway,shift=(\CF_ifempty{#4}{225}{#4}+\CF_arrowcurrentangle:\CF_ifempty{#5}{5pt}{#5})](line@start)%
        coordinate[midway,shift=(\CF_ifempty{#4}{45}{#4}+\CF_arrowcurrentangle:\CF_ifempty{#5}{5pt}{#5})](line@end)%
        coordinate[midway,shift=(\CF_ifempty{#4}{135}{#4}+\CF_arrowcurrentangle:\CF_ifempty{#5}{5pt}{#5})](line@start@i)%
        coordinate[midway,shift=(\CF_ifempty{#4}{315}{#4}+\CF_arrowcurrentangle:\CF_ifempty{#5}{5pt}{#5})](line@end@i);
    \draw(line@start)--(line@end);%
    \draw(line@start@i)--(line@end@i);%
    \CF_arrowdisplaylabel{#1}{0.5}+\CF_arrowstartnode{#2}{0.5}-\CF_arrowendnode
}
\catcode`\_=8

\usepackage{pict2e,picture,graphicx}
\usepackage[overload]{empheq}
\usepackage{microtype}
\usepackage{graphicx}
\usepackage{subcaption}
\usepackage{booktabs}
\usepackage{amsmath}
\usepackage{cases}
\usepackage{mathtools}
\usepackage{amsthm}
\usepackage{thm-restate}
\usepackage{enumitem}
\usepackage[ruled]{algorithm2e} %

\SetAlFnt{\small}
\SetAlCapFnt{\small}
\SetAlCapNameFnt{\small}
\SetAlCapHSkip{0pt}
\IncMargin{-\parindent}
\allowdisplaybreaks
\usepackage{xcolor}
\usepackage{nicefrac, xfrac}
\usepackage{xparse}
\usepackage{hyperref}
\usepackage[capitalize, noabbrev]{cleveref}
\usepackage{wrapfig}
\usepackage{bbm}
\usepackage{yfonts}
\usepackage{nicefrac} 
\usepackage{multirow}
\usepackage{bm}
\usepackage{bbm}
\usepackage{mdframed}
\usepackage{subcaption}
\usepackage{makecell}

\usepackage[T1]{fontenc}

\hypersetup{
	colorlinks = true,
	linkcolor = black,
	citecolor = black,
	linktocpage = true,
	urlcolor = mgreen
}

\usepackage{comment}

\usepackage{colortbl}
\usepackage{cellspace}

\colorlet{darkgreen}{green!70!black}

\newcommand{\CFont}[1]{{\textup{\textsc{#1}}}\xspace}
\newcommand{\PPAD}{\CFont{PPAD}}
\newcommand{\PLS}{\CFont{PLS}}
\newcommand{\CLS}{\CFont{CLS}}

\newcommand{\FP}{\CFont{FP}}
\newcommand{\FNP}{\CFont{FNP}}
\newcommand{\coNP}{\text{co-}\CFont{NP}}

\newcommand{\TFNP}{\CFont{TFNP}}
\newcommand{\EOTL}{\CFont{End-of-the-Line}}

\newcommand{\MINMAX}{\CFont{LocalMinMax}}
\newcommand{\LRMINMAXJC}{\CFont{LR-LocalMinMax-JC}}
\newcommand{\LRMINMAX}{\CFont{LR-LocalMinMax}}
\newcommand{\LRMINMAXCBC}{\CFont{LR-LocalMinMax-Indep}}

\newcommand{\VI}{\textsc{VI}\xspace}
\newcommand{\GDAFP}{\CFont{GDAFixedPoint}}
\newcommand{\GDAFPJC}{\CFont{GDAFixedPoint-JC}}
\newcommand{\GDAFPJCinfty}{\CFont{GDAFixedPoint-JC${}_\infty$}}
\newcommand{\GDAFPCBC}{\CFont{GDAFixedPoint-Indep}}
\newcommand{\Kakutani}{\CFont{Kakutani}}

\newcommand{\Brouwer}{\CFont{Brouwer}}
\newcommand{\POLY}{\CFont{PolyMatrix}}
\newcommand{\QVI}{\CFont{QuasiVI}}
\newcommand{\LINVI}{\CFont{LinVI}}

\newcommand{\GDA}{\CFont{GDA}}
\newcommand{\SGDA}{\CFont{sGDA}}
\newcommand{\HDSperner}{\CFont{HighD-Sperner}}

\newcommand{\tg}{\widetilde{\nabla}}

\newcommand{\poly}{\textnormal{poly}}
\newcommand{\size}{\textnormal{size}}
\def\ie{\textit{i.e.}\@\xspace}
\def\eg{\textit{e.g.}\@\xspace}
\newcommand{\Reals}{\mathbb{R}}
\newcommand{\Rationals}{\mathbb{Q}}
\newcommand{\Naturals}{\mathbb{N}}

\newcommand{\True}{\texttt{True}\xspace}
\newcommand{\False}{\texttt{False}\xspace}

\usepackage{color}
\definecolor{mygreen}{rgb}{0.0, 0.5, 0.0}
\definecolor{myorange}{rgb}{0.55, 0.62, 1}

\newcommand{\minipara}[1]{\vskip1mm\indent$\triangleright$ \textbf{\emph{#1.}}~}

\theoremstyle{plain}

\newcommand{\figorange}[1]{\textbf{\textcolor{orange2}{#1}}}

\theoremstyle{plain}
\newtheorem{theorem}{Theorem}[section]
\newtheorem{proposition}[theorem]{Proposition}

\newtheorem{lemma}[theorem]{Lemma}

\newtheorem{corollary}[theorem]{Corollary}
\theoremstyle{definition}
\newtheorem{definition}[theorem]{Definition}

\newtheorem{problem}{Problem}

\theoremstyle{remark}

\definecolor{niceRed}{RGB}{190,38,38}
\definecolor{Red2}{RGB}{219, 50, 54}
\definecolor{mgreen}{RGB}{160, 200, 140}
\definecolor{blueGrotto}{RGB}{5,157,192}
\definecolor{limeGreen}{HTML}{81B622}
\definecolor{myellow}{rgb}{0.88,0.61,0.14}
\definecolor{darkGreen}{HTML}{2E8B57}
\definecolor{navyBlueP}{HTML}{03468F}
\definecolor{Sepia}{HTML}{7F462C}
\definecolor{red2}{HTML}{1F462C}
\definecolor{orange2}{HTML}{FF8000}
\definecolor{mgray}{HTML}{ABB3B8}
\definecolor{lgray}{HTML}{E5E8E9}
\definecolor{myPurple}{RGB}{175,0,124}
\definecolor{mypurple2}{rgb}{0.8,0.62,1}
\definecolor{royalBlue}{HTML}{057DCD}
\definecolor{mpink}{HTML}{FC6C85}
\definecolor{lblue}{RGB}{74,144,226}
\definecolor{peagreen}{RGB}{152,193,39}
\definecolor{typ_navy}{HTML}{001f3f}
\definecolor{typ_blue}{HTML}{0074d9}
\definecolor{typ_aqua}{HTML}{7fdbff}
\definecolor{typ_teal}{HTML}{39cccc}
\definecolor{typ_eastern}{HTML}{239dad}
\definecolor{typ_purple}{HTML}{b10dc9}
\definecolor{typ_fuchsia}{HTML}{f012be}
\definecolor{typ_maroon}{HTML}{85144b}
\definecolor{typ_red}{HTML}{ff4136}
\definecolor{typ_orange}{HTML}{ff851b}
\definecolor{typ_yellow}{HTML}{ffdc00}
\definecolor{typ_olive}{HTML}{3d9970}
\definecolor{typ_green}{HTML}{2ecc40}
\definecolor{typ_lime}{HTML}{01ff70}
\definecolor{newgreen}{HTML}{83c702}
\definecolor{newpurp}{RGB}{97,96,121}

\usepackage[style=alphabetic,natbib=true,maxcitenames=2, maxbibnames=10,backend=bibtex]{biblatex}
\usepackage{biblatex}
\allowdisplaybreaks

\title{Improved Hardness Results for Min-Max Optimization with Coupled Constraints}

\author{
    Martino Bernasconi$^\dagger$ \quad
    Matteo Castiglioni$^\ddagger$ \quad
    Andrea Celli$^\dagger$ \quad
    Gabriele Farina$^\ast$  \vspace{6mm}\\
    $^\dagger$\ Bocconi university\\
    $^\ddagger$\ Politecnico di Milano\\
    $^\ast$\ Massachusetts Institute of Technology\vspace{2mm}\\
    {\textcolor{black}{\small\texttt{\{martino.bernasconi,andrea.celli2\}@unibocconi.it}, \quad \texttt{matteo.castiglioni@polimi.it,}}}\\
    {\textcolor{black}{\small\texttt{gfarina@mit.edu}}}
}
\hypersetup{
	colorlinks = true,
	linkcolor = royalBlue,
	citecolor = darkGreen,
	linktocpage = true,
	urlcolor = mgreen
}

\date{}

\begin{document}

\maketitle
\begin{abstract}
We investigate the computational complexity of min-max optimization under coupled constraints. 
The work of \citet*{daskalakis2021complexity} was the first to study min-max optimization through the lens of computational complexity, showing that min-max problems with nonconvex-nonconcave objectives are \PPAD-hard under coupled constraints. %
By carefully exploiting the coupled constraints rather than the structure of the objective function, we are able to significantly simplify and strengthen the proof of the hardness result.
More precisely, the first contribution of this paper is a fundamentally new proof of their main result, which improves it in multiple directions: it holds for degree-$2$ polynomials which are quadratic-linear, it improves the dependence on the parameters of the problem (also yielding constant inapproximability for gradient descent-ascent in $\ell_\infty$-norm), and it is much simpler than previous approaches.
Second, we show that with general constraints (\textit{i.e.},~the min player and max player have different constraints), even convex-concave (bilinear) min-max optimization becomes \textsc{PPAD}-hard.
Along the way, we also provide \textsc{PPAD}-membership of a general problem related to quasi-variational inequalities, which has applications beyond our problem.
\end{abstract}

\setcounter{tocdepth}{1} %
\tableofcontents

\section{Introduction}

In this paper, we study the complexity of min-max optimization in the presence of constraints.
This class of problems plays a key role in the development of game theory \citep{v1928theorie}, adversarial robustness in optimization, statistics, machine learning \citep{ben2002robust,huber2011robust,mkadry2017towards,sinha2018certifying}, and generative models such as Generative Adversarial Networks \citep{goodfellow2014generative,arjovsky2017wasserstein}.
In its simplest form, a constrained min-max optimization problem can be informally written as
\begin{equation}\label{eq:minmax general}
\begin{array}{l}
\displaystyle
\min_{x\in\Reals^d}\,\max_{y\in\Reals^d}\,\, f(x,y) \quad
 \textnormal{s.t.}\hspace{.3cm} g(x,y) \le 0,
\end{array}
\end{equation}
where $f:\Reals^{d}\times \Reals^d\to \Reals$, and $g:\Reals^{d}\times \Reals^d\to \Reals$. Due to the strong connection with zero-sum games, $x$ and $y$ are often thought of as selected by \emph{players} seeking to maximize and minimize the payoff function in a game.
Although the benign setting in which $f$ is convex-concave and $g$ is convex is tractable and well understood, many fundamental questions remain open regarding more general settings. In this paper, we are specifically interested in these more general settings.
Akin to regular (\ie, non-saddle-point) optimization, in nonconvex domains it is too much to ask for a \emph{global} solution. Instead, \emph{local min-max solutions} (also known as \emph{local equilibria}) serve as the target solution concept. Specifically, given $\epsilon,\delta>0$, we will call \emph{$(\epsilon,\delta)$-solution} of Problem \eqref{eq:minmax general} any pair $(x^\star,y^\star)$ such that $g(x^\star,y^\star)\le 0$, and such that each player cannot improve their utility through any \emph{feasible}, local, unilateral deviation, that is,
\begin{equation}\label{eq:intro:local minmax}
\begin{array}{l}
f(x^\star,y^\star) \le f(x,y^\star)+\epsilon \,\,\textnormal{ for all } x \textnormal{ s.t. } \|x-x^\star\|\le \delta \textnormal{ and } g(x,y^\star)\le 0\textnormal{; and}
\\
f(x^\star,y^\star) \ge f(x^\star,y)-\epsilon \,\,\textnormal{ for all } y \textnormal{ s.t. } \|y-y^\star\|\le \delta \textnormal{ and } g(x^\star,y)\le 0.
\end{array}
\end{equation}

Our work is not the first to study  min-max optimization beyond the convex-concave setup
(see, \eg, \citep{jin2020local,jin2021nonconvex,daskalakis2023stay} and pointers in \Cref{sec:related}). For example, several papers have focused on \emph{nonconvex-concave} objectives under constraints $g$ that induce \emph{product feasible sets} of the form $\mathcal{X} \times \mathcal{Y}$, where numerous algorithms find approximate solutions efficiently \citep{nouiehed2019solving,lin2020gradient,lin2020near,kong2021accelerated,fiez2021global,ostrovskii2021efficient}. 
On the other hand, in a recent landmark result, \citet*{daskalakis2021complexity} proved that finding solutions to \eqref{eq:intro:local minmax}, in the case of nonconvex-concave objective, is \PPAD-complete.

In this paper, we consider a broader class of constraints $g$. In particular, we consider the case in which the min and max players may have different constraints. By focusing on the constraints, more than on the structure of the objective function, we show a much simpler and stronger hardness result with respect to \citet{daskalakis2021complexity}. We obtain a significantly simpler reduction which holds for better approximation factors and for explicit degree-$2$ polynomial. In particular, our result holds when $f(x,y)$ is quadratic in $x$ and linear in $y$.
En route, we show that when the min and max players have different constraints, the problem is \PPAD-hard even when $f(x,y)$ is bilinear. This result is not only a stepping stone for understanding the proof of our main theorem, but it has direct implications on the complexity of quasi-variational inequalities with \emph{monotone} operators \citep{facchinei2003finite}. %
We discuss our contributions and their implications in more depth in the next subsection.
\subsection{Contributions and Implications}

\begin{figure}[!t]
    \centering\scalebox{.75}{\hspace{1.5cm}\begin{tikzpicture}
        \input{settings/colors}
        \def\dX{100pt}
        \def\xo{42pt}
        \def\dY{\dX*1/2/1.618}
        \def\yo{42pt}
        \def\dW{\dX}
        \def\dH{\dY}
        \def\de{20pt}
        \tikzstyle{entry}=[minimum width=\dW, minimum height=\dH, text width=\dW,
        inner sep=0mm,align=center,fill opacity=.8,text opacity=1]
        \draw[-{Latex[]}, very thick] (\xo-\dX/2-1, \yo-\dY/2-1) --node[below=.7cm]{\bf Structure of constraints} (\xo+2.5*\dX+\de, \yo-\dY/2-1);
        \draw[-{Latex[]}, very thick] (\xo-\dX/2-1, \yo-\dY/2-1) --node[xshift=-2.5cm,yshift=-2mm,rotate=90]{\bf Structure of utilities} (\xo-\dX/2-1, \yo+4.5*\dY+\de);

        \draw[dashed] (\xo-\dX/2-2.2cm, \yo+\dY/2) --node[below=.7cm]{} (\xo+2.5*\dX+\dW*0.75, \yo+\dY/2);
        \draw[dashed] (\xo-\dX/2-2.2cm, \yo+5*\dY/2) --node[below=.7cm]{} (\xo+2.5*\dX+\dW*0.75, \yo+5*\dY/2);

        \draw[dashed, black!80] (\xo+2*\dX, .7) node[below] {Independent} +(0, 4.7);
        \draw[dashed, black!80] (\xo+\dX, .7) node[below,text width=5cm,align=center] {Jointly convex} +(0, 4.7);
        \draw[dashed, black!80] (\xo+0*\dX, .7) node[below] {Product}  +(0, 4.7);
        \draw[dashed, black!80] (\xo-\dX/2, \yo) node[left,text width=2cm,align=right] {Convex concave}  +(13.7, 0);
        \draw[dashed, black!80] (\xo-\dX/2, \yo+3*\dY) node[left,text width=2cm,align=right] {Quadratic quadratic}  +(13.7, 0);
        \draw[dashed, black!80] (\xo-\dX/2, \yo+\dY) node[left,text width=2cm,align=right] {Quadratic linear}  +(13.7, 0);
        \draw[dashed, black!80] (\xo-\dX/2, \yo+2*\dY) node[left,text width=2cm,align=right] {Nonconvex linear}  +(13.7, 0);
        \draw[dashed, black!80] (\xo-\dX/2, \yo+4*\dY) node[left,text width=2cm,align=right] {Nonconvex nonconcave} +(13.7, 0);
        \draw[dashed, black!80] (\xo+3*\dX+2, \yo+1.5*\dY) node[fill=white, text width=\dW, align=left] {Nonconvex\\concave};

        \node[entry,fill=niceRed!25, minimum height=1*\dH] at (\xo+0*\dX, \yo+3.*\dY) {Open question};
        \node[entry,fill=lblue!10, minimum height=1*\dH] at (\xo+0*\dX, \yo+4.*\dY) {\PPAD-hard\\ \cite{bernasconi2026complexity}$^\dagger$};
        \node[entry,fill=lblue, opacity=0.1] at (\xo+\dX, \yo+4*\dY) {$\in\PPAD$\\ \cite{daskalakis2021complexity}};
        \node[entry,fill=lblue, opacity=0.3] at (\xo+2*\dX, \yo+4*\dY) {$\in\PPAD$\\ \textbf{\Cref{th:qvippad}}};
        \node[entry,fill=lblue, opacity=0.3] at (\xo+\dX ,\yo+\dY) {\PPAD-hard\\ \textbf{\Cref{th:nonconvexconcavehardness}}};
        \node[entry,fill=lblue, opacity=0.1] at (\xo+\dX ,\yo+3*\dY) {\PPAD-hard\\ \cite{anagnostides2025complexity}};
        \node[entry,fill=mgreen, opacity=0.3, minimum height=2*\dH] at (\xo+0*\dX, \yo+1.5*\dY) {Easy ($\in$ \FP)\\ \cite{ostrovskii2021efficient}};
        \node[entry,fill=lblue, opacity=0.1] at (\xo+\dX ,\yo+2*\dY) {\PPAD-hard\\ \citep{daskalakis2021complexity}};
        \node[entry,fill=lblue, opacity=0.1, minimum height = 3*\dH] at (\xo+2*\dX, \yo+2*\dY) {\PPAD-complete};

        \node[entry,fill=mgreen, opacity=0.3] at (\xo+0*\dX, \yo+0*\dY) {Easy ($\in$ \FP)\\(Folklore)};
        \node[entry,fill=mgreen, opacity=0.3] at (\xo+\dX, \yo+0*\dY) {Easy ($\in$ \FP)\\(Folklore)};
        \node[entry,fill=lblue, opacity=0.3] at (\xo+2*\dX, \yo+0*\dY) {\PPAD-hard\\ \textbf{\Cref{th:qvihardness}}};
    \end{tikzpicture}}
    \caption{Summary of known results regarding the complexity of min-max optimization for a different combination of structures of utilities and constraints. The arrows point toward the direction of increased generality and thus hardness. $\dagger$: \cite{bernasconi2026complexity} is subsequent and largely based on the insights of a preliminary version of this work.}
    \label{tab:result summary}
\end{figure}

A key factor in our taxonomy of optimization problems is the structure of the constraints. We consider the following classes, listed on the x-axis of \cref{tab:result summary} in order of increasing generality.
\begin{itemize}[nosep]
\item \emph{Product constraints}: The feasible set $g(x, y) \le 0$ is equal to $K_1 \times K_2$ for some appropriate $K_1, K_2 \subset \Reals^d$. In other words, there is \emph{no coupling} between the domains of the players.

\item \emph{Jointly convex constraints}: Constraints for both the maximizing and minimizing player are encoded by a function $g$ that is jointly convex, that is, convex as a function of $(x, y)$. This implies that $g(x, y) \le 0$ is a convex subset of $\Reals^{d} \times \Reals^d$. We remark that for our hardness result highlighted in \cref{tab:result summary}, it is enough to restrict to functions $g$ such that $g(x,y) \le 0$ defines a \emph{convex polytope}. This is the class of constraints used in the results of \citet{daskalakis2021complexity}.

\item \emph{Independent constraints}: Finally, to be able to capture problems corresponding to \emph{quasi}-variational inequalities, we also consider a more general setting in which the set of feasible local deviations of the players might differ and is controlled by two constraint functions $g_1, g_2 : \Reals^d\times\Reals^d \to \Reals$, one for each player. This leads to the modified notion of $(\epsilon,\delta)$-solution (or local equilibrium) given as 
\begin{equation}\label{eq:intro:local minmax g1g2}
\begin{array}{l}
f(x^\star,y^\star) \le f(x,y^\star)+\epsilon \,\,\textnormal{ for all } x \textnormal{ s.t. } \|x-x^\star\|\le \delta \textnormal{ and } g_1(x,y^\star)\le 0\textnormal{; and}
\\
f(x^\star,y^\star) \ge f(x^\star,y)-\epsilon \,\,\textnormal{ for all } y \textnormal{ s.t. } \|y-y^\star\|\le \delta \textnormal{ and } g_2(x^\star,y)\le 0.
\end{array}
\end{equation}
and where $g_i(x^\star,y^\star)\le 0$ for $i\in\{1,2\}$.
In particular, we are interested in the case of \emph{bilinear functions} $g_1, g_2$. Due to their bilinear nature, each of these functions defines a local convex polytope of feasible deviations for each player.
\end{itemize}
We will also consider different classes of objective functions: convex-concave, nonconvex-concave (further subdivided in nonconvex-linear and quadratic-linear), and nonconvex-nonconcave. We analyze how the problem's complexity evolves as the structure of the constraints changes. We categorize our results into hardness and membership results. Here, we focus especially on hardness results; more details about membership results are deferred to \Cref{sec:existence}.

\paragraph{Hardness results}

The main contribution of this paper is the following hardness result.

\begin{theorem}[Informal version of \Cref{th:nonconvexconcavehardness}]\label{th:main informal} The problem of computing approximate local min-max equilibria with jointly convex constraints, and nonconvex-concave, $G$-Lipschitz, and $L$-smooth utilities is \PPAD-hard for $\epsilon,L=O(1)$, $\delta=\Omega(1)$, $G=\poly(d)$ and $G_\infty=\sup_{x,y\in[0,1]^{2d}}\|\nabla f(x,y)\|_\infty=O(1)$. Moreover the function $f$ is a degree-$2$ polynomial quadratic in $x$ and linear in $y$.
\end{theorem}
This result has several implications with respect to the hardness result by \citet[Theorem 4.4]{daskalakis2021complexity}. 
Indeed, it is easy to prove that when $\delta\le\frac \epsilon G $ then every point is a solution to the min-max problem and thus the problem is trivially in \FP.  
On the other hand, \citet{daskalakis2021complexity} proved that the problem is hard with $\delta= \Omega(d^{30} \frac\epsilon  G)$. Our results improve on this result by showing \PPAD-hardness already for $\delta = \Omega(\sqrt{d} \frac\epsilon G)$, significantly improving the gap between the polynomial regime and the hard one (we refer to \Cref{fig:complexity plot} for an illustration). This follows since we can set $\delta,\epsilon=\Theta(1)$ and $G=O(\sqrt d)$. 

While we leave the question of showing hardness for the regime $\delta = \Omega(\epsilon/ G)$ as an interesting open question, we remark that our reduction (thanks to the fact that it holds for constant $G_\infty$) shows tight (constant inapproximability) lower bounds for the related problem of computing approximate fixed points of gradient descent ascent with respect to the $\ell_\infty$ norm under joint constraints. We expand on this below and in \Cref{sec:constant_inapprox}. We conjecture that similar tight lower bounds can be derived for the problem of computing approximate local min-max equilibria with jointly convex constraints, when the size of the local deviation is measured in norms other than the Euclidean one.

Moreover, the degree of the polynomial objective function $f$ cannot be improved compared to what we obtain in our reduction. Indeed, it is well known that the case of bilinear---and thus convex-concave---objective admits polynomial time algorithms even with jointly convex constraints (\Cref{tab:result summary} second columns, last row). Thus, we need $f$ to be a quadratic polynomial to hope to show hardness. We do this by constructing a function $f$ that is quadratic only in one of the two players.

\begin{figure}[t]
    \centering
    \scalebox{0.6}{\tikzset{every picture/.style={line width=1pt}} %
\tikzstyle{every node}=[font=\large]

\begin{tikzpicture}[x=0.75pt,y=0.75pt,yscale=-1,xscale=1]
\input{settings/colors}

\def\xA{0}
\def\xB{110}
\def\xG{140}
\def\xC{290}
\def\xD{460}
\def\xE{670}
\def\xF{710}

\def\yA{0}
\def\yB{30}
\def\yC{60}
\def\yD{90}
\def\yCm{45}

\draw  [draw opacity=0][fill=lgray, fill opacity=0.9 ] (\xA,\yA) -- (\xE,\yA) -- (\xE,\yD) -- (\xA,\yD) -- cycle ;

\draw  [draw opacity=0][fill=lblue, fill opacity=0.23 ] (\xC,\yB) -- (\xE/2+\xF/2,\yB) -- (\xE/2+\xF/2,\yD) -- (\xC,\yD) -- cycle ;

\draw  [draw opacity=0][fill=lblue, fill opacity=0.23 ] (\xG,\yB) -- (\xE/2+\xF/2,\yB) -- (\xE/2+\xF/2,\yD) -- (\xG,\yD) -- cycle ;

\draw  [draw opacity=0][fill=mgray, fill opacity=0.6 ] (\xD,\yCm) -- (\xE/2+\xF/2,\yCm) -- (\xE/2+\xF/2,\yD) -- (\xD,\yD) -- cycle ;

\draw  [draw opacity=0][fill=mgreen  ,fill opacity=0.48 ] (\xA,\yC) -- (\xB,\yC) -- (\xB,\yD) -- (\xA,\yD) -- cycle ;

\draw   [ thick] (\xB,\yD) -- (\xB,\yC) -- (\xB/2+\xA/2,\yC) ;
\draw [shift={(\xB/2+\xA/2-2,\yC)}, rotate = 360] [fill={rgb, 255:red, 0; green, 0; blue, 0 }  ][line width=0.08]  [draw opacity=0] (12,-3) -- (0,0) -- (12,3) -- cycle    ;

\draw   [ thick] (\xE,\yD) -- (\xE,\yA) -- (\xE/2-\xA/2,\yA) ;
\draw [shift={(\xE/2-\xA/2-2,\yA)}, rotate = 360] [fill={rgb, 255:red, 0; green, 0; blue, 0 }  ][line width=0.08]  [draw opacity=0] (12,-3) -- (0,0) -- (12,3) -- cycle    ;

\draw [ thick]   (\xC,\yD) -- (\xC,\yB) -- (\xC/2+\xD/2,\yB) ;
\draw [shift={(\xC/2+\xD/2+2,\yB)}, rotate = 180][fill={rgb, 255:red, 0; green, 0; blue, 0 }  ] [line width=0.08]  [draw opacity=0] (12,-3) -- (0,0) -- (12,3) -- cycle    ;

\draw [ thick]   (\xG,\yD) -- (\xG,\yB) -- (\xG/2+\xC/2,\yB) ;
\draw [shift={(\xG/2+\xC/2+2,\yB)}, rotate = 180][fill={rgb, 255:red, 0; green, 0; blue, 0 }  ] [line width=0.08]  [draw opacity=0] (12,-3) -- (0,0) -- (12,3) -- cycle    ;

\draw  [ thick]  (\xD,\yD) -- (\xD,\yCm) -- (\xD/2+\xF/2,\yCm) ;
\draw [shift={(\xD/2+\xF/2+2,\yCm)}, rotate = 180] [fill={rgb, 255:red, 0; green, 0; blue, 0 }  ][line width=0.08]  [draw opacity=0] (12,-3) -- (0,0) -- (12,3) -- cycle    ;

\draw [line width=.5mm]   (\xA,\yD) -- (\xF,\yD) ;
\draw [shift={(\xF+2,\yD)}, rotate = 180] [fill={rgb, 255:red, 0; green, 0; blue, 0 }  ][line width=0.08]  [draw opacity=0] (8.93,-4.29) -- (0,0) -- (8.93,4.29) -- cycle    ;

\draw (\xF,\yD) node [anchor=west][inner sep=2mm]    {$\displaystyle\delta $};
\draw (\xB,\yD) -- +(0,2mm) node [below=1mm,xshift=-2mm] [inner sep=1.3pt]    {$\displaystyle\frac{\epsilon }{G}$};

\draw (\xG,\yD) -- +(0,2mm) node [below=1mm,xshift=+4mm] [inner sep=1.3pt]    {$\displaystyle\Omega\left(\frac{\epsilon }{G_\infty}\right)$};
\draw (\xE,\yD) -- +(0,2mm) node [below=1mm] [inner sep=1.3pt]    {$\displaystyle\sqrt{\frac{2\epsilon }{L}}$};
\draw (\xC,\yD) -- +(0,2mm) node [below=1mm,xshift=5.5mm][inner sep=1.3pt]   {$\displaystyle\Omega\left(\sqrt d\frac\epsilon G\right)$};
\draw (\xD,\yD) -- +(0,2mm) node [below=1mm] [inner sep=1.3pt]    {$\displaystyle \Omega\left(d^{30}\frac\epsilon G\right)$};
\draw (\xA,\yD) -- +(0,2mm) node [below=1mm] [inner sep=1.3pt]    {$0$};

\draw (\xA/2+\xE/2,\yA/2+\yB/2) node [anchor=center][inner sep=0.75pt,align=center]  {$\in$ \PPAD \citep{daskalakis2021complexity}};
\draw (\xA/2+\xB/2,\yC/2+\yD/2) node [anchor=center][inner sep=0.75pt,align=center]    {$\in$ \FP};
\draw (\xC/2+\xD/2,\yB/2+\yD/2) node [anchor=center][inner sep=0.75pt,align=center,text width=4cm]  {\PPAD-hard [\Cref{th:nonconvexconcavehardness}]};
\draw (\xD/2+\xE/2,\yCm/2+\yD/2) node [anchor=center][inner sep=0.75pt,align=center]    {\PPAD-hard \\ \citep{daskalakis2021complexity}};

\draw (\xG/2+\xC/2,\yB/2+\yD/2) node [anchor=center][inner sep=0.75pt,align=center, text width = 4cm]    {\PPAD-hard$^*$ [\Cref{sec:constant_inapprox}]};

\end{tikzpicture}}
    \caption{Our results nearly settle the complexity of finding local equilibria of nonconvex-nonconcave objective functions under jointly convex constraints as a function of the parameter $\delta$. *The constant inapproximability result that holds for $\delta=\Omega(\epsilon/G_\infty)$ is for a natural variant of the problem in distances are measured in $\ell_\infty$ norm (which is polynomial for $\delta=O(\epsilon/G_\infty)$). See \Cref{sec:constant_inapprox} for a detailed discussion.}
    \label{fig:complexity plot}
\end{figure}

In \Cref{sec:ideafirst}, we give a first high-level overview of our new approach.
In essence, we show that coupled constraints enable a simple embedding of a generic vector field into \Cref{eq:intro:local minmax}, thereby solving a \emph{variational inequality} (VI) associated with this vector field.\footnote{A VI problem  consists of finding a point $z\in Q$, for some convex set $Q\subset \Reals^d$, such that $F(z)^\top(z'-z)\ge 0$ for all $z'\in Q$, where $F:\Reals^d\to\Reals^d$ \citep{stampacchia1964formes}. VIs are a powerful framework for modeling optimization problems across diverse fields, including game theory \citep{rockafellar1970monotone}, economics \citep{facchinei2003finite}, and machine learning \citep{goodfellow2014generative}, and they generalize classical optimization problems like complementarity problems \citep{cottle1968complementary}.}

We remark that in subsequent work, \citet{bernasconi2026complexity} resolved the question about the complexity with product constraints, showing \PPAD-hardness. A key idea was to use the specific instances in this paper (without enforcing coupled constraints) and to hide them within a Boolean \PPAD-hard problem.

We complement the above result by proving \PPAD-hardness for the setting where the objective function is restricted to only be convex-concave, but the constraints class is made more complex by allowing independent constraints. This marks a second ``phase change'' in the complexity of the problem due to constraints, since the convex-concave problem is clearly solvable with jointly convex constraints by using the monotone VI formulation of the problem (in \Cref{tab:result summary} this would correspond to the second and third column and last row).

\begin{theorem}[Informal version of \Cref{th:qvihardness}]\label{th:hardness_cc_informal}
 The problem of computing approximate local min-max equilibria with independent constraints, and convex-concave, $G$-Lipschitz, and $L$-smooth utilities is \PPAD-hard for $\epsilon,L=O(1)$, $\delta=\Omega(1)$, $G=\poly(d)$ and $G_\infty=\sup_{x,y\in[0,1]^{2d}}\|\nabla f(x,y)\|_\infty=O(1)$. Moreover, the function $f$ is linear in $x$ and $y$ (separately).
\end{theorem}

Furthermore, our reduction has an interesting consequence on the complexity of \emph{quasi-variational inequalities} (QVIs). These are a generalization of VIs, whereby the VI domain $Q$ is no longer a fixed set but rather a \emph{correspondence}, which depends on the optimization variable.
In this space, we show that even with linear constraints, monotone QVIs are unlikely to be polynomial-time solvable in general.

\begin{corollary}[Informal version of \Cref{cor:qvihard}]\label{th:qvi_informal}
    The problem of computing solutions to QVIs is \PPAD-hard even when the set-valued fixed-point problem encoded by the QVI constraints is linear (and thus solvable in polynomial time) and the operator is monotone.
\end{corollary}

\paragraph{Constant Inapproximability of Gradient Descent-Ascent in $\ell_\infty$-norm}

We are able to obtain a reduction that holds for $G_\infty=\sup_{x,y\in[0,1]^d}\|\nabla f(x,y)\|_\infty=O(1)$. This is significant, as it establishes tight (\ie constant-factor inapproximability) lower bounds for the problem of computing fixed points of gradient descent-ascent under joint constraints, in the setting where the players feasibility sets (given the strategy of the other player) are products of intervals.

Local min-max points are tightly connected with fixed points of the gradient descent-ascent dynamics. In particular, \citet{daskalakis2021complexity} proved a two-way reduction between these two problems, in the Euclidean norm.
Our reduction proves an inapproximability result for the problem of computing a fixed point of gradient descent-ascent in the $\ell_\infty$-norm. In particular, consider the problem of computing $(x,y)$ such that 
\[
\|x-\Pi_{\{x':g(x',y)\le 0\}}(x-\eta\nabla_x f(x,y))\|_\infty+\|y-\Pi_{\{y':g(x,y')\le 0\}}(y+\eta\nabla_y f(x,y))\|_\infty\le \alpha.
\]
Letting $G_\infty = \sup_{x,y\in[0,1]^d}\|\nabla f(x,y)\|_\infty$, the problem becomes trivially solvable in polynomial time for ``small'' $\eta$. In particular, in the special case in which the sets $\{x':g(x',y)\le 0\}, \{y':g(x,y')\le 0\}$ are product of intervals for each $(x,y)$, and $\eta= O(\frac{\alpha}{G_\infty})$, then every $(x,y)$ is a solution to the problem above.
In particular this holds when the ``unit-free'' ratio $\eta/(\epsilon/G_\infty)=O(1)$. %
However, our reduction holds for $\eta,\alpha,G_\infty=O(1)$, and thus when $\eta/(\alpha/G_\infty)=\Omega(1)$, showing that there is no polynomial time algorithm for the problem in $\ell_\infty$. 
It is a natural question whether one could obtain the same inapproximability result in the $\ell_2$ norm. We expand on this discussion in \Cref{sec:constant_inapprox}.

\paragraph{\PPAD-membership results}

\PPAD-membership for the case of jointly convex constraints is due to \citet[Theorem 5.2]{daskalakis2021complexity}. We look at the case of independent constraints. 
To do so, we prove \PPAD-membership of a very general problem related to computing solutions to quasi-variational inequalities. 
Then, our \PPAD-membership result for local min-max equilibria follows from the connection between QVIs, fixed points of gradient descent-ascent dynamics, and approximate local min-max equilibria.

\begin{theorem}[Informal version of \Cref{th:qvippad}]%
    The problem of finding an approximate solution to quasi-variational inequalities is in \PPAD. %
\end{theorem}

This result may be of independent interest, as the generality of the QVI problem could enable new \textsc{PPAD}-membership results.
For example, we can easily give \PPAD membership for generalized equilibrium \cite{rosen1965existence}, which are more general than those recently considered in \cite{filos2024ppad} (see \Cref{sec:existence}).

\subsection{Overview of Techniques}\label{sec:ideafirst}

This section provides a high-level overview of the primary techniques employed throughout the paper by describing the main steps of the proof of our main result (\Cref{th:main informal}). This is a concise summary of the discussion provided in \Cref{sec:ideaproof}.

\paragraph{Linear variational inequalities} The first step in our reduction is to prove the hardness of an intermediate problem related to VI with linear operators defined on the hypercube, which we call \LINVI. Specifically, given an affine operator $F(z)=Dz+c$ we consider the problem of finding a point $z$ such that $F(z)^\top(z'-z)\ge 0$. We show that \LINVI is \PPAD-hard through a reduction from the problem of finding Nash equilibria in polymatrix games. 
This choice is instrumental in obtaining a hard problem on the hypercube and a stronger dependence on the parameters.

Then, we connect our problem to find fixed points of gradient descent-ascent dynamics, and we reformulate it as solving a variational inequality (VI) problem. Unfortunately, gradient descent-ascent dynamics are related to VI with a specific operator $F$, constructed by stacking the gradients $\nabla_x f$ and $-\nabla_y f$, which correspond to the utilities of the two players.
In game theory this is sometimes called ``pseudo-gradient'' and we indicate it with the symbol 
\[
\tg f(x,y)\coloneqq(\nabla_x f(x,y), -\nabla_y f(x,y)).
\]
It is not hard to see that a fixed point of the gradient descent-ascent map defined on $f$ is a solution of the VI instance defined with the operator $\tg f$ (\Cref{th:fromVItoMinMax}).

\begin{figure}[t!]
\centering
\begin{subfigure}{0.45\textwidth}
    \centering
    {\includegraphics[scale=0.7]{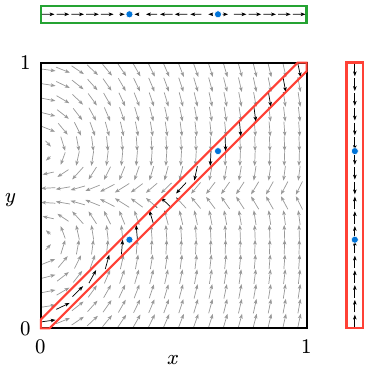}}
    \caption{Embedding of $F(y)=(y-\frac13)\cdot(y-\frac23)$ with $f(x,y)=x^\top F(y)$.}
    \label{fig:embedding1}
\end{subfigure}
\hfill
\begin{subfigure}{0.45\textwidth}
    \centering
    {\includegraphics[scale=.7]{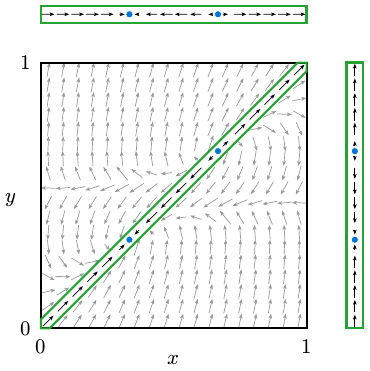}}
    \caption{Embedding of $F(y)=(y-\frac13)\cdot(y-\frac23)$ with $f(x,y)=(x-y)^\top F(y)$.}
    \label{fig:embedding2}
\end{subfigure}
\caption{
The two embeddings of the field $F:\Reals^d\to\Reals^d$ into the pseudo-gradient of $f:\Reals^{2d}\to\Reals$. The box on the right is the field $x\mapsto\nabla_x f$, while the one on top represents the field $y\mapsto -\nabla_y f$. The embedded field is $F(y)=(y-\frac13)\cdot(y-\frac23)$; the blue dots are in correspondence with its zeros and are inserted only for reference. In both figures we have that the $x$-projection of the pseudo-gradient field on the $x=y$ subspace corresponds to the field $F$, while only in \Cref{fig:embedding2}, both the $x$ and $y$ projections are aligned with the field $F$.
}
\label{fig:embeddings}
\end{figure}

\paragraph{Pseudo-gradient fields} It is clear that if we had to deal with standard gradients, then the problem would be much easier (and \PPAD-hardness would be hopeless), as any local minima would satisfy the first-order optimality conditions given by the VI. Therefore, our approach hinges on the distinction between ``pseudo-gradient'' fields (vector fields generated by applying the pseudo-gradient operator to a differentiable function $f$) and gradient fields. Pseudo-gradient fields are more general than gradient fields, and allow for more complex behavior, \eg gradient lines might form connected paths, while for conservative vector fields the gradient lines decrease the potential.

\paragraph{Perfect imitation and independent constraints} 
Independent constraints provide a simple means of implementing the linear operator via imitation. Given the linear operator $z\mapsto Dz+c$ defining the \LINVI instance, we embed it into the first half components $x\in[0,1]^d$ of the function $f$, while using the constraints to force the $y\in[0,1]^d$ components to lie on a specific $d$-subdomain of $[0,1]^{2d}$. Specifically, if we want to embed the operator $[0,1]^d\ni z\mapsto F(z)\in [0,1]^d$ into a pseudo-gradient $\tg f$, we can consider the following function $f(x,y)=x^\top F(y)$ and force $y$ to imitate $x$ (for a illustration with $d=1$ see \Cref{fig:embedding1}). Indeed, with independent constraints, we can build instances in which the $y$ player must choose $y\in\{x\}$ while the $x$ player is only constrained to play in the hypercube. Now, we have that $\nabla_x f(x,y)=F(y)=F(x)$ and thus, if $x$ is a fixed point of gradient descent-ascent, we have that $F(x)^\top(x'-x)\ge 0$, implying that $x$ is a solution to \LINVI. This idea is enough to prove hardness of the convex-concave case with independent constraints (\Cref{th:hardness_cc_informal}).

\paragraph{Approximate imitation and jointly convex constraints}
Attempting to apply the same construction as in the case of jointly convex constraints quickly leads to seemingly insurmountable challenges. First, the function used in the previous construction was convex-concave, and equilibria for such functions are easy to compute on a jointly convex set $K$. Therefore, we must increase the complexity of $f$ to be at least quadratic in $x$ or $y$.
Second, the most natural way to extend the imitation gadget from independent constraints to jointly convex ones is to consider the set $K\coloneqq\{x,y:\|x-y\|_\infty\le\Delta\}$, for ``small'' values of $\Delta$. However, this comes at a high cost: we can no longer disregard player $y$ and consider only the optimality condition of the $x$ player. What if the solution $x,y$ found by solving the min-max problem is on the boundary of the set $K$?
Intuitively, we can see that the constraint $\{x,y:\|x-y\|_\infty\le\Delta\}$ can only limit one player at a time, and either the $x$ player or the $y$ player can move towards any deviation $x'$.\footnote{Formally,
this is not entirely accurate; we need to decompose a deviation direction into two distinct components, one corresponding to the $x$-player and the other to the $y$-player. See \Cref{sec:ideaproof} for more details.} 
However, now we have to design $f$ to align the gradient of the $y$ player to the operator $F(z)=Dz+c$ of the \LINVI instance we are reducing from. Otherwise, an optimality condition of the $y$ player on $f$ would be meaningless in terms of its implications for the original \LINVI instance. A better embedding of an operator $F(z)$ is to consider the function $f(x,y)=(x-y)^\top F(x)$.
We can see that this is a promising choice since $\nabla_x f(x,y)=F(x)+J_F(x)^\top(x-y)$, where $J_F(x)$ is the Jacobian of $F$ at $x$, and $-\nabla_y f(x,y)=F(x)$. Given that the constraints enforce that $\|x-y\|\le \Delta$, we have that approximately both gradients are aligned with $F(x)$, and thus a solution to min-max would be a solution to the original \LINVI instance (\Cref{fig:embedding2}).
Incidentally, the set $K$ is the same considered in the reduction by \citet{daskalakis2021complexity}. We analyze the connections with their reduction in the following paragraph.

\paragraph{Relationship with the reduction of \citet{daskalakis2021complexity}}

Our reduction follows a completely different approach compared to prior work.
For example, we reduce from a continuous optimization problem, naturally defined on the hypercube, whereas \citet{daskalakis2021complexity} reduces from a discrete optimization problem related to Sperner's lemma. This allows us to work with simple degree-two polynomials instead of building functions that are defined by circuits of some discrete problem. This simpler construction provides stronger results, such as simpler functions, constant smoothness, and hardness for larger approximations.
More in detail, the reduction of \citet{daskalakis2021complexity} starts from a problem (\textsc{HighD-BiSperner}) related to a generalization of Sperner's lemma \citep{sperner1928neuer}. They partition the $d$-dimensional hypercube in cublets and color each vertex with $d$ colors out of a total of $2d$ colors (instead of the $d+1$ usually used in Sperner's lemma). This problem is shown to be \PPAD-hard by a reduction from \Brouwer. Then they build a function $f$ such that the fixed points of gradient descent-ascent dynamics are close to panchromatic cubelets of the simplicization, in which they interpret the $d$ colors of each vertex as directions of the pseudo-gradient of $f$. To do so, they give a function value and direction based on the local coloring and then interpolate the function with a high-degree smooth step function. Their ingenious construction is extremely technical and requires careful techniques to ensure that no extra fixed points of gradient descent-ascent are created outside of those corresponding to panchromatic cubelets. It is precisely this last requirement that forces a coupling between the strategies of the minimizer and maximizer variables since the coefficients of the interpolation produce perturbations in the pseudo-gradient, which depend on the difference $x-y$. 
In contrast, our construction sidesteps any need for interpolating (or, in fact, the use of discrete Sperner-type problems altogether) by constructing a problem instance with an explicit quadratic-linear objective, and relying on the constraints to drive hardness from the start.

\subsection{Further Related Work}\label{sec:related}
\paragraph{Complexity of optimization problems}

We contribute to the line of work that shows lower bounds of optimization problems employing tools from computational complexity.
This new line of research has achieved remarkable results with significant implications across complexity theory \citep{fearnley2022complexity}, game theory \citep{babichenko2021settling}, and machine learning \citep{daskalakis2021complexity, hollender2023computational}.
Optimization problems belong naturally to the class of \emph{total search problem} (\TFNP) \citep{johnson1988easy,megiddo1991total,papadimitriou1994complexity}.
The bridge between optimization and search problems is not new and, for instance, it has been the main motivation for the definition of the \PLS class \citep{johnson1988easy, schaffer1991simple, krentel1989structure}.
\citet{fearnley2022complexity} showed that the problem of computing fixed points of gradient descent is complete for the class $\PLS\cap\PPAD$, and this result also proves that such class is equal to \CLS. Therefore, it lies much lower in the hierarchy of \TFNP problems then \PPAD. 
Moreover, while we work inside the framework of ``constrained'' optimization (meaning that we work inside the hypercube rather than on the entire Euclidean space), there are some very recent works that consider the computational complexity of optimization in the unconstrained setting \citep{hollender2023computational, kontogiannis2024computational}.

\paragraph{Imitation}
The idea of imitation in recent years has been instrumental in various \PPAD-hardness results \citep{mclennan2005imitation,rubinstein2015inapproximability,babichenko2016query,babichenko2020communication,babichenko2021settling}. These settings mainly focus on general-sum games, and imitation is attained through carefully designed payoffs of one (or more) player. However, our setting is zero-sum, and adding an imitation component to the payoff would discourage one player from imitating just as much as it would encourage the other. Thus, our result can be seen as enforcing imitation through constraints rather than payoffs.

Most recently, \citet{anagnostides2025complexity} proposed a reduction from symmetric equilibria in symmetric games, using ideas similar to our paper (\emph{cf.} \cref{sec:ideafirst}), to prove a weaker version of \Cref{th:nonconvexconcavehardness} via a simpler reduction. 
Several key differences set our result apart from theirs. First, our result shows hardness already for nonconvex-concave functions (more precisely, quadratic-linear), while their result only applies to nonconvex-nonconcave quadratic polynomials.
Moreover their problem is for min-max optimization over the simplex (rather that on the hypercube), which admits a natural quasi-polynomial-time approximation scheme and thus can only lead to weaker hardness results.

\paragraph{Solving min-max problems} A substantial body of recent research has focused on developing practical first-order and low-order methods for min-max optimization problems.
A wide range of works illustrate divergent or cycling behavior when extending beyond minimization problems \citep{mertikopoulos2018cycles,hsieh2021limits}. However, efficient algorithms are available for well-behaved objectives.
In the convex-concave setup, solutions can be efficiently computed using convex programming techniques \citep{korpelevich1976extragradient,azizian2020tight,golowich2020last,mazumdar2020gradient,hamedani2021primal,daskalakis2019last,mokhtari2020unified,abernethy2021last} and via algorithms for monotone VIs \citep{bruck1977weak,eckstein1992douglas,tseng1995linear,nemirovski2004prox,chen2017accelerated,gorbunov2022extragradient}.

Nonconvex-concave problems with ``simple'' constraints can also be solved efficiently \citep{daskalakis2018training,nouiehed2019solving,lin2020gradient,lin2020near,kong2021accelerated,ostrovskii2021efficient}.
In the nonconvex-nonconcave setting, motivated by the complexity results of \citep{daskalakis2021complexity}, research has focused on identifying structural assumptions on the objective (\eg, assuming the weak Minty variational inequality holds) that can help overcome the computational hardness barriers \citep{diakonikolas2021efficient,pethick2022escaping}, and identifying different notions of local solutions and studying convergence to these points \citep{jin2020local,mangoubi2021greedy,keswani2022convergent}.
Finally, a more recent stream of works considers different notions of stationarity, such as Goldstein’s stationarity to handle nonsmooth objectives \citep{zhang2020complexity,jordan2022complexity,kornowski2024hardness}.

\section{Preliminaries}\label{sec:preliminaries}

We study local min-max solutions, or \emph{equilibria}, of a function $f:[0,1]^d\times[0,1]^d\to[-B,B]$ with $B\in \Reals_+$ in which the two variables are subject to functional constraints $g_1,g_2:[0,1]^d\times[0,1]^d\to[-1,1]$. Formally, for any $\epsilon>0$ and $\delta>0$, an $(\epsilon,\delta)$-equilibrium is a tuple of points $(x^\star,y^\star)\in[0,1]^d\times[0,1]^d$ such that   it holds $g_i(x^\star,y^\star)\le 0$ for each $i\in\{1,2\}$, and
\begin{align*}
f(x^\star, y^\star)&\le f(x,y^\star)+\epsilon \quad \forall x\in[0,1]^d\text{ such that } \|x-x^\star\|\le \delta \text{ and } g_1(x,y^\star)\le 0,\\
f(x^\star, y^\star)&\ge f(x^\star,y)-\epsilon \quad \forall y\in[0,1]^d\text{ such that } \|y-y^\star\|\le \delta \text{ and } g_2(x^\star,y)\le 0.
\end{align*}
In this paper, we use the symbol $\|\cdot\|$ to denote the Euclidean norm.
We study the complexity of computing equilibria under a combination of assumptions on the structure of the payoff function and the constraint functions $g_1, g_2$. We detail the assumptions under consideration next. 
In \Cref{tab:result summary} we can see a table summarizing the payoffs and constraints considered in this paper.

\paragraph{Payoff function}
Throughout the paper, we assume that $f$ is $G$-Lipschitz and $L$-smooth. Formally, a $G$-Lipschitz function satisfies
\[
|f(x,y)-f(x',y')|\le G\left\|(x,y)-(x',y')\right\|\quad\forall (x,y),(x',y')\in[0,1]^{2d},
\]
while $L$-smoothness is defined as the $L$-Lipschitzness of the gradient, \ie,
\[
\|\nabla f(x,y)-\nabla f(x',y')\|\le L\left\|(x,y)-(x',y')\right\|\quad\forall (x,y),(x',y')\in[0,1]^{2d}.
\]
We study three different classes of payoff functions $f$. We present the classes from the most specific to the most general

\begin{itemize}
\item The simplest scenario we address is the \emph{convex-concave} case, where the function $f(x,y)$ is convex in $x$ over $[0,1]^d$ for every fixed $y\in[0,1]^d$ and concave in $y\in [0,1]^d$ for every fixed $x\in[0,1]^d$.
\item Moving to a more complex scenario, we consider the \emph{nonconvex-concave} case, where $f$ remains concave in $y$ over $[0,1]$ for every fixed $x\in[0,1]^d$, without imposing any specific properties on $f$ for a fixed $y$.
\item Finally, we consider the \emph{nonconvex-nonconcave} case where no specific properties are imposed on $f$, other than being $G$-Lipschitz and $L$-smooth.
\end{itemize}
\paragraph{Constraints}
We consider the case in which $g_1$ and $g_2$ are \emph{bilinear} functions. 
Let  $[m]=\{1,\ldots, m\}$. Formally, for each player $i$ the function $g_i$ is defined by a set of $m$ bilinear functions $h^{(i)}_j$:
\begin{equation}\label{eq:constraints}
    g_i(x,y):=\max_{j\in[m]} h^{(i)}_j(x,y).
\end{equation}
More formally, for each $i\in\{1,2\}$ and $j\in[m]$ there exists a matrix $B^{(i)}_j\in\Reals^{d\times d}$, vectors $b^{(i)}_{1,j},b^{(i)}_{2,j}\in\Reals^d$, and constant $c^{(i)}_j\in\Reals$ such that $h_j^{(i)}(x,y):=x^\top B_j^{(i)} y+b^{(i),\top}_{1,j}x+b^{(i),\top}_{2,j}y+c^{(i)}_j$.
The functions $g_{i}$ induce a set of  (approximately) feasible polytopes. In particular, for any $\nu\ge0$ we can define the sets
\begin{align*}
K_1^\nu(y):=\{x'\in[0,1]^{d}: g_1(x',y)\le \nu\} \quad \text{and} \quad K^\nu_2(x):=\{y'\in[0,1]^{d}: g_2(x,y')\le \nu\}.
\end{align*}
The bilinear structure of $g_{1}$ implies that for all $y\in[0,1]^d$ the set $K^\nu_1(y)$ is a polytope defined by $m$ inequalities. Similarly, for all $x\in[0,1]^d$ the set $K^\nu_2(x)$ is a polytope.
In the previous definition, we allow for a slight violation $\nu$ of the constraints. As will become evident in the following, this violation is needed for technical reasons and we will keep this violation negligible (arbitrarily small) or zero in all of our results. When the sets do not include the superscript $\nu$, we imply $\nu=0$. 

We consider three different classes of constraints. We present them from the most specific to the most general:

\begin{itemize}
\item The simplest type of constraints are \emph{product constraints}, where $g_1(x,y)=g_1(x,y')$ for all $x,y,y'\in[0,1]^d$ (\ie, $g_1$ is independent of $y$), and $g_2(x,y)=g_2(x',y)$ for all $x,x',y\in[0,1]^d$ (\ie, $g_2$ is independent of $x$). In this setting there exist two convex sets such that $x\in K_1\subseteq [0,1]^d$ (respectively, $y\in K_2\subseteq [0,1]^d$) if and only if $g_1(x,\cdot)\le 0$ (respectively, $g_2(\cdot,y)\le 0$).
\item A more general class of constraints are \emph{jointly convex constraints} where there exists a \emph{convex} function $g:[0,1]^{2d}\to[-1,1]$ such that $g_1=g_2=g$. In particular, we assume $g$ to be linear in $(x,y)$, \ie, without the bilinear component.\footnote{This is the class of constraints studied in \citet{daskalakis2021complexity}.} In this setting, we can define a polytope $K:=\left \{(x,y)\in[0,1]^{2d}: g(x,y)\le 0\right\}$ and then for all $(x,y)\in K$ it holds:
\begin{align*}
K_1(y)=\{x': (x',y)\in K\},\quad\text{and}\quad K_2(x)=\{y': (x,y')\in K\}.
\end{align*}
\item Finally, we consider a broader class of constraints in which the feasible strategies of one player depend on the strategy chosen by the other player, with no additional conditions beyond that $g_1$ and $g_2$ be defined as in \Cref{eq:constraints}. We refer to these as \emph{independent constraints}.

We remark that with independent constraints, the function $g_i$ defining the coupling between the player $i$ and player $j$ strategies can be chosen independently of the function $g_j$ defining the coupling between the player $j$ and player $i$ strategies.

\end{itemize}

All our hardness results apply with only linear constraints (\ie $g_1$ and $g_2$ are defined without the bilinear components). This makes our results stronger since it is possible to find points that satisfy the constraints in polynomial time.
On the other hand, we restrict ourselves to independent constraints (instead of considering more general ones or sets given implicitly by \eg separation oracles as in \citet{papadimitriou2023computational}) when proving our membership results.
This is mainly done to guarantee some properties that simplify the exposition and analysis. For instance, it is possible to compute in polynomial time the projection on $K^\nu_1(y)$ and $K^\nu_2(x)$, and it is easier to ``enlarge'' the sets with a parameter $\nu$. Our membership results, however, can be adapted for more general classes of constraints assuming the existence of an approximate projection oracle, as in \citet{papadimitriou2023computational}. Indeed, the properties to guarantee such membership results are mainly related to convexity and the existence of strictly feasible points. Considering non-explicit definitions of the sets would add much complexity to the exposition without adding much to our contributions.

\subsection{Gradient Descent-Ascent Dynamics and Min-Max Optimization}
In this section, we introduce the main computational problems we deal with in this paper, while in \Cref{sec:cc} we formally introduce reductions between total problems.
The formal problem of computing a local min-max equilibrium is defined as follows.

\begin{problem}[\MINMAX]
   Given $\epsilon,\delta,L,G,\nu \in \Reals_+$, two circuits implementing a $G$-Lipschitz and $L$-smooth function $f$ and its gradient $\nabla f$, and an explicit representation of functions $g_1$ and $g_2$, find two points $(x^\star,y^\star)\in[0,1]^d$ such that $g_i(x^\star,y^\star)\le \nu$ for all $i\in\{1,2\}$ and:
   \begin{align*}
    f(x^\star, y^\star)&\le f(x,y^\star)+\epsilon \quad \forall x\in[0,1]^d\text{ such that } \|x-x^\star\|\le \delta \text{ and } g_1(x,y^\star)\le \nu,\\
    f(x^\star, y^\star)&\ge f(x^\star,y)-\epsilon \quad \forall y\in[0,1]^d\text{ such that } \|y-y^\star\|\le \delta \text{ and } g_2(x^\star,y)\le \nu,
   \end{align*}
   or output $\bot$ if no such point exists.
\end{problem}

We call a solution $(x^\star,y^\star)$ of the \MINMAX problem an $(\epsilon,\delta,\nu)$-solution to \MINMAX. If $\nu=0$, we simply say that it is an $(\epsilon,\delta)$-solution to \MINMAX.

The problem \MINMAX defined above is not assured to have solutions. 
Following the approach of \citet{daskalakis2021complexity}, we specify \MINMAX into \emph{regimes}, and we focus on the \emph{local regime}.

\begin{problem}[\LRMINMAX]
We define \LRMINMAX (\emph{local regime} min-max) as the class of \MINMAX problems with $\delta<\sqrt{\frac{2\epsilon}{L}}$.
\end{problem}
\paragraph{Relation to local min-max problems}
We observe that \MINMAX generalizes the local min-max optimization problem studied by  \citet{daskalakis2021complexity} along two different directions. First, we consider a more general class of constraints, specifically bilinearly-coupled constraints that may be independent among players, whereas \citet{daskalakis2021complexity} considers only jointly convex constraints. Moreover, we relax the problem allowing for slight violations of the constraints, meaning that we can violate the functions $g_i$ by a margin $\nu\ge0$. 
In most cases we will restrict to $(\epsilon,\delta)$-solutions and zero violation (\ie, $\nu=0$).
However, when considering \MINMAX in its full generality, an arbitrary (even exponentially) small relaxation $\nu>0$ is needed to guarantee the existence of solutions (see \Cref{sec:existence}). 

\subsubsection{Fixed points of Gradient Descent-Ascent} \label{sec:prelimGDA}

Since the problem of verifying the local optimality of solutions is a hard problem, the local min-max problem is not a $\FNP$ problem. 
Therefore, we focus on a problem intimately related to approximate local min-max equilibria whose solutions are easy to verify, and hence the problem formally lies in \TFNP.
Specifically, we investigate the problem of finding a fixed point of gradient descent-ascent dynamics.
This is essentially equivalent to the choice of \citet{fearnley2022complexity} to focus on KKT solutions instead of local minima.\footnote{\citet{daskalakis2021complexity} seems to overlook this technicality and thus treats \LRMINMAX as a \FNP problem. However, this comes with little or no collateral damage as long as one is aware of this issue.}

Formally, given a smooth function $f$ and any $\nu\ge 0$, we can define the gradient descent-ascent update as:
\[
F_{\GDA}^\nu(x,y) = \begin{pmatrix}\Pi_{K^\nu_1(y)}(x-\nabla_xf(x,y))\\ \Pi_{K^\nu_2(x)}(y+\nabla_yf(x,y))\end{pmatrix}.
\]

Then, the computational problem of finding a fixed point of the gradient descent-ascent dynamics is defined as follows.

\begin{problem}
[\GDAFP]\label{def:gdafp}
Given $\alpha$, $L$, $G$, $\nu\in\Reals_+$, two circuits implementing a $G$-Lipschitz and $L$-smooth function $f:[0,1]^d\times[0,1]^d\to [-B,B]$ and its gradient $\nabla f: [0,1]^d\times[0,1]^d\to\Reals^{2d}$, and an explicit representation of functions $g_1$ and $g_2$, find $(x^\star, y^\star)\in[0,1]^{2d}$ such that 
\[ 
g_i(x^\star,y^\star)\le\nu \quad \forall i\in\{1,2\}\quad \textnormal{   and  } \quad \|(x^\star,y^\star)-F_{\GDA}^\nu(x^\star,y^\star)\|\le\alpha,
\] 
or output $\bot$ if no such point exists.
\end{problem}

At the heart of \citet{daskalakis2021complexity} lies the connection between \GDAFP and \LRMINMAX.
Indeed, \citet{daskalakis2021complexity} showed that, in the local regime, any $\alpha$-fixed point of \GDAFP with $\alpha=\poly(\epsilon,\delta,G,L)$ is an $(\epsilon,\delta)$-solution to  \LRMINMAX.
The reduction between \GDAFP and \LRMINMAX was proved by \citet{daskalakis2021complexity} only for the jointly convex case, but the proof extends without changes to bilinear constraints since their proof simply exploits that $K^\nu_2(x)$ is convex for all $x$ and $K^\nu_1(y)$ is convex for all $y$. This is verified when we deal with bilinear constraints.

\begin{theorem}[\protect{\cite[Theorem~5.1]{daskalakis2021complexity}}]\label{th:GDA_eq_minmax}
    Consider a \LRMINMAX problem, and assume that $f$ is $G$-Lipschitz and $L$-smooth. Then:
    \begin{enumerate}[label=\roman*)]
        \item For any $\epsilon>0$, $\delta>0$ and any $\nu\ge0$, define 
        $\alpha:=\frac{1}{2}\left(\sqrt{(G+\delta)^2+4\left(\epsilon-\frac{L}{2}\delta^2\right)}-(G+\delta)\right)$. Any $\alpha$-approximate fixed point $(x,y)$ of \GDAFP is an $(\epsilon,\delta,\nu)$ solution to \MINMAX.
        \item For any $\alpha>0$, define $\epsilon:=\frac{\alpha^2L}{(5L+2)^2}$ and $\delta\le\frac{\alpha}{5L+2}$. Any $(\epsilon,\delta,\nu)$ solution to \LRMINMAX is an $\alpha$-approximate fixed point of \GDAFP.
    \end{enumerate}
\end{theorem}

The previous theorem shows that there are solutions to \LRMINMAX that can be verified in polynomial time. However, this does not imply that \emph{all} of them are checkable in polynomial time.
Additionally, this also shows that there exists a reverse reduction that essentially shows that if finding an $\alpha$-fixed point of $F_\GDA$ is \PPAD-hard, then \MINMAX is also \PPAD-hard. 

\paragraph{Promises} Following the earlier discussion, \MINMAX and \GDAFP are formally defined as promise problems. All inputs satisfy the promise that the arithmetic circuits output function and gradient values that are consistent with the smoothness and Lipschitz constants specified in the input. Moreover, we make some minimal promises about the constraints that depend on whether the constraints are jointly convex constraints or independent constraints (see \Cref{sec:interest}).

\paragraph{Function representation}
When defining a computational problem, we assume as input two circuits: one implementing a $G$-Lipschitz and $L$-smooth function $f$ and the other implementing its gradient $\nabla f$. Notably, this does not require the function $f$ to be represented explicitly.
Differently, we always assume that the functions $g_i$ for each $i\in\{1,2\}$ are given in input by $\{B^{(i)}_j,b^{(i)}_{1,j},b^{(i)}_{2,j},c^{(i)}_j\}_{i\in\{1,2\},j\in[m]}$ defining the constraints as per \Cref{eq:constraints}.

\subsubsection{Problems of interest} \label{sec:interest}

We further partition the instances of \MINMAX depending on $\nu,\epsilon,\delta$ and the shape of the constraints $g_1$ and $g_2$.
We focus on the local regime since, outside of this regime, solutions are not guaranteed to exist. Moreover, in the case of independent constraints,  an arbitrarily small relaxation of the constraints $\nu>0$ is required to guarantee the existence of solutions (details in \Cref{sec:existence}).
Formally, we define the following problems.

\begin{problem}[\LRMINMAXJC]\label{def:lrminmaxjc}
    We define as \LRMINMAXJC (\emph{local regime min-max with jointly convex constraints}) as the class of \MINMAX problems with jointly convex constraints, $K\neq\emptyset$, $\nu=0$, and $\delta<\sqrt{\frac{2\epsilon}{L}}$.
\end{problem}

\LRMINMAXJC corresponds to the \emph{local-regime local min-max equilibrium computation problem} which is the main problem analyzed in \cite{daskalakis2021complexity}. 
Additionally, we define the corresponding problem with independent constraints.

\begin{problem}[\LRMINMAXCBC]\label{def:lrminmaxcbc}
    We define as \LRMINMAXCBC (\emph{local regime min-max with independent constraints}) as the class of \MINMAX problems with $\nu>0$ and $\delta<\sqrt{\frac{2\epsilon}{L}}$. Moreover, we require that $K_1(y)\neq\emptyset$ for all $y\in[0,1]^d$ and $K_2(x)\neq\emptyset$ for all $x\in[0,1]^d$.
\end{problem}

Intuitively, the non-emptiness assumption is required to guarantee that each player has a feasible strategy for each opponent's strategy. This guarantees that the problem is well-posed and it always admits solutions.
Unfortunately, as discussed above, even if there are certifiable solutions to these two problems, it is generally \coNP-hard to verify a generic candidate solution. %
Thus, we consider the following two problems, each respectively related to \LRMINMAXJC and \LRMINMAXCBC. These problems have the advantage of being formally in \FNP, as the definition of \GDAFP directly implies polynomial-time verification.

\begin{problem}[\GDAFPJC]\label{def:gdajc}
    We define as \GDAFPJC (fixed-point with \emph{ jointly convex constraints}) the class of \GDAFP problems with jointly convex constraints, $K\neq\emptyset$, and $\nu=0$. 
\end{problem}

\begin{problem}[\GDAFPCBC]\label{def:gdacbc}
    We define as \GDAFPCBC (fixed point with \emph{independent constraints}) the class of \GDAFP problems with $\nu>0$. Moreover, we require that $K_1(y)\neq\emptyset$ for all $y\in[0,1]^d$ and $K_2(x)\neq\emptyset$ for all $x\in[0,1]^d$.
\end{problem}

While the \PPAD-membership of \GDAFPJC was proven in \cite[Theorem 4.3]{daskalakis2021complexity}, the presence of independent constraints between the players introduces significant challenges to deriving existence results.
We will prove the totality and \PPAD-membership of \GDAFPCBC through the connection with (Quasi-)Variational inequalities (which might be of independent interest).
We will exploit this connection also to show that \GDAFPCBC and \LRMINMAXCBC might not admit solutions for $\nu=0$.

\subsection{Variational and Quasi-Variational Inequalities, and Kakutani Fixed Points}

In this section, we introduce the required tools to understand our membership results. We will prove \PPAD membership for a generalization of the \GDAFP problem, which is related to quasi-variational inequalities (QVIs) \citep{facchinei2003finite}. This effort is not only directed toward our membership result as Variational Inequalities (VI) will also be useful to build our main hardness results. In \Cref{sec:existence} we will provide a clear picture of the relation among the computational versions of VIs, QVIs, and our problems of interests related to min-max optimization.

QVIs are a generalization of variational inequalities. A classical variational inequality problem is the problem of finding a vector $z\in Q$ such that:
\[
F(z)^\top(z'-z)\ge 0\quad\forall z'\in Q,
\]
where $F:\Reals^d\to\Reals^d$ and $Q\subset \Reals^d$ is a convex set. 
QVIs generalize the problem by considering a non-fixed feasible set $Q$ defined through a correspondence.
\begin{definition}[Correspondence]
    A correspondence between two topological spaces $X$ and $Y$ is a set-valued function that associates every point of $X$ to an element of $2^Y$ (set of subsets of $Y$). We write $Q:X\rightrightarrows Y$ to denote a correspondence.
\end{definition}

Given a correspondence $Q:\Reals^d\rightrightarrows\Reals^d$ and an operator $F:\Reals^d\to\Reals^d$, the QVI problem corresponds to finding a vector $z$ such that
\begin{enumerate}[label=\arabic*.]
    \item $z\in Q(z)$ and 
    \item $F(z)^\top(z'-z)\ge 0\quad\forall z'\in Q(z)$.
\end{enumerate}

The fact that feasible deviations $Q(z)$ change as $z$ varies adds non-trivial complexity issues, and QVIs are, in general, much more challenging than VIs. Note that QVIs embed a set-valued fixed-point problem, \ie find $z$ such that $z\in Q(z)$. The existence of such fixed points is guaranteed, under suitable assumptions, by Kakutani's fixed-point theorem \citep{kakutani1941generalization}.\footnote{We defer all the details about continuity of correspondences and the statement of Kakutani theorem to \Cref{app:prooffriends}.}

Now, we are ready to introduce some computationally meaningful problems related to QVIs. 

\begin{problem}[\QVI]
    Given $G,L,\epsilon>0,\nu> 0$, a circuit implementing a $G$-Lipschitz function $F:\Reals^d\to\Reals^d$, and two circuits implementing a $L$-Lipschitz continuous matrix valued function $A:[0,1]^d\to\Reals^{n\times d}$ and a $L$-Lipschitz continuous vector valued function $b:\Reals^n\to\Reals^d$ defining the  correspondence $Q_\nu(\tilde z):=\{z\in[0,1]^d:A(\tilde z)z\le b(\tilde z)+\nu1_d\}$, find a point $z\in Q_\nu(z)$ such that:
    \[
        F(z)^\top(\tilde z-z)\ge -\epsilon,
    \]
    for all $\tilde z\in Q_\nu(z)$. We assume that $Q(z)\neq \emptyset$ for all $z\in [0,1]^d$.\footnote{We say that a matrix-valued function $A:[0,1]^d\to\Reals^{n\times d}$ is $L$-Lipschitz if its application to every $z\in[0,1]^d$ is $L$-Lipschitz, \ie, if the function $\tilde z\mapsto A(\tilde z)z$ is $L$-Lipschitz for every $z\in[0,1]^d$.}
\end{problem}

\begin{problem}[\Kakutani] \label{prob:kakutani}
    Given $L,\nu\ge0$, %
    and two circuits implementing a $L$-Lipschitz continuous matrix valued function $A:[0,1]^d\to\Reals^{n\times d}$ and a $L$-Lipschitz continuous vector valued function $b:\Reals^n\to\Reals^d$ defining the correspondence $Q_\nu(\tilde z):=\{z\in[0,1]^d:A(\tilde z)z\le b(\tilde z)+\nu 1_d\}$, find a point $z$ such that $z\in Q_\nu(z)$. We assume that $Q(z)\neq \emptyset$ for all $z\in [0,1]^d$.
\end{problem}
Finally, we consider a problem that simplifies \QVI problem along two directions.
First, we consider a fixed set, more specifically the hypercube, hence restricting to a VI problem.
Second, we force the function $F$ to be affine.\footnote{In this work, we sometimes refer to functions with affine components as linear for simplicity.}

\begin{problem}[\LINVI]
    Given a matrix $D\in[-1,1]^{d\times d}$, a vector $c \in  [-1,1]^d$, and an approximation parameter $\epsilon>0$, find a point $z\in [0,1]^d$ such that:
    \[
    (Dz+c)^\top(z'-z)\ge -\epsilon\quad\forall z'\in [0,1]^d.
    \]
\end{problem}

\paragraph{Relation of QVIs to VIs and Kakutani}
Here we expand our discussion on how \QVI generalizes \Kakutani and VIs. Clearly \Kakutani is \PPAD-hard, by a trivial reduction from \Brouwer. This proves that \QVI is hard even when the operator $F$ is the constant $0$.
Second, QVIs are a generalization of VIs, which are already hard by a trivial reduction from the problem of computing Nash equilibria in two-player games. Thus, it is straightforward to conclude that QVIs are hard even when the correspondence $Q$ is fixed.
An intriguing question arises here:

\vskip3mm
\begin{quote}\centering
\begin{mdframed}[outerlinewidth=4, backgroundcolor=gray!0, leftmargin=30, linecolor=white]\centering
    \emph{Is \QVI hard even when both the corresponding VI and \Kakutani problem are solvable in polynomial time?}
\end{mdframed}
\end{quote}
\vskip3mm

En route to proving our main result, we answer this affirmatively.
In \Cref{th:qvihardness} and \Cref{cor:qvihard}, we show that \QVI is \PPAD-complete even restricting to instances in which: $i)$ the set-valued fixed-point problem is easy (a vector such that $z\in Q(z)$ can be found in polynomial time) and $ii)$ the operator $F$ with the \QVI problem is monotone, \ie
\[
(F(z)-F(z'))^\top(z-z')\ge 0\quad\forall z,z'\in\Reals^d.
\]
We recall that a monotone operator implies that any VI over a convex set $Q$ can be solved in polynomial time \citep{nemirovski2004prox}.

\paragraph{Representation of sets}
Note that our correspondences are more restrictive than those employed by \citet{papadimitriou2023computational}, which, to our knowledge, is the only work studying a computational version of Kakutani's fixed-point theorem. Our primary goal is to demonstrate that these fixed-point problems are challenging even in basic settings, so using simpler correspondences aligns with this objective.

\section{Hardness of Local Min-Max Optimization} \label{sec:Harness}

In this section, we provide two fundamental hardness results on the problem of solving local min-max problems in the presence of constraints. For missing details about this section, we refer to \Cref{app:omitted hard}.

\paragraph{Nonconvex-concave function and jointly-convex constraints}
The first result shows that both \LRMINMAXJC and \GDAFPJC are \PPAD-hard, for much better dependency on the parameters of the problem, and even constant inapproximability for the problem of computing fixed points of gradient descent-ascent in $\ell_\infty$. %
\Cref{sec:ideaproof} provides an overview of the proof, which is formally described in \Cref{sec:formalproof}.
\begin{restatable}{theorem}{nonconvexconcavehardness}\label{th:nonconvexconcavehardness}
    Given a $\delta>0$, \LRMINMAXJC with nonconvex-concave objective is \PPAD-hard for $G=O(\sqrt{d})$, $L,\epsilon=O(1)$,$\delta=\Omega(1)$, and $\sup_{x,y\in[0,1]^d}\|\nabla f(x,y)\|_\infty=G_\infty=O(1)$. Moreover, \GDAFPJC with nonconvex-concave objective is \PPAD-hard for $\alpha_\GDA,L=O(1)$, $G=O(\sqrt{d})$ and $\sup_{x,y\in[0,1]^d}\|\nabla f(x,y)\|_\infty=G_\infty=O(1)$. This holds even when $f$ is quadratic in $x$ and linear in $y$.
\end{restatable}

\paragraph{Regime of the hardness result}

As pointed out by \citet{daskalakis2021complexity}, for $\delta\le \epsilon/G$ the problem becomes easily solvable in polynomial time (indeed, when $\delta\le \epsilon/G$ any pair $(x,y)$ is a solution to \LRMINMAX).
Our proof shows that the problem is \PPAD-hard for any $\delta=\Omega\left(\sqrt{d}\frac{\epsilon}{G}\right)$.
This significantly reduces the ratio between  $\delta$ and $\epsilon/G$ for which the problem remains hard. Indeed, in previous result, the problem is hard for $\delta=\Omega(d^{30}\frac\epsilon G)$.
Moreover our reduction employs function with $G_\infty=O(1)$. The most important consequence of this result is that we are able to show constant inapproximability for the problem of computing fixed points of gradient descent-ascent when distances are measured in $\ell_\infty$. We point to \Cref{sec:constant_inapprox} for more details about this point.

\paragraph{Convex-concave function and independent constraints}

The next result shows \PPAD-hardness for the simplest case of payoff functions (\ie, convex-concave, and more precisely linear-linear) and the most challenging case of constraints (\ie, independent constraints). This highlights even more how constraints are capable of injecting hardness in otherwise computationally amenable problems since the convex-concave problem is solvable with jointly convex
constraints. 
We remark that our result holds for linear constraints $g_1$ and $g_2$. This guarantees that the hardness result is not related only to solving the Kakutani-like problem. Indeed, a solution to the corresponding \Kakutani problem can be found through linear programming.
The proof of this result is provided in \Cref{sec:proofqvihard}.

\begin{restatable}{theorem}{qvihardness}\label{th:qvihardness}
     Given a $\delta>0$, \LRMINMAXCBC with convex-concave (bilinear) objective and linear constraints is \PPAD-hard for $G=O(\sqrt{d})$, $\sup_{x,y\in[0,1]^{d}}\|\nabla f(x,y)\|_\infty=O(1)$, $L,\epsilon=O(1)$, and $\delta=\Omega(1)$. Moreover, \GDAFPCBC
    with convex-concave objective and linear constraints is \PPAD-hard for $\alpha_\GDA=O(1)$, $G=O(\sqrt{d})$, $L=O(1)$ and $\sup_{x,y\in[0,1]^{d}}\|\nabla f(x,y)\|_\infty=O(1)$.
\end{restatable}

\paragraph{Hardness of monotone QVIs} The result holds when $f(x,y)$ is a bilinear function of $x$ and $y$. Then, as a direct corollary of the previous theorem, we have that finding a solution to a monotone QVI is \PPAD-hard. As we already remarked, the constraints in our reduction are linear. Hence, this result holds even when the set-valued fixed-point problem defined by the \QVI constraints is solvable in polynomial time, and the VI associated with the QVI operator is monotone.

\begin{corollary}\label{cor:qvihard}
    \QVI is \PPAD-hard even if the operator $F$ is monotone, the constraints are linear.
\end{corollary}

\subsection{High-level Idea of the Proof: From Conservative Vector Fields to VIs}\label{sec:ideaproof}

Given the well-known relationship between VIs and fixed points of gradient maps (we recall this relationship formally in \Cref{th:fromVItoMinMax}), we could potentially prove hardness for \MINMAX by using its corresponding VI reformulation, \ie find a point $z\in [0,1]^d$ such that
\[
\tg f(z)^\top(z'-z)\ge 0\quad\forall z'\in [0,1]^d,
\]
where $\tg f=(\nabla_x f, -\nabla_y f)$ is the ``pseudo-gradient'' operator and $z=(x,y)$.
Clearly, this idea has a chance of succeeding only because of the inverted sign in the $\tg$ operator. If instead of $\tg$ we had the classic $\nabla$ operator, this would doom any approach since any (approximately) local minima of $f$ would (approximately) satisfy the VI reformulation.

Moreover, observe that the reduction would be straightforward if we were free to design whatever VI operator we desire.
Unfortunately, we will see in the following that such an approach is over-optimistic.
However, we should not be entirely discouraged.
Indeed, we know that we need to implement very simple operators $F$. A reduction from computing approximate Nash equilibria in polymatrix games shows that even linear vector fields $F(z)=Dz+c$ with $\|D\|_1,\|D\|_\infty\le1$ are \PPAD-hard.\footnote{For a matrix $D$, the $\|D\|_1$ norm represents the maximum of the $\ell_1$ norms of its columns, while the $\|D\|_\infty$ norm corresponds to the maximum of the $\ell_1$ norms of its rows. Additional details are provided in \Cref{app:basic}.}${}^{,}$\footnote{The reader can safely skip momentarily the second part of the statement, as it will only be used for obtaining better dependence on the parameters of the problem.}

\begin{restatable}{theorem}{linvihard}\label{th:linearvihard}
There exists a constant $\rho^*$ such that finding an $\rho^*$-approximation to \LINVI is \PPAD-complete. Moreover, the same result holds even when restricting to deviations along a single component and restricting to $\lVert D\rVert_{1}\le 1$ and $\lVert D\rVert_{\infty}\le 1$, \ie, finding a $z$ such that
\[
    (Dz+c)_i \cdot (z'_i-z_i)\ge -\rho^*\quad \forall i\in[d],\, z_i'\in[0,1].
\]
\end{restatable}

\paragraph{A first attempt}
The simplest construction would be to choose $f$ such that $\tg f(z)=Dz+c$, where $D$ and $c$ define the linear vector field of a \LINVI instance and $K=[0,1]^d$. However, this simple approach leads to a dead end.
The linear form of $Dz+c$ suggests that a second-degree $f$ should be sufficient. Indeed, the operator $\tg$ still maps $n$-degree polynomials into $n-1$ polynomials, exactly as the classic gradient $\nabla$. However, the operator $\tg$ can only produce ``pseudo-gradient fields'', which are not conservative vector fields. Nevertheless, their structure remains too restrictive to implement any arbitrary linear field defined by $D$.
Indeed, it is easy to check that there is no choice of second-degree polynomial $f$ that makes $\tg f(z)$ equal to $(Dz+c)$ for a generic matrix $D$. There are settings in which this equivalence can be attained, for example if $D=-D^\top$, but with this condition \LINVI is not hard as it becomes monotone.\footnote{Any second degree polynomial can be written as $f(z)=z^\top Mz+h^\top z$. Applying to this form $\tg$ we obtain $\tg f(z)=\begin{bmatrix}
    M_{11}+M_{11}^\top & M_{12}+M_{21}^\top\\
    -(M_{12}^\top+M_{21}) & -(M_{22}+M_{22}^\top)
\end{bmatrix}z+\begin{bmatrix}
    h_1\\-h_2
\end{bmatrix}$, where we highlighted the block components of $M$ and $h$. Solving the system $\tg f(z)=Dz+c$ it is easy to realize that there is no obvious class of $D$ and $c$ that can be implemented by $\tg$ and it is large enough to guarantee that \LINVI is still hard.}

\paragraph{A path forward: Imitation gadgets}

In this paragraph, we temporarily relax the assumption of jointly convex constraints and instead consider independent.
Our main idea to address the mentioned complications is to increase the dimensionality relative to the \LINVI instance. 
In particular, we consider functions $f(z,w)$, where $z$ matches the variables of the original \LINVI instance, and $w$ are additional auxiliary variables. The aim is to leverage the additional variables $w$ to implement \LINVI only in the equilibrium constraints of the minimizing player of $f$. The key ingredient here is the idea of \emph{imitation}, according to which we force one player to imitate the other. However, to make this idea effective, we need to exploit the independence of the constraints between $z$ and $w$. The construction works as follows.

Consider the function $f(z,w)=z^\top (Dw + c)$. We have that $\tg f(z,w)=(Dw+c, -D^\top z)$. Then, we can exploit the presence of constraints in our construction. Define the correspondence $\Omega(z,w):= K\times \{z\}$, meaning that the constraints of the $z$ player are independent of $w$, \ie, the $z$ player is free to deviate to any $z'\in K$, while the only available strategy for the $w$ player is to imitate $z$, \ie, the player $w$ is forced to choose $w'=z$. Essentially, this means that the second player cannot deviate. 
Take any point that satisfies the VI, which operator is the pseudo-gradient of $f$, and consider only the component relative to $z$, \ie,
\begin{align*}
&\tg f(z,w)^\top(z'-z,w'-w)\ge 0\quad \forall(z',w')\in\Omega(z,w)\implies
(Dw+c)^\top(z'-z)\ge 0\quad\forall z'\in K.
\end{align*}
Finally, noticing that $w=z$ for each local min-max, \ie, $(w,z)\in \Omega(z,w)$, we can conclude that $z$ is a solution to \LINVI.
This result is interesting in itself, as it proves that monotone QVIs are hard even if the set inclusion problem $z\in Q(z)$ is easy and the operator is monotone. The formal statement of the theorem can be found in \Cref{th:qvihardness}

Our result shows that the use of imitation gadgets is promising, but, to make it work in the setting of jointly convex constraints, we need to refine our construction. In particular, with independent constraints, we could completely ignore the player associated with the extra variables $w$. However, critically, this approach fails with joint convex constraints. Crucially, we will circumvent the related difficulties by exploiting deviations of both players to prove the soundness of our reduction from \LINVI.

\begin{figure}[t!]
\centering
\begin{subfigure}{0.495\textwidth}
    \centering
    \scalebox{1}{\tikzset{every picture/.style={line width=0.75pt}} %

\begin{tikzpicture}[x=0.75pt,y=0.75pt,yscale=-1,xscale=1]

\draw  [fill=mgray  ,fill opacity=0.3 ] (190,120) -- (290,120) -- (290,220) -- (190,220) -- cycle ;
\draw [color=newgreen  ,draw opacity=1 ]   (330,120) ;
\draw [shift={(330,120)}, rotate = 0] [color=newgreen  ,draw opacity=1 ][fill=newgreen  ,fill opacity=1 ][line width=0.75]      (0, 0) circle [x radius= 3.35, y radius= 3.35]   ;
\draw [color=typ_blue  ,draw opacity=1 ][line width=1.5]    (190,120) -- (199.03,83.88) ;
\draw [shift={(200,80)}, rotate = 104.04] [fill=typ_blue  ,fill opacity=1 ][line width=0.08]  [draw opacity=0] (11.61,-5.58) -- (0,0) -- (11.61,5.58) -- cycle    ;
\draw [color=orange2  ,draw opacity=1 ][line width=1.5]    (240,170) -- (316.12,189.03) ;
\draw [shift={(320,190)}, rotate = 194.04] [fill=orange2  ,fill opacity=1 ][line width=0.08]  [draw opacity=0] (11.61,-5.58) -- (0,0) -- (11.61,5.58) -- cycle    ;
\draw  [dash pattern={on 4.5pt off 4.5pt}]  (160,150) -- (140,170) ;
\draw  [dash pattern={on 4.5pt off 4.5pt}]  (260,120) -- (330,120) ;
\draw [color=niceRed  ,draw opacity=1 ][line width=1.5]    (190,120) -- (162.83,147.17) ;
\draw [shift={(160,150)}, rotate = 315] [fill=niceRed  ,fill opacity=1 ][line width=0.08]  [draw opacity=0] (11.61,-5.58) -- (0,0) -- (11.61,5.58) -- cycle    ;
\draw [color=niceRed  ,draw opacity=1 ]   (140,170) ;
\draw [shift={(140,170)}, rotate = 0] [color=niceRed  ,draw opacity=1 ][fill=niceRed  ,fill opacity=1 ][line width=0.75]      (0, 0) circle [x radius= 3.35, y radius= 3.35]   ;
\draw    (240,170) ;
\draw [shift={(240,170)}, rotate = 0] [color={rgb, 255:red, 0; green, 0; blue, 0 }  ][fill={rgb, 255:red, 0; green, 0; blue, 0 }  ][line width=0.75]      (0, 0) circle [x radius= 3.35, y radius= 3.35]   ;
\draw [color=newgreen  ,draw opacity=1 ][line width=1.5]    (190,120) -- (226,120) ;
\draw [shift={(230,120)}, rotate = 180] [fill=newgreen  ,fill opacity=1 ][line width=0.08]  [draw opacity=0] (11.61,-5.58) -- (0,0) -- (11.61,5.58) -- cycle    ;
\draw    (190,120) ;
\draw [shift={(190,120)}, rotate = 0] [color={rgb, 255:red, 0; green, 0; blue, 0 }  ][fill={rgb, 255:red, 0; green, 0; blue, 0 }  ][line width=0.75]      (0, 0) circle [x radius= 3.35, y radius= 3.35]   ;

\draw (192,223.4) node [anchor=north west][inner sep=0.75pt]    {$||z-w||_{\infty } \leq \Delta $};
\draw (222,123.4) node [anchor=north west][inner sep=0.75pt]    {$z$};
\draw (238,173.4) node [anchor=north east] [inner sep=0.75pt]    {$w$};
\draw (142,173.4) node [anchor=north west][inner sep=0.75pt]    {$\tilde{z}'$};
\draw (349,122.4) node [anchor=north east] [inner sep=0.75pt]    {$\tilde{z}$};
\draw (202,83.4) node [anchor=north west][inner sep=0.75pt]    {$Dw+c$};
\draw (322,186.6) node [anchor=south west] [inner sep=0.75pt]    {$-D^{\top } z-c$};

\end{tikzpicture}}
    \caption{$f(z,w)=z^\top (Dw+c)$}
    \label{fig:directions}
\end{subfigure}
\hfill
\begin{subfigure}{0.495\textwidth}
    \centering
    \scalebox{1}{\tikzset{every picture/.style={line width=0.75pt}} %

\begin{tikzpicture}[x=0.75pt,y=0.75pt,yscale=-1,xscale=1]
\input{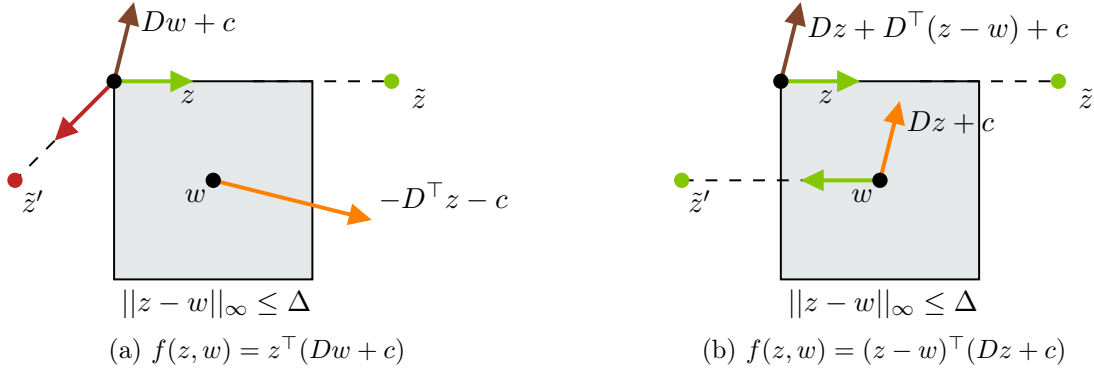}
\draw  [fill=mgray  ,fill opacity=0.3 ] (210,140) -- (310,140) -- (310,240) -- (210,240) -- cycle ;
\draw [color=newgreen  ,draw opacity=1 ]   (350,140) ;
\draw [shift={(350,140)}, rotate = 0] [color=newgreen  ,draw opacity=1 ][fill=newgreen  ,fill opacity=1 ][line width=0.75]      (0, 0) circle [x radius= 3.35, y radius= 3.35]   ;
\draw [color=typ_blue  ,draw opacity=1 ][line width=1.5]    (210,140) -- (219.03,103.88) ;
\draw [shift={(220,100)}, rotate = 104.04] [fill=typ_blue  ,fill opacity=1 ][line width=0.08]  [draw opacity=0] (11.61,-5.58) -- (0,0) -- (11.61,5.58) -- cycle    ;
\draw [color=orange2  ,draw opacity=1 ][line width=1.5]    (260,190) -- (269.03,153.88) ;
\draw [shift={(270,150)}, rotate = 104.04] [fill=orange2  ,fill opacity=1 ][line width=0.08]  [draw opacity=0] (11.61,-5.58) -- (0,0) -- (11.61,5.58) -- cycle    ;
\draw  [dash pattern={on 4.5pt off 4.5pt}]  (160,190) -- (260,190) ;
\draw  [dash pattern={on 4.5pt off 4.5pt}]  (280,140) -- (350,140) ;
\draw [color=newgreen  ,draw opacity=1 ][line width=1.5]    (210,140) -- (246,140) ;
\draw [shift={(250,140)}, rotate = 180] [fill=newgreen  ,fill opacity=1 ][line width=0.08]  [draw opacity=0] (11.61,-5.58) -- (0,0) -- (11.61,5.58) -- cycle    ;
\draw    (210,140) ;
\draw [shift={(210,140)}, rotate = 0] [color={rgb, 255:red, 0; green, 0; blue, 0 }  ][fill={rgb, 255:red, 0; green, 0; blue, 0 }  ][line width=0.75]      (0, 0) circle [x radius= 3.35, y radius= 3.35]   ;
\draw [color=newgreen  ,draw opacity=1 ][line width=1.5]    (260,190) -- (224,190) ;
\draw [shift={(220,190)}, rotate = 360] [fill=newgreen  ,fill opacity=1 ][line width=0.08]  [draw opacity=0] (11.61,-5.58) -- (0,0) -- (11.61,5.58) -- cycle    ;
\draw    (260,190) ;
\draw [shift={(260,190)}, rotate = 0] [color={rgb, 255:red, 0; green, 0; blue, 0 }  ][fill={rgb, 255:red, 0; green, 0; blue, 0 }  ][line width=0.75]      (0, 0) circle [x radius= 3.35, y radius= 3.35]   ;
\draw [color=newgreen  ,draw opacity=1 ]   (160,190) ;
\draw [shift={(160,190)}, rotate = 0] [color=newgreen  ,draw opacity=1 ][fill=newgreen  ,fill opacity=1 ][line width=0.75]      (0, 0) circle [x radius= 3.35, y radius= 3.35]   ;

\draw (212,243.4) node [anchor=north west][inner sep=0.75pt]    {$||z-w||_{\infty } \leq \Delta $};
\draw (238,143.4) node [anchor=north east] [inner sep=0.75pt]    {$z$};
\draw (258,193.4) node [anchor=north east] [inner sep=0.75pt]    {$w$};
\draw (162,193.4) node [anchor=north west][inner sep=0.75pt]    {$\tilde{z} '$};
\draw (369,142.4) node [anchor=north east] [inner sep=0.75pt]    {$\tilde{z}$};
\draw (222,103.4) node [anchor=north west][inner sep=0.75pt]    {$Dz+D^{\top }( z-w) +c$};
\draw (272,153.4) node [anchor=north west][inner sep=0.75pt]    {$Dz+c$};

\end{tikzpicture}}
    \caption{$f(z,w)=(z-w)^\top (Dz+c)$}
    \label{fig:directions2}
\end{subfigure}
\caption{(a) Two possible deviations of $z$ to $\tilde z$ and $\tilde z'$. A deviation towards $\tilde z$ is possible without violating the constraint $\lVert z-w\rVert_{\infty}\le \Delta$, while a deviation toward $\tilde z'$ is not feasible.
(b) The deviation toward $\tilde z$ satisfying the constraint is achievable by either deviating from $z$ or by deviating from $w$. The \textcolor{typ_blue}{\bf{blue}} arrow represents the gradient $\nabla_z f(z,w)$ while the \figorange{orange} arrow represents the gradient $-\nabla_w f(z,w)$.}
\label{fig:directions_tot}
\end{figure}
\paragraph{Overcoming pseudo-gradient fields using approximate imitation gadgets}
A natural imitation gadget for jointly convex constraints is obtained by defining the joint space $\Omega\coloneqq\{(z,w)\in K\times K: \|z-w\|_\infty\le\Delta\}$ for some small $\Delta$. Differently from the case of independent constraints, also the deviations $z'$ are forced to be close to $w$, \ie, we can consider only deviations of the kind $(z',w)\in\Omega$. Therefore, there are some points of $K$ toward which $z$ cannot move.
Indeed, if we try to use $f(z,w)=z^\top(Dw + c) $ as in the previous paragraph (which leads to $\tg f(z,w)=(Dw +c,-D^\top z)$), we inevitably fail to cover some directions.
\Cref{fig:directions} shows two examples of possible deviations. Deviation $\tilde z$ is covered by a deviation of the $z$ player, but deviation $\tilde z'$ cannot be covered. This is because the $z$ player cannot move even slightly in the direction of $\tilde z'$ due to the constraint that $\|z'-w\|_\infty\le\Delta$ for any deviation $z'$. This means that a solution to the local min-max problem does \emph{not} guarantee that $(Dz+c)^\top(\tilde z'-z)\ge0$.

A closer look at \Cref{fig:directions} suggests a new idea. 
It is inevitable that there are directions $\tilde z - z$, with $\tilde z \in K$, along which $z$ cannot move.
However, it is possible that even if $z$ cannot, $w$ might be able to move in that direction.
We see that by choosing $f(z,w)=z^\top (Dw+ c)$ the interests of $w$ player (\figorange{orange} arrow in \Cref{fig:directions}) are misaligned to the interests of the $z$ player (\textcolor{typ_blue}{\bf{blue}} arrow in \Cref{fig:directions}). Indeed, $\tg f(z,w)=(Dw+c,-D^\top z)$.
Even if $w\approx z$, clearly $Dw+c\not\approx -D^\top z$ and the optimality condition of player $w$ along the direction $\tilde z'$ does \emph{not} imply that optimality conditions of the \LINVI instance are satisfied along that direction.
This is unsurprising; indeed, this approach will fail unless the function $f$ is modified. In this case, $\tg f$ yields a monotone \LINVI, which is solvable in polynomial time. Therefore, any function $f$ we aim to construct must be at least quadratic (nonconvex) in either $z$ or $w$. This is expected since the convex-concave problem can be solved in polynomial time over jointly convex domains \citep{facchinei2003finite}.

A possible idea to realign the players' incentives is to design $f$ so that $\tg f(z,w)\approx F_{\text{ideal}}(z,w)\coloneqq(Dz+c, Dw+c)$. While it is clear that inducing this pseudo-gradient exactly is not possible, we can attempt to obtain a good approximation by exploiting the imitation gadget implemented by the constraints. For example, the function $f(z,w)=(z-w)^\top (Dz+c)$ induces the operator $\tg f=(Dz+D^\top(z-w)+c,Dz+c)$ (see \Cref{fig:directions2}). This, together with the imitation gadget obtained through the constraints $\Omega$ that forces $w\approx z$, yields $\tg f(z,w)\approx F_{\text{ideal}}(z,w)$. 

\paragraph{The importance of dealing with the hypercube}
\begin{wrapfigure}[18]{r}{0.4\textwidth} 
    \vspace{-0.2cm}
    \centering
    \scalebox{1}{\tikzset{every picture/.style={line width=0.75pt}} %

\begin{tikzpicture}[x=0.75pt,y=0.75pt,yscale=-1,xscale=1]
\input{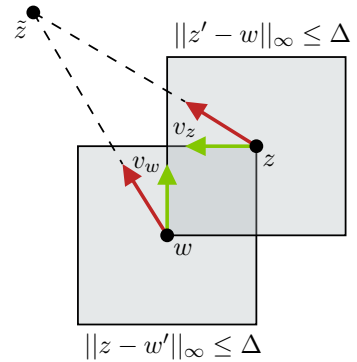}
\draw  [fill=mgray  ,fill opacity=0.3 ] (150,125) -- (250,125) -- (250,225) -- (150,225) -- cycle ;

\draw  [fill=mgray  ,fill opacity=0.3 ] (200,75) -- (300,75) -- (300,175) -- (200,175) -- cycle ;

\draw    (125,50) ;
\draw [shift={(125,50)}, rotate = 0] [color={rgb, 255:red, 0; green, 0; blue, 0 }  ][fill={rgb, 255:red, 0; green, 0; blue, 0 }  ][line width=0.75]      (0, 0) circle [x radius= 3.35, y radius= 3.35]   ;

\draw [color=niceRed  ,draw opacity=1 ][line width=1.5]    (250,125) -- (213.39,102.12) ;
\draw [shift={(210,100)}, rotate = 32.01] [fill=niceRed ,fill opacity=1 ][line width=0.08]  [draw opacity=0] (11.61,-5.58) -- (0,0) -- (11.61,5.58) -- cycle    ;

\draw [color=niceRed  ,draw opacity=1 ][line width=1.5]    (200,175) -- (177.12,138.39) ;
\draw [shift={(175,135)}, rotate = 57.99] [fill=niceRed  ,fill opacity=1 ][line width=0.08]  [draw opacity=0] (11.61,-5.58) -- (0,0) -- (11.61,5.58) -- cycle    ;
\draw  [dash pattern={on 4.5pt off 4.5pt}]  (125,50) -- (175,135) ;
\draw  [dash pattern={on 4.5pt off 4.5pt}]  (125,50) -- (210,100) ;
\draw [color=newgreen  ,draw opacity=1 ][line width=1.5]    (250,125) -- (214,125) ;
\draw [shift={(210,125)}, rotate = 360] [fill=newgreen  ,fill opacity=1 ][line width=0.08]  [draw opacity=0] (11.61,-5.58) -- (0,0) -- (11.61,5.58) -- cycle    ;
\draw [color=newgreen  ,draw opacity=1 ][line width=1.5]    (200,175) -- (200,139) ;
\draw [shift={(200,135)}, rotate = 90] [fill=newgreen  ,fill opacity=1 ][line width=0.08]  [draw opacity=0] (11.61,-5.58) -- (0,0) -- (11.61,5.58) -- cycle    ;
\draw    (200,175) ;
\draw [shift={(200,175)}, rotate = 0] [color={rgb, 255:red, 0; green, 0; blue, 0 }  ][fill={rgb, 255:red, 0; green, 0; blue, 0 }  ][line width=0.75]      (0, 0) circle [x radius= 3.35, y radius= 3.35]   ;
\draw    (250,125) ;
\draw [shift={(250,125)}, rotate = 0] [color={rgb, 255:red, 0; green, 0; blue, 0 }  ][fill={rgb, 255:red, 0; green, 0; blue, 0 }  ][line width=0.75]      (0, 0) circle [x radius= 3.35, y radius= 3.35]   ;

\draw (202,178.4) node [anchor=north west][inner sep=0.75pt]    {$w$};
\draw (252,128.4) node [anchor=north west][inner sep=0.75pt]    {$z$};
\draw (123,53) node [anchor=north east] [inner sep=0.75pt]    {$\tilde{z}$};
\draw (152,228.4) node [anchor=north west][inner sep=0.75pt]    {$||z-w'||_{\infty } \leq \Delta $};
\draw (202,71.6) node [anchor=south west] [inner sep=0.75pt]    {$||z'-w||_{\infty } \leq \Delta $};
\draw (198,135) node [anchor=east] [inner sep=0.75pt]    {$v_{w}$};
\draw (210,121.6) node [anchor=south] [inner sep=0.75pt]    {$v_{z}$};

\end{tikzpicture}}
    \caption{Neither $w$ nor $z$ can be moved towards $\tilde z$. However, we can decompose $\tilde z-z$ into $v_w$ and $v_z$ such that $v_w+v_z$ is parallel to $\tilde z-z$.}
    \label{fig:jointdev}
\end{wrapfigure}
In the proof of \Cref{th:nonconvexconcavehardness} we will rely heavily on the fact that \LINVI remains \PPAD-hard even when restricted to the hypercube. This has several important consequences for our reductions. The most important one is that we can focus on the individual components of a deviation $\tilde z$.
Indeed, we only have to consider constraints that are independent among components, namely the hypercube constraint and the approximate imitation gadget constraint implemented by the infinity norm.
In our reductions, for every deviation $\tilde z$, we analyze each of its $i$-th components independently. For all $i\in[d]$, we show that either $z_i$ or $w_i$ can be moved towards $\tilde z_i$ since at least one of the two deviations is feasible. Then, we identify two \emph{orthogonal} deviations $v_z$ and $v_w$, which are summed to define a vector along the direction $\tilde z-z$ (see \Cref{fig:jointdev} for an intuitive picture). This approach would not have been feasible with constraints that are not independent among components. Indeed, this would impose joint constraints between variables and make it impossible to construct the deviations component-wise.

\paragraph{Single-components deviations and constant approximations for \GDAFP}

Now we leverage another key feature of our hardness result for \LINVI, specifically that the problem remains hard when considering single-component deviations (second part of \Cref{th:linearvihard}). This will be key to obtain constant inapproximability for \GDAFP in $\ell_\infty$-norm, as discussed in \Cref{sec:constant_inapprox}.

The main objective of this section is to show how we can obtain a reduction with $G_\infty=O(1)$ and a that works even for $\delta=\Omega(\sqrt{d}\frac\epsilon G)$ instead of $\delta=\Omega(d\frac\epsilon G)$.

Let's assume that for a deviation $\tilde z$ the direction $\tilde z-z$ is feasible for the $z$-player. The locality condition implies that the furthest we can move in this direction is $O(\delta/\|\tilde z-z\|)$ (which is $O(\delta/\sqrt{d})$ in the worst case). This is obtained by considering the local deviation $z'=z+O(\delta/\sqrt{d})(\tilde z-z)$. 
Hence, the locality condition would add a $\sqrt{d}$ factor in the approximation results:
\[
f(z,w)\le f(z',w)+\epsilon \implies (Dz+c)^\top(\tilde z-z)\ge -\frac{\epsilon\sqrt{d}}\delta.
\]
Then, we would have to set $\delta=\Omega(\epsilon\sqrt{d})$ and thus $\delta=\Omega(d\frac{\epsilon}{G})$ to obtain a constant approximation for the \LINVI instance.
However, we can employ a simple trick that sidesteps this problem. If we consider only single-component deviations $\tilde z=(\tilde z_{i},z_{-i})$, then $\| \tilde z-z\|\le \delta$. Hence, the amount we can move $z$ in the direction $\tilde z-z$ is exactly $\delta$ by taking $z'=z+\delta(\tilde z_i-z_i)$ (assuming as above that the direction $z'-z$ is feasible). This allows us to recover the instance-dependent factor, as simple calculations show that
\[
f(z,w)\le f(z',w)+\epsilon \implies (Dz+c)_i\cdot(\tilde z_i-z_i)\ge -\frac{\epsilon}\delta,
\]
even for $\epsilon,\delta = O(1)$ and thus $\delta=\Omega(\sqrt d\frac\epsilon G)$.

Finally, we have to transfer this result to \GDAFP. A direct application of the reduction from \LRMINMAX to \GDAFP (introduced by \cite{daskalakis2021complexity} and rephrased in \Cref{th:GDA_eq_minmax}), would yield $\alpha=O(\frac{1}{G}(\epsilon-\frac{L}{2}\delta^2))$. Such $\alpha$ is not constant in our instance as $G=O(\sqrt{d})$. However, we can design a specialized reduction from \LRMINMAX to \GDAFP that exploits that \LRMINMAX is \PPAD-hard even for single-component deviations. Intuitively, the parameter $\alpha$ in our specialized reduction from \LRMINMAX to \GDAFP (\Cref{lem:from_GDA_to_MM}; see \Cref{app:omitted hard} for its proof) depends on $G_\infty=\sup_{x,y}\|\nabla f(x,y)\|_\infty=O(1)$ instead of $G=\sup_{x,y}\|\nabla f(x,y)\|=\Theta(\sqrt{d})$.
This makes it possible to obtain hardness results for $\alpha=O(\frac{1}{G_\infty}(\epsilon-\frac{L}{2}\delta^2))$ which is constant since $L,G_\infty,\epsilon,\delta=O(1)$ and $\delta/(\epsilon/G)=\sqrt{n}$.

\begin{restatable}{lemma}{lemmagdamm}\label{lem:from_GDA_to_MM}
    Assume that for all $y\in[0,1]^d$ there exist $d$ intervals $\{K_{1,j}(y_j)\}_{j\in[d]}$ such that $K_1(y)=\bigtimes_{j\in[d]}K_{1,j}(y_j)$ and $d$ intervals $\{K_{2,j}(x_j)\}_{j\in[d]}$ such that $K_2(x)=\bigtimes_{j\in[d]}K_{2,j}(x_j)$. Let $(x^\star,y^\star)\in K_1(y^\star)\times K_2(x^\star)$ satisfy 
    \[
    \|(x^\star,y^\star)-F^\nu_\GDA(x^\star,y^\star)\|\le \alpha
    \]
    Then, for all $i\in[d]$ and all single-component deviation $x'=(x_i',x^\star_{-i}), y'=(y_i',y^\star)$ it holds
    \begin{align*}
    f(x^\star,y^\star) &\le f(x',y^\star)+ \alpha(1+G_\infty)+\frac{L}{2}\delta^2\quad \forall x'\in K_1^\nu(y^\star)\quad\text{and}\quad\|x'-x^\star\|\le\delta, \\
    f(x^\star,y^\star) &\ge f(x^\star,y')  -\left.\alpha(1+G_\infty)-\frac{L}{2}\delta^2\right. \quad \forall y'\in K_2^\nu(x^\star)\quad\text{and}\quad\|y'-y^\star\|\le\delta,
    \end{align*}
    where $ G_\infty=\sup_{x,y\in[0,1]^d} \|\nabla f(x,y)\|_\infty$.
    
    The statement holds even if we only have $\|(x^\star,y^\star)-F^\nu_\GDA(x^\star,y^\star)\|_\infty\le \alpha$.
\end{restatable}

\subsection{Constant Inapproximability of \GDAFPJCinfty}\label{sec:constant_inapprox}

We consider the following problem related to computing fixed points of $F_\GDA$ in $\ell_\infty$-norm.

\begin{problem}
[\GDAFPJCinfty]\label{def:gdafpinfty}
Given $\eta>0$, $\alpha$, $L$, $G_\infty$, $\nu\in\Reals_+$, a circuit implementing a $G$-Lipschitz and $L$-smooth function $f:[0,1]^d\times[0,1]^d\to [-B,B]$ and its gradient $f:[0,1]^d\times[0,1]^d\to \Reals^{2d}$, where the $\ell_\infty$-norm of its gradients are upper bounded by $G_\infty$. Moreover we are given a polytope $K\subset[0,1]^{2d}$ such that the slices $K_1(y)$ and $K_2(x)$ are product of intervals for every $x$ and $y$. The problem is to find $(x^\star,y^\star)\in[0,1]^{2d}$ such that $(x^\star,y^\star)\in K$ and 
\[ 
\|x^\star-\Pi_{K_1(y^\star)}(x^\star-\eta \nabla_x f(x^\star,y^\star))\|_\infty\le\alpha\quad   \textnormal{and}\quad\|y^\star-\Pi_{K_2(x^\star)}(y^\star+\eta \nabla_y f(x^\star,y^\star))\|_\infty\le\alpha,
\]
or output $\bot$ if no such point exists.
\end{problem}

The problem becomes trivially solvable in polynomial time ---since any $(x,y)$ is a solution--- if $\eta=0$. More generally, even when $\eta=O(\alpha/G_\infty)$ any $(x,y)$ is a solution to \GDAFPJCinfty provided that the slices $K_1(y)$ and $K_2(x)$ are product of intervals, \ie $K_1(y)=\bigtimes_{j\in[d]} K_{1,j}(y_j)$ for some intervals $\{K_{1,j}(y_j)\}_{j\in[d]}$ and analougusly for the $y$ player.
Indeed, by the non-expansiveness of the projection (which applies since the slices $K_1(y)$ are a product of intervals), we have:
\[
\|x-\Pi_{K_1(y)}(x-\eta \nabla_x f(x,y))\|_\infty\le \eta\|\nabla_x f(x,y)\|_\infty\le \alpha,
\]
and an analogous bound holds for the $y$ player.

The instance used in the proof of \Cref{th:nonconvexconcavehardness} satisfies all the conditions of \Cref{def:gdafpinfty}. Moreover, it is clear from the proof that \Cref{lem:from_GDA_to_MM} still holds when the fixed points of $F_\GDA$ are measured in $\ell_\infty$-norm. Therefore, by slightly modifying the final part of the proof of \Cref{th:nonconvexconcavehardness}, we can also prove that \GDAFPJCinfty is \PPAD-hard when $\eta/(\epsilon/G_\infty)=\Omega(1)$. This result is tight since, as per the discussion above,  the problem is in \FP  for $\eta/(\epsilon/G_\infty)=O(1)$.

\section*{Acknowledgments}

This work is supported by an ERC grant (Project 101165466 — PLA-STEER). MB, MC are partially supported by the FAIR (Future Artificial Intelligence Research) project PE0000013, funded by the NextGenerationEU program within the PNRR-PE-AI scheme (M4C2, investment 1.3, line on Artificial Intelligence). MC is also partially supported by the EU Horizon project ELIAS (European Lighthouse of AI for Sustainability, No. 101120237). GF was supported in part by the National Science Foundation award CCF-2443068, the Office of
Naval Research grant N000142512296, and an AI2050 Early Career Fellowship.

\printbibliography
\newpage
\appendix
\section*{Appendix}
\section{Complexity Classes and Reductions}\label{sec:cc}

\paragraph{Model of computation}

We work in the standard Turing model of computation in which inputs are represented by irreducible fractions and natural numbers are represented in binary \citep{arora2009computational}. Functions are represented by algebraic (also known as arithmetic) circuits. 
An algebraic circuit is a directed acyclic graph in which each node represents one of the operations $\{+,*,\max,\times c\}$ where for $c\in\Rationals$ the symbol $\times c$ denotes the multiplication by some constant $c$.
All nodes have in-degree $2$ and out-degree $1$ except for the $*c$ nodes with in- and out-degree $1$. The tree's leaves are the inputs, and nodes without children are the outputs. 
For rational numbers $x$, we denote with $\size(x)$ the number of bits used to write down the numerator and denominator, while for a circuit $\mathcal{C}$, we use $\size(\mathcal C)$ to denote the number of bits used to write down its gates and constants.\footnote{
We adopt the assumption from \citet{fearnley2022complexity} that algebraic circuits are \emph{well-behaved}, meaning that any directed path leading to an output contains at most $\log(\size(\mathcal{C}))$ multiplication gates. This condition ensures efficient evaluation of the circuit by preventing iterated squaring, which could otherwise produce doubly exponential numbers with an exponential representation size relative to the input and circuit size.
}

\paragraph{Total search problems}
A search problem is defined by a binary relation $P$, where for each input $x\in\{0,1\}^*$ the objective is to find a $y\in\{0,1\}^*$ such that $P(x,y)$ evaluates to \True, and returns ``$\bot$'' if $P(x,y)$ evaluates to \False for all $y\in\{0,1\}^*$. \FNP is the class of search problems in which $P(x,y)$ can be evaluated in polynomial time in $\size(x)+\size(y)$, and there
exists some polynomial $p$ such that if there exists $ y$ such that $P(x,y)=\True$ then there
exists $z$ such that $P(x,z)=\True$ and $\size(z) \le p(\size(x))$.

We will work with a subset of problems in \FNP for which a solution is guaranteed to exist, specifically within the class \TFNP (\emph{total search problems in NP}). This class contains the search problems in \FNP for which, for every input $x$, there exists a $y$ such that $P(x,y)$ evaluates to \True. 

Let $\Pi_1$ and $\Pi_2$ be two problems in \TFNP.  We say that \emph{$\Pi_1$ reduces to $\Pi_2$} if there exist two polynomial-time functions $a$ and $b$ such that for any instance $x$ of $\Pi_1$, if we can find a solution $y$ to instance $a(x)$ of $\Pi_2$, then $b(y)$ yields a solution to instance $x$ for $\Pi_1$. In symbols, $\Pi_1$ reduces to $\Pi_2$ if, for all $x,y$,
\[ 
\Pi_2(a(x),y)=\True \quad \implies \quad \Pi_1(x, b(y))=\True.
\]
This indicates that $\Pi_2$ is ``more difficult'' than $\Pi_1$. Moreover, if we could efficiently find a solution $y$ for $a(x)$, we would also be able to efficiently find a solution $b(y)$ for $\Pi_1$.

\paragraph{The class \PPAD} It is widely believed that \TFNP has no complete problems \citep{megiddo1991total}.
Hence, its analysis has been advanced by identifying and studying interesting subclasses. These subclasses are characterized by the type of existence argument used to prove their membership in \TFNP. One of the most extensively studied subclasses is \PPAD \citep{papadimitriou1994complexity}. The complexity class $\PPAD\subseteq\TFNP$ consists of all problems which can be reduced to the \EOTL problem, which we recall next.

\begin{problem}[\EOTL]
Given in input two boolean circuits $S:\{0,1\}^n\to\{0,1\}^n$ and $P:\{0,1\}^n\to\{0,1\}^n$, such that $S(0^n)\neq0^n$ and $P(0^n)=0^n$, find $\omega\in\{0,1\}^n$ such that $P(S(\omega))\neq\omega$ or $S(P(\omega))\neq\omega\neq 0^n$.
\end{problem}
The two circuits $P$ and $S$ implicitly define a directed graph with an exponential number of vertices, each corresponding to an element of $\{0,1\}^n$, where each vertex has in- and out-degree of at most $1$. The goal of \EOTL is to find a non-trivial source or a sink.

An important \PPAD-complete problem is finding fixed points of continuous functions, whose existence is guaranteed by Brouwer's fixed point theorem \citep{brouwer1911abbildung}. 

\begin{problem}[\Brouwer]
    Given a precision parameter $\epsilon\ge0$ and a circuit implementing an $L$-Lipschitz function $F:[0,1]^d\to[0,1]^d$, find a point $z\in[0,1]^d$ such that
        $\|F(z)-z\| \le \epsilon.$
\end{problem}

Another \PPAD-complete problem that we will use extensively in this paper is related to computing equilibria in polymatrix games. In these games, agents are represented as nodes of an undirected graph, where each agent plays a bi-matrix game with every other node to whom they are connected.

\begin{problem}[\POLY]\label{pr:polymatrix}
Given a graph $(V,E)$, a matrix $A_{i,j}\in [0,1]^{n\times n}$ for each $(i,j)\in E$, and an approximation $\epsilon>0$, find vectors $x_i\in\Delta_n$, $i \in V$ such that:
\[
\sum_{j\in V:(i,j)\in E} x_i^\top A_{i,j} x_j \ge \sum_{j\in V:(i,j)\in E} (A_{i,j} x_j)_k-\epsilon\quad\forall  i \in V,k\in [n].
\]
\end{problem}
\POLY was at the center of a landmark result of \citet{rubinstein2015inapproximability}, which first proved constant inapproximability of equilibria. 

\begin{proposition}[\protect{\cite[Theorem~1]{rubinstein2015inapproximability}}]\label{th:polymatrix}
    There exists a constant $\epsilon^*$ such that \POLY is \PPAD-complete for $\epsilon^*$, even when the graph has degree $3$, and each player has $n=2$ actions.
\end{proposition}

\noindent Recently, \citet{deligkas2024pure} improved the constant to $\epsilon^*\approx 0.088$.\footnote{The result by \citet{deligkas2024pure} applies even in the case of normalized payoffs. Since our payoffs are not normalized, we could, in principle, use a larger constant. We choose not to, as we do not focus on optimizing constants.}

\paragraph{Promise problems}

Formally, problems in \TFNP are expected to guarantee the existence of a solution for any input. However, many of our problems admit solutions only when certain conditions are met, placing them in \TFNP only for a restricted subset of inputs. The community refers to these conditions as \emph{promises}. For instance, we may promise that a circuit indeed defines an $L$-Lipschitz function.
One recent approach to circumvent this issue and obtain problems that are formally in \TFNP is to add a certificate of promise violations to the solutions of the problems \citep{daskalakis2011continuous}.
However, verifying violations of the promises may introduce additional complexities in solving the problem. For instance, verifying whether a circuit computes an $L$-Lipschitz function could be significantly more challenging than finding a solution to the original problem.
A principled approach proposed by \citet{fearnley2022complexity} to address this issue is to utilize \emph{promise-preserving reductions}, where any violation of certain promises corresponds to a violation of the promise of the original problem from which we are reducing. In our specific setting, while this may be more formally accurate, it complicates the statements and proofs without significantly adding merits to our results. Therefore, we concentrate on the promise version of our problems.
Moreover, when proving hardness results, the instances generated by our reductions will enforce the promises syntactically, and thus our hardness results will not rely on the hardness of checking the satisfaction of the promises.

\section{\PPAD Membership}\label{sec:existence}

There are a few recent works that give general techniques to prove \PPAD-membership of \TFNP problems. \citet{papadimitriou2023computational} is the first to define a meaningful computational problem related to Kakutani's theorem. Their main contribution is to enable a highly general representation of sets (\eg, via separation oracles). Their main interest is Kakutani's problem in the context of concave games, whereas our focus is on QVIs, which are more general. More details about this relationship can be found in \Cref{app:proofqvippad}. 
More recently, \citet{filos2024ppad} extended the ``pseudogate'' technique of \citet{filos2023fixp} (originally used for \textsc{FIXP}-membership results) to the \PPAD setting. They designed a general technique, based on pseudogates, that can be used to easily prove membership of an impressive number of problems with exact rational solutions.
Independently and concurrently, \citet{kapron2024computational} extended the result by \citet{papadimitriou2023computational}, and showed \PPAD-membership of QVIs when the correspondences are described via separation oracles (and thus capture more general settings than ours). One difference is that \citet{kapron2024computational} assumes that the set-valued function of the QVI is Lipschitz with respect to the Hausdorff distance, whereas we do not. However, as part of our proof, we show that a particular relaxation of our QVI instance is Lipschitz with respect to the Hausdorff distance. See \Cref{app:irrational} for more details.

In this section, we describe the \PPAD-membership results for the case of independent constraints. 
The \PPAD membership of \GDAFPCBC is proved by showing the \PPAD membership of  \QVI, which is very general and can be of independent interest.\footnote{For the reasons discussed in \Cref{sec:prelimGDA}, we cannot formally provide membership results for \LRMINMAX since such problems are not in \TFNP.}

\subsection{Existence of Solutions With Jointly Convex Constraints}\label{sec:existenceJoint} 

We begin by briefly reviewing how \citet{daskalakis2021complexity} established \PPAD membership for \MINMAX under jointly convex constraints and explaining why this approach does not readily extend to independent constraints. \citet{daskalakis2021complexity} introduced the \emph{safe gradient descent-ascent} map $F_\SGDA$ in which the projection step is joint and thus ``safe'', \ie, the resulting point is guaranteed to lie in $K$. Formally,  $F_\SGDA$ is defined as follows: 
\[
F_{\SGDA}(x,y) = \Pi_{K}\begin{pmatrix}x-\nabla_xf(x,y)\\ y+\nabla_yf(x,y)\end{pmatrix}.
\]
Note that this joint projection step can only be defined for jointly convex constraints. In \Cref{app:QVIandVI}
we give a thorough overview of the relationship of QVIs, VIs, and fixed points of various gradient descent-ascent maps.

The membership proof by \citet{daskalakis2021complexity} proceeds roughly as follows: fixed points of $F_{\SGDA}$ are fixed points of $F_\GDA$, which are, by \Cref{th:GDA_eq_minmax}, solutions to \LRMINMAXJC. Additionally, the problem of computing fixed points of $F_{\SGDA}$ is in \PPAD through a standard reduction to \Brouwer.

\begin{theorem}[\protect{\cite[Theorem~5.2]{daskalakis2021complexity}}] \label{thm:daskalakiscomplexity}
\GDAFPJC  is in \PPAD.
\end{theorem}

Extending this result to the case of independent constraints introduces fundamental complications, which can be understood by looking at the well-known equivalence between VIs and approximate fixed points of the safe gradient steps (see \citet{facchinei2003finite} and \Cref{th:VI_eq_sgda} for details).
In particular, when the constraints are jointly convex, the problem reduces to a single player optimizing a ``complicated'' function on the joint set $K$. However, with independent constraints, there is no straightforward single-player formulation or VI formulation of the problem. As we already pointed out, $F_{\GDA}$ does not guarantee the resulting points to be feasible. Moreover, even with jointly convex constraints, solutions to $F_{\GDA}$ and $F_{\SGDA}$ are not equivalent as fixed points of $F_{\GDA}$ are not necessarily fixed points of $F_{\SGDA}$ (see \Cref{ex:eq_not_vi}). 

This suggests that we should look to a larger class of problems to provide existence results for independent constraints.
As we demonstrate in the following section, min-max problems with independent constraints are equivalent to QVIs rather than VIs, making the original approach inapplicable.

\subsection{Existence of Solutions With Independent Constraints}

In this section, we discuss the existence of solutions and \PPAD-membership of \GDAFPCBC. Moreover, we show why it is necessary to relax the constraint satisfaction. In particular, we can easily show that, without relaxing the constraints, a solution to \QVI and \MINMAX might not exist even for independent constraints (the proof of this proposition and the proof of the other results on the \PPAD-membership with independent constraints is provided in \Cref{app:irrational}). 

\begin{restatable}{proposition}{irrationalandfriends}\label{th:irrationalandfriends}
    \QVI, \LRMINMAX, and \GDAFP with $\nu=0$ do not always admit a solution. On the other hand, \Kakutani always admits a (potentially irrational) solution even when $\nu=0$.\footnote{We have slightly abused notation, as \QVI is formally defined only for $\nu > 0$.}
\end{restatable}

This should not come as a surprise. Indeed, the standard result regarding the existence of solutions to QVI \citep{chan1982generalized} requires the correspondence \( Q \) to be continuous.\footnote{Different definitions of continuity for correspondences are recalled in \Cref{app:basic}.}
However, our correspondences are of the form $Q(z)=\{z'\in[0,1]^d:A(z)z'-b(z)\le0\}$ for an $L$-Lipschitz continuous matrix $A:[0,1]^d\to[0,1]^{m\times d}$ and an $L$-Lipschitz continuous vector $b:[0,1]^d\to[0,1]^{m}$. This generates upper semicontinuous correspondences but not continuous ones. In contrast, Kakutani theorem \citep{kakutani1941generalization} requires only upper semicontinuous correspondences to guarantee the existence of solutions. This is precisely the reason why, in \Cref{th:irrationalandfriends}, we could prove that \Kakutani always admits solutions. The dichotomy between \QVI and \Kakutani highlighted in \Cref{th:irrationalandfriends} shows that the class of correspondences chosen in our problems is not trivial, and has interesting structural properties.

The following theorem relates specific \QVI problems to \LRMINMAX problems and, by \Cref{th:GDA_eq_minmax}, to \GDAFP (the proof is provided in \Cref{sec:qvi eq minmax}). We observe that even an arbitrarily small relaxation of the constraints is enough to establish the result.

\begin{restatable}[Two-way reduction between \QVI and \LRMINMAX]{theorem}{qvi}\label{th:MM_eq_QVI}
Consider a \LRMINMAX problem where $f$ is $G$-Lipschitz and $L$-smooth. Then:
\begin{enumerate}[label=\roman*)]
    \item For any $\epsilon,\delta>0,$ and $\nu\ge 0$, define $\alpha= \frac{1}{8}\Big(\sqrt{(G+\delta)^2+4(\epsilon-\frac{L}{2}\delta^2)}-(G+\delta)\Big)^2$. Any $(x,y)$ such that $g_i(x,y)\le\nu$ for $i\in\{1,2\}$ and
    \[
    \left(\nabla_x f(x,y), -\nabla_y f(x,y)\right)^\top\,\left(x'-x, y'-y\right)\ge -\alpha \,\textnormal{ for all }(x',y')\in K^\nu_1(y)\times K^\nu_2(x),
    \]
    is an $(\epsilon,\delta,\nu)$-solution to \LRMINMAX.
    \item Any  $(\epsilon,\delta,\nu)$-solution $(x,y)$ to \LRMINMAX satisfies $g_i(x,y)\le\nu$ for $i\in\{1,2\}$ and
    \[
    \left(\nabla_x f(x,y), -\nabla_y f(x,y)\right)^\top\hspace{-2.6pt}\left(x'-x,y'-y\right)\ge -\sqrt{d} \left(3 L\delta+2\frac{\epsilon}{\delta}\right)\textnormal{ for all }(x',y')\in K^\nu_1(y)\times K^\nu_2(x).
    \] 
 \end{enumerate}
\end{restatable}

Our main membership result shows that \QVI is in \PPAD, and hence the same result holds for \GDAFPCBC by the previous discussion (the proof is provided in \Cref{app:proofqvippad}).

\begin{restatable}{theorem}{qvippad}\label{th:qvippad}
    \QVI and \GDAFPCBC are in \PPAD.
\end{restatable}

The above theorem is proven by showing a reduction from \QVI to a high-dimensional version of Sperner's Lemma. 
In \Cref{app:proofqvippad}, we provide all the details of this result. In particular, we discuss in detail why we need to reduce to Sperner rather than to \Brouwer. Intuitively, we cannot find a continuous selector function for the correspondence $Q$ that is Lipschitz and for which fixed points are approximate solutions of \QVI.

\Cref{th:MM_eq_QVI}, together with \Cref{th:qvippad}, also shows that there exists a reduction from \LRMINMAXCBC to \EOTL (or any other \PPAD-complete problem). However, as already discussed in \Cref{sec:prelimGDA}, this alone is not enough to conclude that the problem is in \PPAD, as we lack \FNP membership, which we ruled out. Notably, the existence of this reduction highlights that certain solutions to \LRMINMAXCBC have polynomial witnesses.

\subsection{Globalization} 
To be concise, we only gave existence results for nonconvex-nonconcave problems. Clearly, these results can hold for the local regime (\ie, $\delta\le\sqrt{2\epsilon/L}$). However, as expected, when the payoff function of the minimizing player is convex (or, equivalently, when it's concave for the maximizing player), we can ``globalize'' the solution and extend the existence result to $\delta\ge\sqrt{d}$, which is the diameter of the hypercube and thus encodes global deviations. In \Cref{app:proofqvippad} we provide the proof for the case in which the minimizer (\ie, the $x$ player) has a convex function (a similar statement would hold for $y$ if $f(x,\cdot)$ was concave).

\begin{restatable}{proposition}{globalization}\label{th:globalization}
    For any $(x^\star,y^\star)$ which is an $(\epsilon,\delta,\nu)$-solution to \MINMAX, if the function $f(\cdot,y)$ is convex for any $y\in[0,1]^d$, then 
    $f(x^\star,y^\star)\le f(x',y^\star)+\frac{\epsilon\sqrt{d}}{\delta}\quad\forall x'\in K_1^\nu(y^\star).$
\end{restatable}

In particular, in the case of convex-concave objectives, given any $\tilde \epsilon>0$, we can take  $\epsilon=\frac{\tilde \epsilon^2}{dL}$ and $\delta=\frac{\tilde\epsilon}{L\sqrt{d}}$. Then, since there exists  an $(\epsilon,\delta,\nu)$-solution for any $\delta<\sqrt{\frac{2\epsilon}{L}}$, we have that there exists an $(\tilde\epsilon, \tilde \delta, \nu)$-solution where $\tilde \delta=\sqrt{d}$. The same reasoning can be easily extended to cases where only one between the minimizer or maximizer is convex/concave.

\subsection{Application to Generalized Equilibria}\label{sec:gnep}

Our \PPAD membership result has implications beyond local min-max equilibria. It shows that we can work with highly general correspondences while still ensuring the existence of \emph{rational} (and verifiable) solutions. At a high level, we show that a slight relaxation of the correspondence is sufficient to guarantee the existence of solutions when no solution exists in the non-relaxed problem. Moreover, this ensures the existence of rational solutions even when all solutions of the original correspondence are irrational. Alternatively, by looking at our proof, we observe that this relaxation is not strictly required and can be replaced by the assumption that the correspondence $Q(z)$  has always non-zero volume. To illustrate the generality of our approach, we include an application to generalized equilibria.

In a recent work about a general convex optimization approach to \PPAD membership results \citep{filos2024ppad}, the authors considered a natural computational version of Generalized Equilibrium Problems (GNEPs) \citep{debreu1952social, facchinei2010generalized}. Using a reduction to \QVI, we can extend the membership result of \citet[Theorem 4.4]{filos2024ppad} to a broader class of constraints by relaxing the correspondences.

The problem is defined by a set of $n$ players, and each player $i$ has a set of actions $X_i\subseteq [0,1]^l$, and its utility is represented through a circuit computing a concave $L_u$-Lipschitz and $G$-smooth function $u_i:\bigtimes_{j\in[n]}X_j\to\Reals$, its gradient $\nabla_{x_i}u_i:\bigtimes_{j\in[n]}X_j\to \Reals^l$, and a correspondence $\Gamma_i:X_{-i}\rightrightarrows X_i$.\footnote{We denote with $x_{-i}$ the tuple $(x_1,\ldots, x_{i-1},x_{i+1},\ldots, x_n)$ and, similarly, we denote with $X_{-i}$ the set $\bigtimes_{j\in[n],j\neq i} X_j$.}
We assume that the correspondences have the form
\(
\Gamma^\nu_i(x_{-i})=\left\{x_i\in X_i: A_i(x_{-i}) x_i\le b(x_{-i})+\nu \cdot1_{m_j}\right\}
\)
for each $i\in[n]$.
These are polytopes with $m_i$ inequalities for each $x_{-i}$.
We assume that the matrix $A_i:X_{-i} \to \Reals^{m_i\times l}$ and vector $b_i:X_{-i} \to \Reals^{m_i}$ are generic circuits implementing $L$-Lipschitz functions and $\nu>0$ is the relaxation of the correspondences.
As standard in GNEPs we assume that, for all $i\in[n]$ and $x_{-i}\in[0,1]^{l(n-1)}$, the set of available strategies is not empty, \ie, $\Gamma^\nu_i(x_{-i})\neq \emptyset$.

A generalized equilibrium is a tuple $x=(x_i)_{i\in[n]}$ that  is feasible for each player $i\in[n]$, and such that each player is  best-responding, \ie 
\[
x_i\in\bigtimes_{j\in[n]}\Gamma^\nu_j(x_{-j})\quad\forall i\in[n]\quad\text{and}\quad u_i(x)\ge u_i(x'_i,x_{-i})-\epsilon\quad \forall x_i'\in \Gamma^\nu_i(x_{-i}),
\]
for a given approximation $\epsilon>0$.

We can reduce this problem to \QVI by mimicking the existence proof of GNEP in \citep{chan1982generalized}.
Define the search space as $[0,1]^d$, where $d=nl$, and the correspondence as \(Q_\nu(z)=\bigtimes_{j\in[n]} \Gamma^\nu_j(z_{-j}).\)
Moreover, define the  operator as $F(z)=(-\nabla_1 u_1(z),\ldots, -\nabla_n u_n(z))$.

We now prove that the constructed problem is indeed a \QVI problem.
First, since each $u_i$ is smooth, the operator $F$ is $nG$-Lipschitz. Moreover, $Q(z)$ is nonempty for all $z$ since it is the Cartesian product of non-empty sets. 
Finally, each set $Q(z)$ is the Cartesian product of polytopes and thus is itself a polytope and can be written in the form $Q(z)=\{z'\in[0,1]^d:\tilde A(z) z'\le \tilde b(z)\}$.\footnote{Formally, $\tilde A(z)$ is the block diagonal matrix with $\{A_i(z_{-i})\}_{i\in[n]}$ on the diagonal and $\tilde b(z)$ is the concatenation of the vectors $\{ b_i(z_{-i})\}_{i\in[n]}$.} 
This proves the \PPAD-membership of GNEP.

The ingenious framework developed by \citet{filos2024ppad} only considers exact problems with exact rational solutions. This allows them to create a ``pseudogate'' that, when inserted into arithmetic circuits, automatically computes fixed points and provides \PPAD membership results by reduction to \Brouwer.\footnote{Formally, their main result is that any circuit with pseudogates can be mimicked with normal gates, by adding extra variables, in the sense that from a fixed point of the normal circuit one can build a fixed point of the one with pseudogates.} 
This approach is extremely powerful, as it proves the \PPAD-membership of a large class of problems with relative ease. However, it can only work with problems for which exact rational solutions exist.
For example, in the particular case of generalized equilibria, the authors considered the case in which $\Gamma_i=\{x_i\in X_i: A_ix_i\le b(x_{-i})\}$, \ie, the case in which the dependency on the other players in the correspondence only appears in the right-hand side vector of the polytope, which guarantees the existence of exact rational solutions \citep[Remark 3.4]{filos2024ppad}. We remark that, in our definition, the matrix $A_i$ is not fixed, and we cannot guarantee the existence of rational solutions (also see \Cref{th:irrationalandfriends}). We overcome this difficulty at the expenses (which is necessary, also by \Cref{th:irrationalandfriends}) of relaxing the correspondence (or equivalently assuming positive volume). It seems that our \QVI problem manages to handle approximate fixed points naturally. Even if our main interest in this paper was about hardness results, it would be interesting to explore whether our approach could be adapted to show membership for problems without exact rational solutions. For example, this argument seems to be extendable to the class of Piecewise-Linear best response oracles (PLBRO) games defined in \citet{filos2024ppad}. 

\section{Reductions among \MINMAX, \GDAFP, QVI  and VI}\label{app:QVIandVI}

In this section, we give a clear overview of the relationships between QVIs, VIs, \MINMAX, and fixed points of gradient descent-ascent dynamics. For the sake of completeness, we include some folklore results. Here we consider the computational versions of the problems, and, therefore, such results can be considered as polynomial-time reductions between problems. 
\Cref{fig:qvi} gives a summary of these results. 

Finally, \Cref{app:prooffriends} provides a proof for \Cref{th:irrationalandfriends}, showing why constraints have to be relaxed to guarantee the existence of a solution. 

\begin{figure}[!htp]
    \centering
    \hspace{-2.5cm}
    \scalebox{0.8}{\input{settings/colors}
\scalebox{.9}{\begin{tikzpicture}[thick,every node/.style={scale=1.0}, shorten >=1mm, shorten <=1mm]
	\def\dhorizontal{5.1cm}
	\def\dvertical{3.3cm}
	\centering

	\node[draw] (QVI) at (3,6) {QVI};
	\node[draw] (MM) at ($(QVI)+(\dhorizontal,0)$) {$\LRMINMAX$};
	\node[draw, text width=3.1cm,align=center] (GDA) at ($(MM)+(1.2*\dhorizontal,0)$) {$\GDAFP$};
	\node[draw] (VI) at ($(MM)+(\dhorizontal*0.945,-\dvertical)$) {$\VI$};
        \node[draw, text width=2cm,align=center] (SGDA) at ($(VI)+(\dhorizontal,0)$) {Fixed point of $F_\SGDA$};
    
    \draw[rounded corners] ($(MM)-(-0.9,\dvertical+22.5)$) rectangle ($(SGDA)+(1.4,\dvertical*0.7)$) node[inner sep=1mm, pos=0.5, below=36pt, fill=white, align=center] {Jointly Convex Constraints};

    \draw[implies-implies,double equal sign distance] 
        (SGDA)
        -- node[above=1mm] {\Cref{th:VI_eq_sgda}}
        (VI);
    \draw[-implies,double equal sign distance] 
 	(VI)
        -- node[inner sep=1mm, pos=0.5, fill=white, align=center]  {\Cref{th:fromVItoMinMax}} 
        (GDA);
    \draw[-implies,double equal sign distance,black!30]
        (MM)
        -- node[inner sep=1mm, pos=0.15,sloped,niceRed] {\huge{\boldmath$\times$}} node[inner sep=1mm, pos=0.5,sloped,niceRed] {\huge{\boldmath$\times$}} node[inner sep=1mm, pos=0.85,sloped,niceRed] {\huge{\boldmath$\times$}}node[black,inner sep=1mm, pos=0.5, sloped, above=-7mm,fill=white, align=center] {\Cref{ex:eq_not_vi}}
        (VI);
    \draw[implies-implies,double equal sign distance]
        (MM)
        -- node [inner sep=1mm, pos=0.5, above=1mm, fill=white, align=center] {\Cref{th:GDA_eq_minmax}}
        (GDA);
    \draw[implies-implies,double equal sign distance] 
        (MM)
        -- node [inner sep=1mm, pos=0.5, fill=white, align=center, above=1mm] {\Cref{th:MM_eq_QVI}}
        (QVI);
\end{tikzpicture}}}
    \caption{A connection between two problems means that they can be reduced to one another in polynomial time. The crossed connection between \LRMINMAX and VI means that there are solutions of \LRMINMAX that are not solutions to VI.
    We remark that in the picture, VI and QVI do not refer to the computational problems but refer to the specific instances associated with the local min-max problem. \Cref{th:fromVItoMinMax}, \Cref{ex:eq_not_vi} and \Cref{th:VI_eq_sgda} holds only for jointly convex constraints, while \Cref{th:GDA_eq_minmax} and \Cref{th:MM_eq_QVI} holds even for independent constraints.}
    \label{fig:qvi}
\end{figure}

\subsection{Separation Between Fixed Points of $F_\GDA$ and Fixed Points of $F_\SGDA$}\label{app:sep gda sgda}
It is interesting to analyze the relation between the gradient updates of $F_{\GDA}$ and $F_{\SGDA}$ in the case of jointly convex constraints (\ie,  when both are well-defined). %
One might be induced to believe that fixed points of $F_\SGDA$ and $F_\GDA$ are equivalent. We show that this is not the case, and only one direction holds. In particular, all fixed points of $F_\SGDA$ are fixed points of $F_\GDA$ (\Cref{th:fromVItoMinMax}), but the opposite does not hold (\Cref{ex:eq_not_vi}).
This observation highlights the fundamental difference between VIs and QVIs.
Indeed, while we can associate both a VI and a QVI to a \LRMINMAXJC problem, the QVI has a strictly larger set of solutions. 

\Cref{th:fromVItoMinMax}, together with \Cref{th:GDA_eq_minmax}, states that we can look for solutions of \VI in order to find a solutions to \LRMINMAXJC. For instance, if the payoff function is convex-concave, then there are polynomial-time algorithms that find solutions to the VI since it becomes monotone \citep{nesterov2006solving}. 

Before proving the theorem, we provide two technical lemmas.

\begin{lemma}[\protect{\citep[Theorem~3.1.1]{hiriart2004fundamentals}}]\label{lem:FP_proj}
    For any $x\in \Reals^d$, and convex set $K$ in $\Reals^d$ it holds:
    \[
    (y-\Pi_K(x))^\top(\Pi_K(x)-x)\ge 0\quad\forall y\in K.
    \]
\end{lemma}

\begin{lemma}\label{lem:VIareFP}
    Consider a pair $(x,y) \in K_1^\nu(y)\times K_2^\nu(x) $ and a $\beta\ge 0$.
    If $\nabla_x f(x,y)^\top(x'-x)\ge -\beta$ for all $x'\in K_1^\nu(y)$ then:
    \[
   \|x-\Pi_{K^\nu_1(y)}(x-\nabla_x f(x,y))\|^2\le \beta.
    \]
    Similarly, if $-\nabla_yf(x,y)^\top(y'-y)\ge -\beta$ for all $y'\in K_2^\nu(x)$ then:
    \[
    \|y-\Pi_{K^\nu_2(x)}(y+\nabla_y f(x,y))\|^2\le \beta.
    \]
\end{lemma}

\begin{proof}
    We are going to prove only the first part of the statement, as the second can be proven with a symmetric argument.

    Take $x'=\Pi_{K^\nu_1(y)}(x-\nabla_x f(x,y))$ which clearly lies in $K^\nu_1(y)$. By \Cref{lem:FP_proj} we know that for all $\tilde x\in K_1(y)$ and $\hat x \in \Reals^d$
    \[
    (\tilde x-\Pi_{K^\nu_1(y)}(\hat x))^\top(\Pi_{K^\nu_1(y)}(\hat x)-\hat x)\ge 0.
    \]
    By taking $\tilde x=x$, $\hat x= x-\nabla_xf(x,y)$, it holds $x'= \Pi_{K^\nu_1(y)}(\hat x)$ and
    \begin{equation}\label{eq:tmp7}
    (x-x')^\top(x'-(x-\nabla_xf(x,y)))\ge 0.
    \end{equation}
    Thus it follows that:
    \begin{align*}
        \beta&\ge \nabla_x f(x,y)^\top(x-x')\\
        &=(x-(x-\nabla_x f(x,y))^\top(x-x')\\
        &=(x-x')^\top(x-x')+(x'-(x-\nabla_x f(x,y)))^\top(x-x')\\
        &\ge \|x-x'\|^2,
    \end{align*}
    where the first inequality holds by assumption as $x'\in K_1^\nu(y)$ and in the last inequality we used \Cref{eq:tmp7}.
\end{proof}

\begin{restatable}[From VIs to \GDAFPJC]{theorem}{minmaxVI}\label{th:fromVItoMinMax}
    Consider an instance of \GDAFPJC with jointly convex constraints $K:=\{x,y\in[0,1]^d: g(x,y)\le 0\}$. Then, any $(x,y)\in K$ such that:
    \[
    \begin{pmatrix}\nabla_x f(x,y)\\ -\nabla_y f(x,y)\end{pmatrix}^\top\begin{pmatrix}x'-x\\y'-y\end{pmatrix}\ge -\frac{\beta^2}{2},\text{ for all }(x',y')\in K,
    \]
    is a $\beta$-fixed point of \GDAFP.
\end{restatable}
\begin{proof}

   The statement follows easily from \Cref{lem:VIareFP}. Indeed by assumption, we have that for $y'=y$ it holds that:
   \[
   \nabla_xf(x,y)^\top(x'-x)\ge \frac{-\beta^2}{2}
   \]
   for all $(x',y)\in K$. The inequality holds for all $(x',y)$ such that $g(x',y)\le 0$. Thus we can apply \Cref{lem:VIareFP} by suitably redefining $\beta$ and find that:
   \[
   \frac{\beta^2}{2}\ge\|x-\Pi_{K^\nu_1(y)}(x-\nabla_x f(x,y))\|^2.
   \]
   We can repeat the same argument for the second player and obtain:
    \begin{align*}
        \|(x,y)-F_\GDA(x,y)\|^2=\|y-\Pi_{K_2(x)}(y+\nabla_yf(x,y))\|^2+\|x-\Pi_{K_1(y)}(x-\nabla_x f(x,y))\|^2\le\beta^2,
    \end{align*}
    which concludes the proof.
\end{proof}

The following example demonstrates that the converse of \Cref{th:fromVItoMinMax} does not hold. Indeed, the set of solutions of \VI is strictly smaller than the set of solutions to \LRMINMAXJC.

\begin{theorem}\label{ex:eq_not_vi} There exists a payoff function $f$ and a jointly convex constraint function $g$ such that:
  \begin{enumerate}[label=\roman*)]
\item there exists a point $(x^\star,y^\star)$ which is a solution to \MINMAX for any $\epsilon>0$ and $\delta>0$,
\item $(x^\star,y^\star)$ is not a $\bar \beta$-approximate solution of the associated VI even for a constant $\bar \beta$, \ie, there exists a pair $(x',y')\in K$ such that 
\begin{align}
        (\nabla_x f(x^\star,y^\star), -\nabla_y f(x^\star,y^\star))^\top(x'-x^\star,y'-y^\star)<-\bar \beta.
    \end{align}  
\end{enumerate}
\end{theorem}

\begin{proof}
    Consider the payoff function $f:[0,1]^2\to[0,1]$ given by:
    \[
    f(x,y)=\frac{4}{5}\left((x-1)^2-\left(y-\frac{1}{2}\right)^2+\frac{1}{4}\right)
    \]
    and the constraint function $g$ given by:
    \[
    g(x,y)=x+y-1.
    \]
    The function $f$ is $8/5$-smooth and $4\sqrt{5}/5$-Lipschitz. Let $(x^\star,y^\star)=(1,0)$. This is a $(\epsilon,\delta)$-solution to \MINMAX for any $\epsilon$ and $\delta$.
    Indeed,
    \[
    f(x^\star,y^\star)=0
    \]
    and for all $x\in [0,1]$ we have that:
    \[
    f(x,y^\star)\ge 0.
    \]
    On the other hand, independently from $\delta>0$, the second player cannot deviate as $g(x^\star,y)\le 0$ if and only if $y=y^\star=0$.
    This proves that, for any $\epsilon\ge0$ and $\delta\ge0$, the point $(x^\star,y^\star)$ is a solution to \MINMAX. 
    
    However, there exists a constant $\bar\beta=\frac{4}{5}$ such that the point $(x^\star, y^\star)$ does not solve the following VI for any $\beta<\bar\beta$:
    \[
    \begin{pmatrix}\nabla_x f(x^\star,y^\star)\\ -\nabla_y f(x^\star,y^\star)\end{pmatrix}^\top\begin{pmatrix}x'-x^\star\\y'-y^\star\end{pmatrix}\ge -\beta,\text{ for all }(x',y')\in K.
    \]
    Indeed, we have $\nabla_x f(x,y)=\frac{4}{5}(2x-2)$ and $\nabla_x f(x^\star,y^\star)=0$. On the other hand, $\nabla_y f(x,y)=-\frac{4}{5}(2y-1)$ and thus: $\nabla_y f(x^\star,y^\star)=\frac{4}{5}$. This yields
    \begin{align}
        \begin{pmatrix}\nabla_x f(x^\star,y^\star)\\ -\nabla_y f(x^\star,y^\star)\end{pmatrix}^\top\begin{pmatrix}x'-x^\star\\y'-y^\star\end{pmatrix}&=-\frac{4}{5}y'.
    \end{align}
    Hence, for $x'=0$ and $y'=1$ we have that $(x',y')\in K$ and 
    \[
    \begin{pmatrix}\nabla_x f(x^\star,y^\star)\\ -\nabla_y f(x^\star,y^\star)\end{pmatrix}^\top\begin{pmatrix}x'-x^\star\\y'-y^\star\end{pmatrix}=-\frac{4}{5}.
    \]
    This concludes the proof.
\end{proof}

\subsection{Equivalence Between QVIs and \MINMAX With Independent Constraints}\label{sec:qvi eq minmax}
We now shift our focus to proving the main result of this section: the equivalence between \QVI and \MINMAX.

\qvi*
\begin{proof}
    Part $i)$:
    Let $\alpha:= \left( \frac{1}{2\sqrt{2}}\left(\sqrt{(G+\delta)^2+4(\epsilon-\frac{L}{2}\delta^2)}-(G+\delta)\right)\right)^2$.
    By applying \Cref{lem:VIareFP} we can show that
    \[
    \|(x,y)-F^\nu_\GDA(x,y)\|\le \sqrt{2\alpha}.
    \]
    Indeed, it holds $ \lVert \Pi_{K^\nu_1(y)}(x-\nabla_xf(x,y)) \rVert^2\le \alpha $ and $\lVert\Pi_{K^\nu_2(x)}(y-\nabla_y f(x,y))\rVert^2 \le \alpha$.
    The proof of $i)$ is concluded by using the first point of \Cref{th:GDA_eq_minmax}.

Part $ii)$: Let $(x,y)$ be a $(\epsilon,\delta,\nu)$ solution to \LRMINMAX. Define $\hat x$ as any solution in 
\[\arg \min_{\tilde x\in K^\nu_1(y)\cap \{\bar x:\lVert \bar x-x\rVert\le \delta\}} f(\tilde x, y).\] Then, the following VI holds for all $x'\in K^\nu_1(y)$ such that $\lVert x'-x\rVert\le \delta$
\begin{equation}\label{eq:qvi2tmp1}
\nabla_x f(\hat x,y)^\top(x'-\hat x)\ge 0.
\end{equation}

Moreover, the $L$-smoothness of $f$ implies that
\begin{align}
\nabla_x f(\hat x, y)^\top(x-\hat x)& \le \frac{L}{2}\|x-\hat x\|^2 + f(x,y)- f(\hat x,y) \nonumber \\
&\le \frac{L}{2}\|x-\hat x\|^2+\epsilon\nonumber\\
&\le \frac{L}{2}\delta^2+\epsilon. \label{eq:qvi2tmp2}
\end{align}

Then, combining \Cref{eq:qvi2tmp1} and \Cref{eq:qvi2tmp2} we have the for all $x'\in K^\nu_1(y)$ such that $\lVert x'-x\rVert\le \delta$ it holds:
\begin{align*}
    0&\le \nabla_x f(\hat x,y)^\top(x'-\hat x)\tag{\Cref{eq:qvi2tmp1}}\\
    &=\nabla_x f(\hat x,y)^\top(x'- x)+\nabla_x f(\hat x,y)^\top(x- \hat x)\\
    &\le \nabla_x f(\hat x,y)^\top(x'- x)+\frac{L}{2}\delta^2+\epsilon\tag{\Cref{eq:qvi2tmp2}}\\
    &=\nabla_x f(x,y)^\top(x'-x) +(\nabla_x f(\hat x,y)-\nabla_x f( x, y))^\top(x'-x)+\frac{L}{2}\delta^2+\epsilon\\
    &\le \nabla_x f(x,y)^\top(x'-x)+\|\nabla_x f(\hat x,y)-\nabla_x f( x, y)\|\cdot\|x'-x\|+\frac{L}{2}\delta^2+\epsilon\tag{Cauchy–Schwarz}\\
    &\le \nabla_x f(x,y)^\top(x'-x) + L\|\hat x-x\|\cdot \|x'-x\|+\frac{L}{2}\delta^2+ \epsilon\tag{$L$-smoothness}\\
    &\le \nabla_x f(x,y)^\top(x'-x) + L\delta^2+\frac{L}{2}\delta^2+\epsilon \tag{$\lVert \hat x-x\rVert$, $\lVert x'-x\rVert\le \delta$}\\
    &= \nabla_x f(x,y)^\top(x'-x)+ \frac{3}{2} L\delta^2+\epsilon.
\end{align*}

Take any $x''\in K^\nu_1(y)$ and consider $x':=x+\delta\frac{x''-x}{\|x''-x\|}$. Clearly, since $K^\nu_1(y)$ is convex, we have that $x'\in K^\nu_1(y)$ and $\lVert x'-x\rVert\le \delta$. Therefore, 

\begin{align*}
- \frac{3}{2} L\delta^2-\epsilon&\le\nabla_x f(x,y)^\top(x'-x)\\
&=\frac{\delta}{\|x''-x\|}\nabla_x f(x,y)^\top (x''-x).
\end{align*}
Thus, for all $x''\in K^\nu_1(y)$:
\[
\nabla_x f(x,y)^\top (x''-x)\ge -\frac{\|x''-x\|}{\delta} \left( \frac{3}{2} L\delta^2+\epsilon\right)\ge -\frac{\sqrt{d}}{\delta} \left(\frac{3}{2} L\delta^2+\epsilon\right).
\]
Similarly we can prove that for all $y''\in K^\nu_2(x)$:
\[
\nabla_y f(x,y)^\top (y''-y)\le \frac{\sqrt{d}}{\delta} \left(\frac{3}{2} L\delta^2+\epsilon\right).
\]
Combining the two inequalities, we complete the proof.
\end{proof}

\subsection{Equivalence Between VIs and Fixed Points Under Jointly Convex Constraints}

In order to complete the picture for jointly convex constraints, we provide the following folklore theorem (adapted from \citet{facchinei2003finite} in order to account for approximations) that states the well-known equivalence between fixed points and solutions to VI.

\begin{restatable}[Equivalence between VIs and Fixed Points]{theorem}{VIeqsgda}\label{th:VI_eq_sgda}
    Given a function $F:[0,1]^d\to[0,1]^d$ and a convex set $K\subseteq[0,1]^d$ the following holds
    \begin{enumerate}[label=\roman*)]
        \item Any point $x\in K$ which is $\alpha$-approximate fixed point of $x\mapsto \Pi_K(x-F(x))$ satisfy the following VI:
        \[
        F(x)^\top(x'-x)\ge-2\alpha\sqrt{d}\text{ for all }x'\in K.
        \]
        \item Any point $x\in K$ which satisfies the following VI:
        \[
        F(x)^\top(x'-x)\ge-\alpha\text{ for all }x'\in K,
        \]
        is a $\sqrt{\alpha}$-approximate fixed point of $x\mapsto \Pi_K(x-F(x))$.
    \end{enumerate}
\end{restatable}

\begin{proof}
    First, we prove $i)$. Define $\hat x:=\Pi_K(x-F(x))$. We know that, by \Cref{lem:FP_proj}, the following holds
    \[
    (x'-\hat x)^\top (\hat x-x+F(x))\ge 0.
    \]

    Simple algebraic manipulations lead to:
    \begin{equation}\label{eq:tmp_vi_eq_sgda}
    F(x)^\top(x'-x)+F(x)^\top (x-\hat x)+(x'-x)^\top(\hat x-x)-\|\hat x- x\|^2\ge 0.
    \end{equation}

    Since $x$ is an $\alpha$-approximate fixed point of $x\mapsto \Pi_K(x-F(x))$ it means that $\|\hat x-x\|\le \alpha$.
    Thus:
    \begin{align*}
        F(x)^\top(x'-x)&\ge \|\hat x-x\|^2+F(x)^\top(\hat x-x)+(x-x')^\top(\hat x-x)\tag{\cref{eq:tmp_vi_eq_sgda}}\\
        &\ge \|\hat x-x\|^2-\|F(x)\|\cdot\|\hat x-x\|-\|x-x'\|\cdot\|\hat x-x\|\tag{Cauchy-Schwartz}\\
        &\ge \|\hat x-x\|^2-\sqrt{d}\cdot\|\hat x-x\|-\sqrt{d}\cdot\|\hat x-x\|\tag{$F(x)_j,x_j,x'_j\in [0,1]$}\\
        &\ge -\sqrt{d}\cdot\|\hat x-x\|-\sqrt{d}\cdot\|\hat x-x\|\tag{$F(x)_j,x_j,x'_j\in [0,1]$}\\
        &\ge -2\alpha\sqrt{d}.
    \end{align*}

    Now we prove $ii)$. Let $\hat x:=\Pi_K(x-F(x))$, again by \Cref{lem:FP_proj}
    \[
    -\|\hat x-x\|+F(x)^\top(x-\hat x)=(x-\hat x)^\top(\hat x-x+F(x))\ge 0,
    \]
    since $x\in K$. Moreover, by assumption we have that $F(x)^\top(x'-x)\ge -\alpha$ for all $x'\in K$ and thus also for $x'=\hat x$.
    Thus, combining the two, we have:
    \begin{align}
        -\|x-\hat x\|^2+\alpha\ge -\|x-\hat x\|^2+F(x)^\top(x-\hat x)\ge 0,
    \end{align}
    which proves that $\|x-\hat x\|\le\sqrt{\alpha}$.
\end{proof}

\citet{daskalakis2021complexity} proved the \PPAD membership of \GDAFPJC, \ie, with \emph{jointly convex} constraints (and therefore also for the case of \emph{box constraints}), for $\nu=0$ and $\delta<\sqrt{2\epsilon/L}$.

Equivalently, we can observe that this result follows since \Cref{th:fromVItoMinMax}, along with the known equivalence between VIs and fixed points (formally restated in \Cref{th:VI_eq_sgda}), implies that every fixed point of $F_{\SGDA}$ is a solution to \GDAFPJC. Moreover, a fixed point of $F_{\SGDA}$ is guaranteed to exist due to a direct reduction from the problem of finding a fixed point of $F_\SGDA$ to \Brouwer (which also proves that the problem of finding fixed points of $F_\SGDA$ is in \PPAD).\footnote{Actually, \citet{daskalakis2021complexity} directly prove that a fixed point of $F_\SGDA$ is a fixed point of $F_\GDA$ without going trough the VI formulation of the problem.}

\section{Missing Proofs for \PPAD-membership}\label{app:irrational}

We present the proofs for our membership results. 
First, in \Cref{app:prooffriends}, we show that without relaxing the constraints, solutions might not exist. Then, in \Cref{app:hausdorff}, we introduce preliminary results on the continuity of the correspondence $Q$, which are then used in \Cref{app:proofqvippad} to establish our main membership result.
Finally, in \Cref{app:global}, we show how our result can be globalized for convex or concave functions.

\subsection{Proof of \Cref{th:irrationalandfriends}} \label{app:prooffriends}

The following statement proves that, without relaxing the constraints, a solution to \QVI and \LRMINMAX might not exist even for independent constraints.

\irrationalandfriends*
\begin{proof}

\minipara{\Kakutani admits a solution} To apply Kakutani's fixed-point theorem (see \Cref{thm:kakutani}) we need to show that $Q(z)=\{z'\in[0,1]^d:(A(z)z'-b(z))_i\le0\quad\forall i\in[n]\}$, given by two circuits $A:[0,1]^d\to[0,1]^{d\times n}$ and $b:[0,1]^d\to[0,1]^{n}$ which are guaranteed to be $L$-Lipschitz and such that $Q(z)\neq \emptyset$ for all $z\in[0,1]^d$, satisfies that:
\begin{itemize}
    \item $Q(z)$ is convex for all $z\in[0,1]^d$;
    \item $Q(z)$ is closed for all $z\in[0,1]^d$;
    \item $Q(z)$ is non-empty for all $z\in[0,1]^d$;
    \item $Q$ is upper-semicontinuous in $[0,1]^d$. 
\end{itemize}
The only non-trivial condition is that $Q$ is upper-semicontinuous (see \Cref{def:semicont}). It is known that if the graph of $Q$
(defined as $\{(z,y)\in[0,1]^{2d}: y\in Q(z)\}$) is closed, then the correspondence $Q$ is upper-semicontinuous \cite[Corollary~of~Theorem~7,~Chapter~VI]{berge1877topological}. In our case, we can define the function:
\[
g(z,z')\coloneqq(A(z)z'-b(z))_i\quad\forall i\in[n],
\]
which is continuous in $[0,1]^{2d}$ and thus $\{z'\in [0,1]^d: g(z,z')\le 0\}$ is closed for each $z$.

\minipara{All solutions might be irrational}
We will show an instance of the problem of finding a point $z^\star\in[0,1]^2$ such that $z^\star\in Q(z^\star)$ has only irrational solutions, where $Q(z)$ is given by the points contained in a polytope $\{z':A(z)z'\le b(z)\}$. 
    
    We write $x$ for the first component and $y$ for the second component of $z\in\Reals^2$.
    Let $A:[0,1]^{2}\to\Reals^{4\times2}$ and $b:[0,1]^2\to\Reals^4$ defined by
    \[
    A(x,y):=\begin{bmatrix}
        +1&0\\
        -1&0\\
        0&+x\\
        0&-x
    \end{bmatrix},\quad     b(x,y):=\begin{bmatrix}
        +y\\
        -y\\
        +\frac{1}{2}\\
        -\frac{1}{2}
    \end{bmatrix}.
    \]
    Clearly a point $(x',y')$ is in $Q(x,y)$ if $x'=y$ and $y'=\frac{1}{2x}$. Thus we are looking for solutions in $[0,1]^2$ to the system 
    \[
        \begin{cases}
            x=y\\
            y=\frac{1}{2x},
        \end{cases}
    \]
    which has as the only solution $x=y=\frac{1}{\sqrt{2}}$, which is irrational.

\minipara{Non-existence of solutions} We will prove the result by providing an instance for \LRMINMAX for which there is no solution. The non-existence of solutions for \QVI  with $\nu=0$ follows by writing the QVI associated with the \LRMINMAX instance. In particular, \Cref{th:MM_eq_QVI} states that from a solution to such a QVI we would been able to find a solution to \LRMINMAX, and thus there is no such solution of the QVI. Similarly, thanks to \Cref{th:GDA_eq_minmax}, we can show that \GDAFP does not admit solutions.

Let $n=d=1$ and let the constraints be defined by $g_1(x,y)=\{x-y\le 0\wedge y-x\le 0\}$, $g_2(x,y)=\{yx\le 0\}$, while the payoff function be given by $f(x,y)=\frac{1}{2}y^2$. The approximations $\epsilon,\delta$ are such that $\delta=\frac{5}{4}\epsilon$.
It can be easily checked that this is a valid \LRMINMAX instance as the function $f$ is $L$-smooth for all $L$ and thus $\delta\le\sqrt{2\epsilon/L}$, is always verified by our choice of $\delta$ and $\epsilon\le 1$.
However, the only point $(x,y)$ for which $g_1(x,y)\le 0$ and $g_2(x,y)\le 0$ is given by $(0,0)$, but this is not an equilibrium, as the second player can deviate to $y'=\delta$ and gain an extra utility $\delta$, and thus, it is not an equilibrium as $\delta>\epsilon$. 
\end{proof}

\subsection{Hausdorff Continuity of the Correspondence $Q$}\label{app:hausdorff}

In this section, we provide some results on the Hausdorff continuity of correspondence $Q_{\beta}(z)$ (see \Cref{prob:kakutani}) both with respect to $z$ and $\beta$. 

\begin{definition}\label{def:Hausdorff}
A correspondence $Q:X\rightrightarrows Y$ between two topological spaces is $K$-Lipschitz with respect to the Hausdorff metric if
\[
d_H(Q(z),Q(z'))\le K\cdot d(z,z'),
\]
where, for two set $A,B\subseteq Y$, the Hausdorff distance $d_H(A,B)$ is defined as
\[
d_H(A,B):=\max\left\{\sup_{a\in A}\inf_{b\in B} d(a,b),\, \sup_{b\in B}\inf_{a\in A} d(b,a)\right\},
\]
and $d(\cdot,\cdot)$ is the distance of the topological space $Y$.
Equivalently, $d_H(A,B)=\inf\{\epsilon>0:A\subseteq B_\epsilon\quad\text{and}\quad B\subseteq A_\epsilon\}$, where $A_\epsilon:=\cap_{a\in A}\{x:d(x,a)\le\epsilon\}$.
\end{definition}

In Euclidean geometry, for a $K$-Lipschitz correspondence $Q$ we have that each point $x$ in the set $Q(z)$ can be projected onto the set $Q(z')$ and the distance between $x$ and $\Pi_{Q(z')}(x)$ is at most $K\|z-z'\|$.

We start by showing the Hausdorff continuity of $Q_{\beta}(z)$ with respect to $\beta$. Notice that our result holds only for (even arbitrary small) relaxations, \ie, $\beta>0$.

\begin{lemma}\label{lem:bit_complexity_LP}
    For any $z\in [0,1]^d$ and for any $\bar \beta>0$, the correspondence $\gamma \to Q_{\gamma}(z)$ is $(\sqrt{d}/\bar \beta)$-Lipschitz in the Hausdorff metric on $\gamma\in[\bar \beta,\infty)$ , \ie,
    \[
    d_H(Q_{\gamma}(z),Q_{\gamma'}(z))\le \frac{\sqrt{d}}{\bar \beta}|\gamma-\gamma'|\quad\forall\gamma,\gamma'\ge \bar\beta.
    \]
\end{lemma}

\begin{proof}
    Consider two $\gamma\ge \gamma'\ge \bar\beta$.
    Clearly, $Q_{\gamma'}(z) \subseteq Q_{\gamma}(z)$.
    Hence, to prove the statement is sufficient to consider any $x \in Q_{\gamma}(z)$ and show that $\|\Pi_{Q_{\gamma'}(z)}(x)-x\|\le \sqrt{d } \frac{\gamma-\gamma'}{\bar \beta}$.

    Consider an $x^0 \in Q(z)$, which exists by our assumption on the non-emptiness of $Q(z)$. 
    Let $x^*= (1-\alpha) x + \alpha x^0 $ where $\alpha$ is set in the following. Then, we have that
    \begin{align*}
        A(z) x^* - b(z) &= A(z) \left( (1-\alpha) x + \alpha x^0 \right) - b(z)\\
        &= (1-\alpha) \left( A(z) x - b(z)\right) + \alpha x^0 \left( A(z) x   - b(z)\right)\\
        & \le (1-\alpha) \gamma.
    \end{align*}
    Hence, for $\alpha= \frac{\gamma-\gamma'}{\gamma} $ we have that 
    \[ 
    A(z) x^* - b(z) \le (1-\alpha) \gamma \le \gamma'.
    \]
    This implies that 
    \[ ||\Pi_{Q_{\beta}(z)}(x)-x||\le ||x^*-x||\le \alpha ||x^0-x|| \le \sqrt{d } \frac{\gamma-\gamma'}{\gamma}\le  \sqrt{d } \frac{\gamma-\gamma'}{\bar \beta} .  \]
    Hence, the function is $\frac{\sqrt{d } }{\bar \beta}$-Lipschitz.
\end{proof}

Now, we show the Hausdorff continuity of $Q_{\beta}(z)$ with respect to $z$. Notice that also in this case our result holds only for (even arbitrarily small) relaxations, \ie, $\beta>0$
We start by recalling the following result about the  Lipschitz continuity of the maximum of Lipschitz functions. We provide a proof for completeness.

\begin{lemma}\label{lem:max_is_Lip}
    Let $h_i:\Reals^d\to \Reals$ be $L_i$-Lipschitz functions for $i\in[n]$. Then $g(\cdot)=\max_{i\in[n]} h_i(\cdot)$ is $L$-Lipschitz, where $L=\max_{i\in[n]}L_i$.
\end{lemma}

\begin{proof}
    For each $i\in[n]$ we have that $h_i(x)\le h_i(y)+L_i\|x-y\|$ by $L_i$-Lipschitzness of $h_i$. Consider now $x$ and $i(x)$ such that $g(x)=h_{i(x)}(x)$. Then 
    \begin{align*}
        g(x)&=h_{i(x)}(x)\\
            &\le h_{i(x)}(y)+L_{i(x)}\|x-y\|\\
            &\le g(y)+L_{i(x)}\|x-y\|\\
            &\le g(y)+L\|x-y\|.
    \end{align*}
    
    Similarly, consider $y$ and $i(y)$ such that $g(y)=h_{i(y)}(y)$. Then 
    \begin{align*}
        g(y)&=h_{i(y)}(y)\\
            &\le h_{i(y)}(x)+L_{i(y)}\|x-y\|\\
            &\le g(x)+L_{i(y)}\|x-y\|\\
            &\le g(x)+L\|x-y\|. 
    \end{align*}
    This proves that $g$ is $L$-Lipschitz.
\end{proof}

We continue by proving some properties related to the constraint function.

\begin{lemma}\label{lem:G}
    Given an $L$-Lipschitz matrix valued function $A:[0,1]^d\to[0,1]^{n\times d}$ and a vector valued $L$-Lipschitz function $b:[0,1]^d\to[0,1]^n$ let $G_i(z|z')$ be defined as:
    \[
        G_i(z|z'):=[A(z')^\top z-b(z')]_i,
    \]
    and let $G(z|z'):=\max_{i\in[n]} G_i(z|z')$. Then the following holds:
    \begin{enumerate}[label=\roman*)]
        \item $G$ is $2L$-Lipschitz in the second argument;
        \item $G$ is convex in the first argument.
    \end{enumerate}
\end{lemma}

\begin{proof}
    First, we prove the first point.
    We prove that for all $i\in[n]$ the function $G_i(z|\cdot):[0,1]^{d}\to\Reals$ is $2L$-Lipschitz for each $z$. Formally:
    \begin{align*}
        |G_i(z|\tilde z_1)-G_i(z|\tilde z_2)|&=|(A(\tilde z_1)z)_i-b_i(\tilde z_1)-(A(\tilde z_2)z)_i+b_i(\tilde z_2)|\\
        &\le |(A(\tilde z_1)z)_i-(A(\tilde z_1)z)_i|+|b_i(\tilde z_1)-b_i(\tilde z_2)|\\
        &\le 2L\|\tilde z_1-\tilde z_2\|.
    \end{align*}
    Applying \Cref{lem:max_is_Lip} we conclude that $G$ is $2L$-Lipschitz in the second argument.

    The second point is proven by noting that $G_i(\cdot|z)$ is a linear function for each fixed $z$, and thus $G(\cdot|z)$ is the maximum over finitely many linear functions, which thus is convex.
\end{proof}

\begin{lemma}\label{lem:Qbeta_is_cnt}
    For any $\beta>0$, the correspondence $Q_\beta:[0,1]^d\to2^{[0,1]^d}$ is $\frac{2L\sqrt{d}}{\beta}$-Lipschitz in the Hausdorff metric.
\end{lemma}

\begin{proof}
    Consider any $z,z'\in[0,1]^d$ and any $\tilde z\in Q_\beta(z)$. We have to prove that 
    \[d(\tilde z,Q_\beta(z')):=\|\tilde z-\Pi_{Q_{\beta}(z')}(\tilde z)\|\le K\|z-z'\|\]
    for some $K>0$.
    This would show that $d(\tilde z,Q_\beta(z'))\le K\|z-z'\|$ for all $\tilde z\in Q_\beta(z)$. By this result, exchanging the roles of $z$ and $z'$ we prove that also $d(\tilde z,Q_\beta(z))\le K\|z-z'\|$ for all $\tilde z\in Q_\beta(z')$ and thus that
    \[
    d_H(Q_\beta(z),Q_\beta(z'))\le K\|z'-z\|.
    \]

    Consider $\bar z:=\alpha \tilde z+(1-\alpha)z_0$ where $z_0$ is any point in $Q(z')$ (and thus $G(z_0|z')\le 0$) and $\alpha\in[0,1]$ will be specified in the following.
    Now consider the following inequalities:
    \begin{align*}
        G(\bar z|z')&\le \alpha G(\tilde z|z')+(1-\alpha)G(z_0|z')\\
        &\le \alpha\left(2L\|z'-z\|+G(\tilde z|z)\right)+(1-\alpha) G(z_0|z')\\
        &\le 2\alpha L\|z'-z\|+\alpha\beta,
    \end{align*}
    where in the first inequality we used ii) of \Cref{lem:G}, in the second inequality we used i) of \Cref{lem:G}, and in the last one $G(\tilde z|z)\le 0$ and $G(z_0|z')\le 0$.
    By taking $\alpha=\beta/(2L\|z'-z\|+\beta)$ we get that $G(\bar z|z')\le \beta$ and thus $\bar z\in Q_\beta(z')$. This implies that:
    \begin{align*}
    \|\tilde z-\Pi_{Q_\beta(z')}(\tilde z)\|&\le \|\tilde z-\bar z\|\\
    &=(1-\alpha)\|z_0-\tilde z\|\\
    &\le(1-\alpha)\sqrt{d}\\
    &=\frac{2L\|z'-z\|}{2L\|z'-z\|+\beta}\sqrt{d}\\
    &\le \frac{2L\sqrt{d}}{\beta}\|z'-z\|.
    \end{align*}
    By the discussion above, this proves that the correspondence $Q_\beta$ is $\left(\frac{2L\sqrt{d}}{\beta}\right)$-Lipschitz in the Hausdorff metric.
\end{proof}

By the previous analysis, we can prove the Lipschitzness-like condition on $\min_{z' \in Q_{\nu}(z)}  F(z)^T z'$ for a fixed $z$ and varying $\nu$.
Formally,

\begin{lemma}\label{lm:lipDown}
 Given $z$, $\nu>0$, and $\beta>0$, it holds
 \[  \min_{z' \in Q_{\nu}(z)} F(z)^T z'\le \min_{z' \in Q_{\nu+\beta}(z)} F(z)^T z' + \frac{d\beta}{\nu}. \]
\end{lemma}

\begin{proof}
Let $\bar z \in \arg \min_{z' \in Q_{\nu+\beta}(z)} F(z)^T z'$, let $\tilde z =\Pi_{Q_{\nu}} (\bar z)$, and let $z^\star \in\arg \min_{z' \in Q_{\nu}(z)} F(z)^T z'$. Then:
    \begin{align*}
        F(z)^\top \bar z &=F(z)^\top (\bar z-\tilde z)+F(z)^\top \tilde z\\
        &\ge F(z)^\top (\bar z-\tilde z) + F(z)^\top  z^\star\\
        & \ge -||F(z)|| \cdot ||\bar z-\tilde z||+ F(z)^\top  z^\star\\
        & \ge -\sqrt{d} \frac{\sqrt{d}}{\nu }\beta + F(z)^\top  z^\star\\
        &= - \frac{d}{\nu }\beta + F(z)^\top  z^\star
    \end{align*}
    where the last follow by \Cref{lem:bit_complexity_LP}.
\end{proof}

\subsection{Proof of Theorem~\ref{th:qvippad}}\label{app:proofqvippad}

In this section, we provide a proof of the following theorem.
\qvippad*

Our main technical contribution is to show that \QVI is in \PPAD. The membership in \PPAD of \GDAFPCBC is then given by \Cref{th:MM_eq_QVI} and \Cref{th:GDA_eq_minmax}.
To see that, let us consider the problem of finding an $\alpha_{\GDA}$-fixed point of \GDAFPCBC with a given $\nu$. Then, by \Cref{th:GDA_eq_minmax} any $(\epsilon,\delta,\nu)$-solution to \LRMINMAX is an $\alpha_{\GDA}$-fixed point of \GDAFPCBC, where $\epsilon= \frac{\alpha_{\GDA}^2L}{(5L+2)^2}$ and $\delta= \frac{\alpha_{\GDA}}{5L+2}$. 
Finally,  let $\size(I)$ be the size of the instance. Then, thanks to \Cref{th:MM_eq_QVI} any solution to \QVI with
\begin{align*}
\alpha&=\left( \frac{1}{2\sqrt{2}}\left(\sqrt{(G+\delta)^2+4(\epsilon-\frac{L}{2}\delta^2)}-(G+\delta)\right)\right)^2 \ge \poly\left(\frac{1}{\size(I)}\right)
\end{align*}
is a $(\epsilon,\delta,\nu)$-solution to \LRMINMAX and thus an $\alpha_{\GDA}$-fixed point of \GDAFPCBC.

We proceede proving the \PPAD-membership of \QVI. We reduce to a problem related to Sperner's Lemma called \HDSperner.

\paragraph{\HDSperner}
\begin{wrapfigure}[20]{r}{0.4\textwidth} 
    \centering
    \vspace{-0.3cm}
    \input{imgs/sperner}
    \caption{Example of a valid coloring of the $2$-dimensional cube. The color \textbf{\textcolor{niceRed}{red}} corresponds to the color $0$ while the color \textbf{\textcolor{mgreen}{green}} and \textbf{\textcolor{blueGrotto}{blue}} corresponds to colors $1$ and $2$, respectively. The box on the boundaries are colored according to the allowed coloring on those edges.
    We highlighted in \textbf{\textcolor{mypurple2}{purple}} a panchromatic simplex.}
    \label{fig:HDSperner}
\end{wrapfigure}
Consider the $d$-dimensional hypercube $[0,1]^d$. A canonical way to partition the hypercube in simplexes is to divide the hypercube into cubelets of side-length and then divide each cublet into simplices with the Coxeter-Freudenthal-Kuhn triangulation \citep{kuhn1960some}. We call $V$ the set of vertices.
A Sperner coloring $T$ of a simplicization is the assignment of one of $d+1$ colors $[d]\cup\{0\}$ to each vertex of the simplicization. The assignment must satisfy the following boundary conditions:
\begin{enumerate}[label=\roman*)]
\item for each $i \in [d]$, none of the vertices $v\in V$ on the face with $v_i=0$ has color $i$;
\item color $0$ is not assigned to a vertex $v\in V$ on the face with $v_i=1$ for some $i\in[d]$.
\end{enumerate}
\Cref{fig:HDSperner} shows a valid application of these rules in the $2$-dimensional hypercube.
A panchromatic simplex induced by $T$ is a simplex whose vertices have all the $d+1$ colors. Such simplex is guaranteed to exist by a parity argument \citep{sperner1928neuer, kuhn1960some}.
It is well known that not only every Sperner coloring of the hypercube has a panchromatic simplex but that the search problem of finding one can be reduced to \EOTL \citep{papadimitriou2023computational,chen2023complexity,etessami2010complexity}.
\begin{theorem}
   Given a boolean circuit $\mathcal{C}:V \rightarrow [d]\cup \{0\}$ that encodes the Sperner's coloring of a Kuhn’s triangulation, the problem of finding a panchromatic simplex is in \PPAD. 
\end{theorem}

Note that the coloring is given implicitly by the circuit $\mathcal{C}$ and that the domain, \ie, set of vertices, is exponential in the input size. Hence, the problem is not trivially easy to solve, as one cannot simply check all the simplexes in the coloring.

\paragraph{Overview}

The high-level idea of the proof is simple: a fixed point of the correspondence that maps $z$ to the set $\tilde Q(z):=\arg\min_{\tilde z\in Q_\nu(z)} F(z)^\top(\tilde z-z)$ would be a solution to \QVI. To show that this lies in \PPAD we would need to find a Lipschitz continuous selector function of the correspondence $\tilde Q$ (a continuous function $f:[0,1]^d\to[0,1]^d$ such that $f(z)\in \tilde Q(z)$) and show a way of constructing efficiently this function as a circuit to feed to \Brouwer. A simple way to define a selector function is to regularize slightly the objective $\tilde z\mapsto F(z)^\top(\tilde z-z)+\frac\eta2\|\tilde z-z\|^2$ to make it strongly convex. 
This makes the correspondence a singleton (convex optimization problems with strongly convex objective functions have a unique solution) without modifying too much the set of solutions (if $\eta$ is small enough). Moreover, by Berge's maximum theorem, we know that this function is continuous~\citep{berge1877topological}. We can thus define the function
\[
p_{\gamma,\eta}(z):=\arg\min_{\tilde z\in Q_\gamma(z)}\left\{F(z)^\top \tilde z+\frac{\eta}{2}\|\tilde z-z\|^2\right\}.
\]
We need to make three important observations about the function $p_{\gamma,\eta}$. First we know that $p_{\gamma,\eta}(z)$ is simply the projection onto the set $Q_\gamma(z)$ of the point $z-\frac{1}{\eta}F(z)$ (this is a standard result about proximal mapping \cite{parikh2014proximal}) and thus we can write
\[
p_{\gamma,\eta}(z)=\Pi_{Q_\gamma(z)}\left(z-\frac{1}{\eta}F(z)\right).
\] 
Second, we know that we can compute exactly and efficiently this projection by means of the ellipsoid method, as the set $Q_\gamma(z)$ is a polytope. The third observation is that, due to the fact that the set onto we project depends on $z$, even though the function $p_{\gamma,\eta}$ is continuous, it is easy to see that it is not Lipschitz continuous and thus we cannot reduce directly to \Brouwer (differently, the function $p_{\gamma,\eta}$ would be trivially Lipschitz continuous if the set $Q_\gamma$ would not depend on $z$).
For this reason, we have to reduce to \HDSperner instead. This approach is not too dissimilar to the one used by \cite{papadimitriou2023computational} as they also reduce to \HDSperner rather than to \Brouwer for similar reasons.

However, the main challenges of our proof are essentially disjoint from \citet{papadimitriou2023computational}. The first challenge is to prove that the correspondence $Q$ is continuous (see \Cref{app:hausdorff}), while continuous correspondences are assumed in \citet{papadimitriou2023computational}. Moreover, it is even more challenging to show that approximate solutions of the Brouwer-like problem, \ie, $(z,z')$ such that $z'=f(z)$ and $\|z-z'\|$ is small, can be used to build solutions of \QVI in polynomial time, \ie, an exact fixed point of the correspondence. A similar result is not presented in \cite{papadimitriou2023computational}.

\begin{wrapfigure}[15]{r}{0.3\textwidth} 
    \centering
    \vspace{-2mm}
    \scalebox{0.75}{\tikzset {_60g91buv4/.code = {\pgfsetadditionalshadetransform{ \pgftransformshift{\pgfpoint{0 bp } { 0 bp }  }  \pgftransformrotate{-225 }  \pgftransformscale{2 }  }}}
\pgfdeclarehorizontalshading{_kmskbclj5}{150bp}{rgb(0bp)=(0.62,0.79,0.55);
rgb(37.5bp)=(0.62,0.79,0.55);
rgb(62.5bp)=(0.02,0.62,0.75);
rgb(100bp)=(0.02,0.62,0.75)}
\tikzset{_06ifg4z2m/.code = {\pgfsetadditionalshadetransform{\pgftransformshift{\pgfpoint{0 bp } { 0 bp }  }  \pgftransformrotate{-225 }  \pgftransformscale{2 } }}}
\pgfdeclarehorizontalshading{_wr42lth3z} {150bp} {color(0bp)=(transparent!40);
color(37.5bp)=(transparent!40);
color(62.5bp)=(transparent!40);
color(100bp)=(transparent!40) } 
\pgfdeclarefading{_aq8gm2g2z}{\tikz \fill[shading=_wr42lth3z,_06ifg4z2m] (0,0) rectangle (50bp,50bp); } 
\tikzset{every picture/.style={line width=0.75pt}} %

\begin{tikzpicture}[x=0.75pt,y=0.75pt,yscale=-1,xscale=1]

\draw  [draw opacity=0][fill={rgb, 255:red, 192; green, 38; blue, 38 }  ,fill opacity=0.6 ] (250,70) -- (330,70) -- (330,150) -- (250,150) -- cycle ;
\draw  [draw opacity=0][shading=_kmskbclj5,_60g91buv4,path fading= _aq8gm2g2z ,fading transform={xshift=2}] (170,150) -- (250,150) -- (250,230) -- (170,230) -- cycle ;
\draw  [draw opacity=0][fill={rgb, 255:red, 158; green, 202; blue, 141 }  ,fill opacity=0.6 ] (250,150) -- (330,150) -- (330,230) -- (250,230) -- cycle ;
\draw  [draw opacity=0][fill={rgb, 255:red, 5; green, 157; blue, 192 }  ,fill opacity=0.6 ] (170,70) -- (250,70) -- (250,150) -- (170,150) -- cycle ;
\draw    (250,150) -- (250,73) ;
\draw [shift={(250,70)}, rotate = 90] [fill={rgb, 255:red, 0; green, 0; blue, 0 }  ][line width=0.08]  [draw opacity=0] (8.93,-4.29) -- (0,0) -- (8.93,4.29) -- cycle    ;
\draw    (250,150) -- (327,150) ;
\draw [shift={(330,150)}, rotate = 180] [fill={rgb, 255:red, 0; green, 0; blue, 0 }  ][line width=0.08]  [draw opacity=0] (8.93,-4.29) -- (0,0) -- (8.93,4.29) -- cycle    ;
\draw    (250,150) ;
\draw [shift={(250,150)}, rotate = 0] [color={rgb, 255:red, 0; green, 0; blue, 0 }  ][fill={rgb, 255:red, 0; green, 0; blue, 0 }  ][line width=0.75]      (0, 0) circle [x radius= 3.35, y radius= 3.35]   ;
\draw    (250,150) -- (288.75,101.56) ;
\draw [shift={(290,100)}, rotate = 128.66] [color={rgb, 255:red, 0; green, 0; blue, 0 }  ][line width=0.75]    (10.93,-3.29) .. controls (6.95,-1.4) and (3.31,-0.3) .. (0,0) .. controls (3.31,0.3) and (6.95,1.4) .. (10.93,3.29)   ;
\draw [color={rgb, 255:red, 0; green, 0; blue, 0 }  ,draw opacity=0.5 ] [dash pattern={on 4.5pt off 4.5pt}]  (170,150) -- (250,150) ;
\draw [color={rgb, 255:red, 0; green, 0; blue, 0 }  ,draw opacity=0.5 ] [dash pattern={on 4.5pt off 4.5pt}]  (250,230) -- (250,150) ;

\draw (248,153.4) node [anchor=north east] [inner sep=0.75pt]    {$v$};
\draw (275,96.6) node [anchor=south west] [inner sep=0.75pt]    {$p_{\gamma,\eta}(v)$};

\end{tikzpicture}}
    \caption{Example of Sperner's coloring induced by $p_{\gamma,\eta}(\cdot)$. The color \textbf{\textcolor{niceRed}{red}} corresponds to color $0$ while the colors \textbf{\textcolor{mgreen}{green}} and \textbf{\textcolor{blueGrotto}{blue}} correspond to colors $1$ and $2$, respectively.}
    \label{fig:HDSperner2}
    \vspace{-1cm}
\end{wrapfigure}
\paragraph{Construction of the Sperner's instance}
Consider an instance of \QVI in $d$ dimensions, with approximation parameters $\nu,\epsilon>0$ and Lipschitzness constant $L>0$. We then build an instance of \HDSperner in which each cublet has side-length $\mu$.
To simplify computations, we define the following parameters that depend on $\eta,\mu,\gamma$:
\[
    \ell:=2d\mu\left(\frac{4L\sqrt{d}}{\gamma}+L+\eta\right),\quad
    \kappa := \frac{2Ld\mu}{\gamma},\quad
    \omega:=\sqrt{d} \left( \mu+\kappa+\sqrt\frac{2\ell}{\eta}\right).
\]
Our construction works for any $\eta,\mu,\gamma$ that satisfy the following inequalities:
\[
\gamma+\sqrt{d}\omega\le \nu,\quad \sqrt{d}(1+\eta)\omega+\frac{d\sqrt{d}\omega}{\gamma}\le \epsilon.
\]

In \Cref{lem:ineqParam}, we prove that there exists a choice for $\eta,\mu,\gamma$ in which all have a value bounded polynomially in $\epsilon,\nu,L,d$ such that the above inequalities hold.
Let $V \subset [0,1]^d$ be the set of vertices of the simplicial decomposition of the hypercube $[0,1]^d$. As is customary in reductions related to Sperner's lemma and fixed-point theorems, we construct the coloring circuit $\mathcal{C}$ by leveraging the fact that the function $p_{\gamma,\eta}$ maps the hypercube to itself. Specifically, the color $i=0$ is assigned to a vertex $v \in V$ only if $(p_{\gamma,\eta}(v) - v)_i \geq 0$ for all $i \in [d]$. Conversely, color $i \in [d]$ is assigned to $v$ only if $(p_{\gamma,\eta}(v) - v)_i \leq 0$. Ties are broken arbitrarily as long as they do not violate Sperner's boundary conditions. Note that this circuit can be constructed in polynomial time using the polynomial-time algorithm that computes $p_{\gamma,\eta}(v)$. \Cref{fig:HDSperner2} illustrates a valid Sperner coloring from the function $p_{\gamma,\eta}(\cdot)$ when $d=2$. Moreover, since $p_{\gamma,\eta}$ maps the hypercube to itself, the boundary conditions for Sperner's coloring are automatically satisfied.
Note that the instance of \HDSperner is polynomial in the size of the \QVI instance. Indeed, $p_{\gamma,\eta}$ can be computed in polynomial time (by means of the ellipsoid algorithm), and the size of the input of the circuit $C$ is $O(d\cdot\log(1/\mu))$.

\paragraph{From a solution to Sperner to a solution of \QVI}

Consider a solution $\{v^i\}_{i \in [d]\cup \{0\}}$ of \HDSperner (without loss of generality we can assume that the vertices of the panchromatic simplex are numbered by their color, \ie $\mathcal{C}(v^i)=i$ for all $i\in[d]\cup\{0\}$).
The difficult part of the proof is to show that a solution to \HDSperner can be used to build solutions to \QVI efficiently.

Let $z^i=p_{\gamma,\eta}(v^i)$ for each $i$.
Then, by the definition of a solution to \HDSperner it holds $(z^0-v^0)_i\ge 0$ for all $i\in [d]$, and $(z^i-v^i)_i\le 0$ for each $i\in [d]$.
In the remaining of the proof, we show that $v^0$ is a solution to \QVI.
We start by showing that all the $z^i$ are close to $z^0$. This is not obvious due to the dependence of the set $Q_\gamma(v^i)$ on $v^i$. Our proof employs that the correspondence $Q_{\gamma}(\cdot)$ is Lipschitz with respect to the Hausdorff metric (\Cref{lem:Qbeta_is_cnt}).

\begin{lemma}\label{lm:proj}
    For all $i\in[d]$ we have that $\|z^0-z^i\|\le \kappa+\sqrt{\frac{2\ell}{\eta}}$.
\end{lemma}
\begin{proof}
    For each $i \in [d]$, let $\bar z^i=\Pi_{Q_{\gamma}(v^0)}(z^i)$ be the projection on $Q_{\gamma}(v^0)$ of $z^i$.
    We bound first $\|z^i-\bar z^i\|$, then we bound $\|z^0-\bar z^i\|$. Finally, we conclude the proof by a simple triangular inequality.
    
    Before proceeding with the proof, we notice that for any two vertexes $v,v'\in \{v^i\}_{i \in [d]\cup \{0\}}$ it holds 
    \begin{equation}\label{eq:C9tmp3}
        \|v-v'\|\le \sqrt{d}\mu
    \end{equation}
    since both vertexes belong to the same cubelet.
    \minipara{Bounding $\|z^i-\bar z^i\|$} First let us bound the difference between $z^i$ and $\bar z^i:=\Pi_{Q_{\gamma}(v^0)}(z^i)$.
    By \Cref{lem:Qbeta_is_cnt} and \Cref{def:Hausdorff}, we have that 
    \begin{align}
        \|z^i-\bar z^i\| & = \|\Pi_{Q_\gamma(v^0)}(z^i)-z^i\|\nonumber\\
        &\le \frac{2L \sqrt{d}}{\gamma} \|v^0-v^i\|\tag{$z^i\in Q_\gamma(v^i)$}\\
        &\le \frac{2L \sqrt{d}}{\gamma} \mu \sqrt{d}\nonumber\\
        &= \frac{2Ld\mu}{\gamma}\nonumber\\
        &:=\kappa\label{eq:C9tmp1}.
    \end{align}

    \minipara{Bounding $\|z^0-\bar z^i\|$}
    We define the function $\Phi_\eta(\tilde z, z):=F(z)^\top \tilde z+\frac{\eta}{2}\|\tilde z-z\|^2$ and thus we can write
    \begin{equation}\label{eq:stronglyconvex_distance}
    p_{\gamma,\eta}(z)=\Pi_{Q_\gamma(z)}(z)=\arg\min_{\tilde z\in Q_\gamma(z)}\Phi_\eta(\tilde z, z).
    \end{equation}
    Now we can use the properties of strongly convex functions to bound the distance between $p_{\gamma,\eta}(z)$ and $\tilde z$ in terms of the residual error of $\Phi_\eta(\tilde z, z)$. Indeed, since the function $\tilde z\mapsto\Phi(\tilde z, z)$ is $\eta$-strongly convex we know that:
    \begin{equation}
    \|p_{\gamma,\eta}(z)-\tilde z\|\le \sqrt{\frac{2}{\eta}(\Phi_\eta(\tilde z,z)-\Phi_\eta(p_{\gamma,\eta}(z),z))}.
    \end{equation}
    
    Moreover, we observe that for all $z\in[0,1]^d$ the function $\tilde z\mapsto \Phi_\eta(\tilde z, z)$ is $\sqrt{d}(1+\eta)$-Lipschitz as $\tilde z\mapsto F(z)^\top\tilde z$ is $(\|F(z)\|\le\sqrt{d})$-Lipschitz and $\tilde z\mapsto\|\tilde z-z\|^2$ is $2\sqrt{d}$-Lipschitz. Similarly the function $z\mapsto \Phi_\eta(\tilde z, z)$ is $\sqrt{d}(L+\eta)$-Lipschitz.

    Consider now the following inequalities:
    \begin{align}
        \Phi_\eta(\bar z^i, v^0) & \le \Phi_\eta(z^i, v^0)+\sqrt{d}(1+\eta)\|\bar z^i-z^i\|\nonumber\\
        &\le \Phi_\eta(z^i, v^i)+\sqrt{d}(1+\eta)\|\bar z^i-z^i\|+\sqrt{d}(L+\eta)\|v^i-v^0\|\nonumber\\
        &\le \Phi_\eta(z^i, v^i)+\frac{2L\mu d^{3/2}}{\gamma}(1+\eta)+d(L+\eta)\mu,\label{eq:C9tmp4}
    \end{align}

    where in the first inequality we used the Lipschitzness of $\tilde z\mapsto \Phi_\eta(\tilde z, v^0)$, in the second inequality we used the Lipschitzness of $ v\mapsto \Phi_\eta(z^i, v)$, and in the last inequality we used \Cref{eq:C9tmp1} and \Cref{eq:C9tmp3}.

    Define $\tilde z^i=\Pi_{Q_\gamma(v^i)}(z^0)$.
    Similarly to the first part of the proof, by \Cref{lem:Qbeta_is_cnt} and \Cref{def:Hausdorff} we have 
    \begin{equation}\label{eq:C9tmp2}
        \|\tilde z^i-z^0\|\le \frac{2L \sqrt{d}}{\gamma} \|v^i-v^0\|\le \frac{2L \sqrt{d}}{\gamma} \mu \sqrt{d}= \frac{2Ld\mu}{\gamma}.
    \end{equation}
    
    Now we combine \Cref{eq:C9tmp4} and the observation that $\Phi_\eta(z^i,v^i)\le \Phi_\eta(\tilde z^i,v^i)$ (since $\tilde z^i\in Q_{\gamma}(v^i)$ by construction) and we    
    use again the Lipschitzness of $\Phi_\eta$ in both its argument to show that
    \begin{align*}
    \Phi_\eta(\bar z^i, v^0)&\le \Phi_\eta(\tilde z^i, v^i) +\frac{2L\mu d^{3/2}}{\gamma}(1+\eta)+d(L+\eta)\mu\\
    &\le \Phi_\eta(\tilde z^i, v^0)+\frac{2L\mu d^{3/2}}{\gamma}(1+\eta)+2d(L+\eta)\mu\\
    &\le \Phi_\eta( z^0, v^0) +\frac{4L\mu d^{3/2}}{\gamma}(1+\eta)+2d(L+\eta)\mu\\
    &\le\Phi_\eta( z^0, v^0) +\ell.
    \end{align*}

    Now, in light of \Cref{eq:stronglyconvex_distance}, we can bound the distance between $z^0$ and $\bar z^i$ as follows:
    \begin{align*}
        \|z^0-\bar z^i\|&=\|p_{\gamma,\eta}(v^0)-\bar z^i\|\\
        &\le \sqrt{\frac{2}{\eta}(\Phi_\eta(\bar z^i,v^0)-\Phi_\eta(\bar z^0,v^0))}\le\sqrt{\frac{2\ell}{\eta}} 
    \end{align*}

    This concludes the proof of the lemma.

\end{proof}

We can exploit the previous lemma to show that $v^0$ is an approximate fixed-point of $p_{\gamma,\eta}(\cdot)$.
\begin{lemma}\label{lm:inclusion}
  It holds that $\|z^0- v^0\|\le \omega$.
\end{lemma}

\begin{proof}
    First, we recall that by the definition of a solution to \HDSperner, it holds $(z^0-v^0)_i\ge 0$ for all $i\in[d]$, and $(z^i-v^i)_i\le 0$ for each $i\in [d]$.
    
    Then, take an $i\in [d]$ and consider the following chain of inequalities:
    \begin{align*}
    (z^0- v^0)_i&\le (z^i- v^0)_i + \kappa+\sqrt{ \frac{2\ell}{\eta}}  \\
    & \le  (z^i- v^i)_i + \mu + \kappa+\sqrt{ \frac{2\ell}{\eta}} \\
    &\le \mu + \kappa+\sqrt{ \frac{2\ell}{\eta}},
    \end{align*}
where in the first inequality we used again \Cref{lm:proj}, in the third one $|(v^i-v^0)_i|\le \mu$, and the last one from $(z^i-v^i)_i\le 0$ by the properties of the Sperner's coloring.  
Hence, $\|z^0- x^0\| \le   \sqrt{d} \left(\mu +  \kappa+\sqrt{ \frac{2\ell}{\eta}} \right):=\omega$.
\end{proof}

Then, we show that $v^0$ belong to the set $Q_{\nu}(v^0)$.

\begin{lemma}
 It holds $v_0 \in Q_{\nu}(v^0)$.
\end{lemma}

\begin{proof}
We have that $z^0=p_{\gamma,\eta}(v^0) \in Q_\gamma(v^0)$. Hence, by definition of the set $Q_\gamma(v^0)$, we have that $A(v^0) z^0 \le b(v^0) + \gamma 1_d$.
This implies that 
\[
A(v^0) v^0 \le b(v^0)+ \gamma 1_d + \sqrt{d}||z^0-v^0|| 1_d \le b(v^0)+ (\gamma+\sqrt{d}\omega) 1_d.
\]
Hence,
\(
v^0 \in Q_{\gamma+\sqrt{d} \omega}(v^0).
\)
Since $\nu\ge\gamma+ \sqrt{d}\omega$, this concludes the proof.
\end{proof}

Now, we can show that $v_0$ is also a approximate solution of $\arg \min_{z \in Q_{\nu}(v^0)} F(v^0)^\top z$.

\begin{lemma}\label{lm:PPADFinal}
    Let $v^0$ be the vertex of a panchromatic simplex of the \HDSperner instance colored with the color $0$. Then,
    \[
    F(v^0)^\top v^0 \le \min_{z \in Q_{\nu}(v^0)} F(v^0)^\top z+\epsilon.
    \]
\end{lemma}

\begin{proof}
It holds 
\begin{align*}
    F(v^0)^\top v^0 & =F(v^0)^\top v^0 +\frac{\eta}{2} \| v^0-v^0\|^2 \\
    &\le \Phi_\eta(z^0,v^0) + \sqrt{d}(1+\eta)\|v^0-z^0\|\tag{$\sqrt{d}(1+\eta)$-Lip.~of $\Phi_\eta(\cdot, v^0)$}\\
    &=F(v^0)^\top z^0+\frac{\eta}{2} \| z^0-v^0\|^2+ \sqrt{d}(1+\eta)\|z^0-v^0\|\\
    &=\min_{z\in Q_{\gamma}(v^0)}\left\{F(v^0)^\top z +\frac{\eta}{2}\|z-v^0\|^2\right\}+ \sqrt{d}(1+\eta)\|z^0-v^0\|\\
    &\le \min_{z\in Q_{\gamma}(v^0)} F(v^0)^\top z +\frac{\eta}{2}\|z^0-v^0\|^2+ \sqrt{d}(1+\eta)\|z^0-v^0\|\\
    &\le \min_{z\in Q_{\gamma}(v^0)} F(v^0)^\top z +\frac{\eta}{2}\omega^2+ \sqrt{d}(1+\eta)\omega\tag{\Cref{lm:inclusion}}\\
    &\le \min_{z\in Q_{\gamma+\sqrt{d}\omega}(v^0)} F(v^0)^\top z + \frac{\eta}{2}\omega^2+\sqrt{d}(1+\eta)\omega+\frac{d\sqrt{d}\omega}{\gamma}\tag{\Cref{lm:lipDown}}\\
    &\le \min_{z\in Q_{\nu}(v^0)} F(v^0)^\top z + \epsilon.
\end{align*}

where in the last inequality we use $\gamma+\sqrt{d}\omega\le \nu$, $\frac{\eta}{2}\omega^2+\sqrt{d}(1+\eta)\omega+\frac{d\nu}{\gamma} \le \epsilon$, and that $Q_{\gamma+\sqrt{d}\omega}(v^0)\subseteq Q_\nu(v^0)$.
\end{proof}

To conclude the proof and show that $v^0$ is a solution to \QVI, we only need to observe that:
\[
F(v^0)^\top (z-v^0) \ge F(v^0)^\top (v^0-v^0) - \epsilon = -\epsilon \quad \forall z \in Q_{\nu}(v^0), 
\]
where the first inequality follows from \Cref{lm:PPADFinal}.

We are only left to show that there exists a set of sufficiently large parameters $\eta,\mu$ and $\gamma$ that satisfy the required inequalities.

\begin{lemma}\label{lem:ineqParam}
    For any given $d\in\Naturals$, any $\epsilon,\nu,L>0$ and any 
    \[
    \eta\le\min\left(1,L,d^{1/4},\sqrt{\frac{\nu}{2}},\sqrt{2dL}, \frac{1}{L^{1/8}},\frac{1}{8d^2\sqrt{2L}},\frac{1}{2(d\sqrt{2dL})^{1/3}},\frac{\epsilon}{24d^3\sqrt{2dL}}\right),
    \]
    let $\mu=\eta^9,\gamma=\eta^2$, and let $\ell$, $\kappa$, and $\omega$ be defined as above.
Then: 
\[
\gamma+\sqrt{d}\omega\le \nu,\quad\frac{\eta}{2}\omega^2+\sqrt{d}(1+\eta)\omega+\frac{d\sqrt{d}\omega}{\gamma} \le \epsilon
\]
\end{lemma}

\begin{proof}
It is easy to see that $\kappa=2Ld\eta^7$ and that
\begin{align*}
    \ell&= 2d\eta^9\left(\frac{4L\sqrt{d}}{\eta^2}+L+\eta\right)\\
    &\le 4d\eta^9L\left(\frac{2\sqrt{d}}{\eta^2}+1\right)\\
    &\le 16\eta^7d^2L,
\end{align*}
where in the first inequality we used that $\eta\le L$ and in the second that $\eta\le d^{1/4}$.
Plugging this upper bounds into $\omega$ we obtain:
\begin{align*}
    \omega=&\sqrt{d} \left(\mu+\kappa+\sqrt{\frac{2\ell}{\eta} } \right)\\
    &=\sqrt{d}\left(\eta^9+2\eta^7dL+4\eta^3d\sqrt{2L}\right)\\
    &\le\sqrt{d}\left(4\eta^7dL+4\eta^3d\sqrt{2L}\right)\\
    &\le 8\eta^3d\sqrt{2dL},
\end{align*}
where in the first inequality we used the upper bound on $\ell$ computed above, in the second inequality we used that $\eta\le\sqrt{2dL}$, the third that $\eta\le \frac{1}{L^{1/8}}$.
Now the first inequality is readily satisfied:
\begin{align*}
    \gamma+\sqrt{d}\omega&\le \eta^2+8\eta^3d^2\sqrt{ 2 L }\\
    &\le 2 \eta^2\\
    &\le \nu,
\end{align*}
where the first inequality holds by the definition of $\gamma$ and the upper bound on $\omega$ computed above, the second follows from the fact that $\eta\le\frac{1}{8d^2\sqrt{2L}}$ and the last from $\eta\le\sqrt{\nicefrac{\nu}{2}}$.
Moreover, the second inequality is also satisfied. First observe that for $\eta\le \frac{1}{2(d\sqrt{2dL})^{1/3}}$ then $\omega\le 1$ and
\begin{align*}
    \frac{\eta}{2}\omega^2+\sqrt{d}(1+\eta)\omega+\frac{d\sqrt{d}\omega}{\gamma}&\le \omega\left(\frac{\eta}{2}+2\sqrt{d}+\frac{d\sqrt{d}}{\gamma}\right)\\
    &=\omega\left(\frac{\eta}{2}+2\sqrt{d}+\frac{d\sqrt{d}}{\eta^2}\right)\\
    &\le 3\omega \frac{d^2}{\eta^2}\\
    &\le 24 \eta d^3\sqrt{2dL},
\end{align*}
where first inequality comes from $\omega\le 1$, the second from $\eta\le d^{1/4}$, the third from the upper bound on $\omega$ and $d\ge 1$, while the last comes from $\eta\le\frac{\epsilon}{24d^3\sqrt{2dL}}$. This concludes the proof of the lemma.
\end{proof}

\subsection{Proof of \Cref{th:globalization}}\label{app:global}
\globalization*
\begin{proof}
    The proof is immediate by convexity. Take any $x'\in K_1^\nu(y)$ and consider $x''=x^\star+\frac{\delta}{\sqrt{d}}(x'-x^\star)$. Note that $\|x''-x^\star\|=\frac{\delta}{\sqrt{d}}\|x'-x^\star\|\le \delta$. Since $(x^\star,y^\star)$ is a $(\epsilon,\delta,\nu)$ solution to \MINMAX, and since $f(\cdot, y^\star)$ is convex, it holds:
    \begin{align*}
        f(x^\star,y^\star)&\le f(x'',y^\star)+\epsilon\le\left(1-\frac{\delta}{\sqrt{d}}\right) f(x^\star,y^\star) +\frac{\delta}{\sqrt{d}} f(x',y^\star)+\epsilon.
    \end{align*}
    Rearranging we obtain
    \(
    f(x^\star,y^\star)\le f(x',y^\star)+\frac{\epsilon\sqrt{d}}{\delta},
    \)
    as desired.
\end{proof}

\section{Omitted Proofs for \Cref{sec:Harness}}\label{app:omitted hard}

\linvihard*
\begin{proof}
    First, we show the membership of \LINVI in \PPAD by reduction to \Brouwer. Then we show hardenss by reducing from \POLY.
    \minipara{Membership} We reduce the problem to \Brouwer. Let $\rho, D$ and $c$ be the inputs defining the \LINVI instance.
    Let $K=[0,1]^d$.
    Consider the function $F(z):=\Pi_K(z-(Dz+c))$. Since the set $K$ is a polytope with rational coefficients, we can compute the $F(z)$ efficiently. Also note that $F:K\to K$ and it is $(1+d)$-Lipschitz. We feed $F$ to \Brouwer with approximation $\alpha:=\frac{\rho}{2\sqrt{d}}$.
    Thus it admits a fixed point $\|z-F(z)\|\le \alpha$.
    By \Cref{th:VI_eq_sgda} we know that for all $z'\in K$ we have
    \[
    (Dz+c)^\top(z'-z)\ge -2\alpha\sqrt{d}=-\rho,
    \]
    thus proving it is a solution to \LINVI.

    \minipara{Hardness} 
    We reduce from \POLY with two actions, graphs with degree $3$, and approximation $\epsilon^\star$.
    Recall that we formally define \POLY in \Cref{pr:polymatrix}.
    Polymatrix games with binary actions are defined by an undirected graph $(V, E)$, where each vertex represents a player. We use $n=|V|$ to denote the number of players in the game.
    Each player $i \in V$ has two actions $a_1$ and $a_2$.
    For each edge $(i, j) \in E$, there is an $2 \times 2$ matrix $A_{i,j}  \in [0,1]^{2\times 2}$ giving the payoffs that player $i$ obtains from their interaction with player $j$. Formally, we let
       \[
    A_{i,j}=\begin{bmatrix}
        a_{i,j}&b_{i,j}\\
        c_{i,j}&d_{i,j}
    \end{bmatrix}.
    \]
    Similarly, a $2 \times 2$ matrix $A_{j,i}$ gives payoffs for player $j$’s interaction with player $i$.

    We represent a mixed strategy of player $i$ as $x_i\in [0,1]$, where $x_i$ is the probability of playing action $a_1$ and $1-x_i$ the probability of playing action $a_2$.
    Then, the utility of player $i$ is given by
    \begin{align*} 
    U_i(x_i,x_{-i})&= \sum_{j:(i,j)\in E}\left( x_i x_{j} a_{i,j} + x_i (1-x_j) b_{i,j}+ (1-x_i) x_j b_{i,j}   + d_{i,j}\right) \\
    &= x_i\cdot \left(\sum_{j:(i,j)\in E} (x_j\alpha_{i,j}+\beta_{i,j})\right)+ \sum_{j:(i,j)\in E} (x_j\gamma_{i,j}+\delta_{i,j}),
    \end{align*}
    where $\alpha_{i,j}=a_{i,j}-b_{i,j}+c_{i,j}-d_{i,j}$, $\beta_{i,j}=d_{i,j}-c_{i,j}$, $\gamma_{i,j}=b_{i,j}-c_{i,j}$ and $\delta_{i,j}=c_{i,j}$.

    Then, the goal is to find an $\epsilon^\star$-approximate solution to \POLY, \ie, an $x\in [0,1]^d$ that for each $i \in V$ satisfies
     \[
    U_i(x_i,x_{-i})\ge U_i(x'_i,x_{-i})-\epsilon^\star \quad \forall x_i'\in[0,1].
    \]
    
    As we recalled in \Cref{th:polymatrix}, there exists a constant $\epsilon^\star$ such that polymatrix is $\PPAD$-hard  even for graphs with degree $3$ \citep[Theorem~1]{rubinstein2015inapproximability}, \citep[Theorem~4.4]{deligkas2024pure}.\footnote{Notice that we do not try to optimize constants. For example, we could exploit that the payoffs of polymatrix games we reduce from are normalized.}

    Consider a polymatrix game with graph $(V,E)$ and matrices $A_{i,j}$. Without loss of generality, we identify the set of vertices $V$ with $[n]$. Then, $x\in[0,1]^{n}$ represents the joint strategy for all players.
    
    Notice that
    $U_i(x_i,x_{-i})\ge U_i(x'_i,x_{-i})-\epsilon^\star$ if and only if
    \[
    x_i\cdot \left(\sum_{j:(i,j)\in E} (x_j\alpha_{i,j}+\beta_{i,j})\right)\ge x'_i\cdot \left(\sum_{j:(i,j)\in E} (x_j\alpha_{i,j}+\beta_{i,j})\right)-\epsilon^\star.
    \]

    Now we define the instance of \LINVI.
    We define $d=n$ variables $z_i$ for each $i\in [n]$. The approximation factor is $\rho^\star=\frac{\epsilon^\star}{6}$. The linear operator is defined by a matrix $D\in[-1,1]^{d\times d}$
    \[
    D_{i,j}=-\frac{\alpha_{i,j}\mathbb{I}((i,j)\in E)}{6}
    \]
    and a vector $c\in [-1,1]^d$

    \[c_{i}= - \frac{\sum_{j:(i,j)\in E}\beta_{i,j}}{6}.\]
 	
 	Notice that since the graph has degree $3$ and  $\beta_{i,j} \in [-1,1]$, $D$ and $c$ have every entry bounded in $[-1,1]$.
    Also, the bounded degree of the graph implies that the sum of the absolute values on each row and each column is bounded by $1$, \ie $\|D\|_\infty,\|D\|_1\le 1$.

    Note that the $i$-th component of $Dz+c$ is equal to
    \begin{equation}\label{eq:Dz_LINVIhard}
       \frac{1}{2d} \left[\sum_{j:(i,j)\in E}z_j\alpha_{i,j}+ \sum_{j:(i,j)\in E} \beta_{i,j}\right].
    \end{equation}

    Now any solution $z\in K$ to \LINVI guarantees that
    \[
    (Dz+c)^\top(z'-z)\ge -\rho^\star\quad \forall z'\in [0,1]^d.
    \]
	
	Our reduction holds even for a weaker requirement. In particular, it is sufficient to consider deviation on a single component. Formally, we show that any $z\in K$ such that
	
	  \[
	(Dz+c)_i\cdot(z'_i-z_i)\ge -\rho^\star \quad \forall i \in [d], z'_i\in [0,1]
	\]

   Indeed,  this clearly implies that for each $i\in [d]$
    \[
    z'_i(Dz+c)_i\ge z_i(Dz+c)_i-\rho^\star \quad \forall z_i'\in[0,1].
    \]
    Using \Cref{eq:Dz_LINVIhard}  we get that the previous inequality implies that for all $i\in[d]$
    \[
    z'_i\left(\sum_{j:(i,j)\in E}z_j\alpha_{i,j}+ \sum_{j:(i,j)\in E} \beta_{i,j}\right)\le z_i\left(\sum_{j:(i,j)\in E}z_j\alpha_{i,j}+ \sum_{j:(i,j)\in E} \beta_{i,j}\right)+6 \rho^\star \quad\forall z_i'\in[0,1],
    \]
    showing that $z$ is a solutions to polymatrix instance since $\rho^\star =\frac{\epsilon^\star}{6}$. 
\end{proof}

\lemmagdamm*
\begin{proof}
    We prove the result for the $x$ player as symmetric arguments hold for the $y$ player.
    Consider the projection $\Pi_{K_1^\nu(y^\star)}(x^\star-\nabla_x f(x^\star,y^\star))$ and any $i\in[d]$. Since $K_1(y)=\bigtimes_{j\in[d]}K_{1,j}(y_j)$ we have that the projection is independent among components \citep[Section~8.1.1]{boyd2004convex}. In particular, its $i$-th component is equal to 
    \[\Pi_{K_{1,i}^\nu(y^\star_i)}(x^\star_i-\partial_{x_i} f(x^\star,y^\star)).\]
    Thus, since $\|(x^\star,y^\star)-F^\nu_\GDA(x^\star,y^\star)\|\le \alpha$ we have that
    \[
    |x^\star_i-\Pi_{K_{1,i}^\nu(y^\star_i)}(x^\star_i-\partial_{x_i} f(x^\star,y^\star))|\le \alpha.
    \]
    Now only have to consider single-dimensional problems and it is easy to see that this implies that either $|\partial_{x_i} f(x^\star,y^\star)|\le \alpha$ (when no projection incurs) or $d(x_i^\star,\partial K_{1,i}^\nu(y^\star_i))\le \alpha$, where we denote with $\partial K_{1,i}^\nu(y^\star_i)$ the boundary of the set $K_{1,i}^\nu(y^\star_i)$.
    Notice that in the second case  $x_i^\star-\partial_{x_i}f(x^\star,y^\star)$ gets projected on the boundary of the interval $K^\nu_{1,j}(y_j)$). Hence, the distance of $x_i^\star$ from the boundary of $K_{1,i}^\nu(y^\star_i)$ must be at most $\alpha$.
    Now we bound $\partial_{x_i}f(x^\star,y^\star)(x^\star_i-x_i')$ with  different arguments depending on the case.
    \minipara{Case 1: $|\partial_{x_i} f(x^\star,y^\star)|\le \alpha$} In this case it is trivial to observe that $\partial_{x_i}f(x^\star,y^\star)(x^\star_i-x_i')\le |\partial_{x_i}f(x^\star,y^\star)|\cdot |x^\star_i-x_i'|\le \alpha$, since by assumption $|\partial_{x_i} f(x^\star,y^\star)|\le \alpha$ and $x_i,x_i'\in[0,1]$.
    \minipara{Case 2: $d(x_i^\star,\partial K_{1,i}^\nu(y^\star_i))\le \alpha$} We denote the interval $K_{1,i}^\nu(y^\star_i))$ with $[a,b]$. We now consider two sub-cases
    \begin{itemize}
        \item $\partial_{x_i} f(x^\star,y^\star)\ge 0$. Then this means that $a\le x_i^\star\le a+\alpha$. If $x'_i\in[a,x_i^\star]$ then $\partial_{x_i}f(x^\star,y^\star)(x^\star_i-x_i')\le \alpha\cdot G_\infty$. If $x_i'> x_i^\star$ then $\partial_{x_i}f(x^\star,y^\star)(x^\star_i-x_i')\le0$. 
        \item $\partial_{x_i} f(x^\star,y^\star)<0$. This means that $b-\alpha\le x_i^\star\le b$. If $x'_i\in[x_i^\star, b]$ then $\partial_{x_i}f(x^\star,y^\star)(x^\star_i-x_i')\le \alpha\cdot G_\infty$. if $x_i'< x_i^\star$ then $\partial_{x_i}f(x^\star,y^\star)(x^\star_i-x_i')\le 0$. 
    \end{itemize}
    In conclusion we have that $\partial_{x_i}f(x^\star,y^\star)(x^\star_i-x_i')\le \max(\alpha,\alpha G_\infty)\le \alpha(1+G_\infty)$.

    Now a simple smoothness argument finishes the proof. Indeed $L$-smoothness of $f$ implies that
    \[
    f(x^\star,y^\star)\le f(x',y^\star)\le \partial_{x_i}f(x^\star,y^\star)(x^\star_i-x_i')+\frac{L}{2}\|x^\star-x'\|^2\le \alpha(1+G_\infty)+\frac{L}{2}\delta^2,
    \]
    concluding the proof.
\end{proof}

\subsection{Proof of \Cref{th:nonconvexconcavehardness}}\label{sec:formalproof}

\paragraph{Construction of the hard instance}

Take any \LINVI instance defined by the matrix $D\in[-1,1]^{d\times d}$, a vector $c\in [-1,1]^d$ and approximation $\alpha_\LINVI=\rho^*$, where $\rho^*$ is the constant in \Cref{th:linearvihard}. We showed that this problem is \PPAD-hard in \Cref{th:linearvihard}. 

\paragraph{Objective function}
The objective function $f:\Reals^d\times\Reals^d\to [-B,B]$ is given by
\[
f(x,y)=(x-y)^\top (Dx+c).
\]
We observe a few facts about the function $f$. This function is linear in $y$ and nonconvex quadratic in $x$, and hence it is a nonconvex-concave function. Moreover, the function $f$ is a second-degree polynomial in $x$ and $y$.
Finally, since in \Cref{th:linearvihard}  we can restrict to $D$ such that  $\|D\|_1,\|D\|_\infty\le 1$, the objective satisfies the following smoothness conditions.

\begin{restatable}{lemma}{lemLip}
    The function $f$ is $G$-Lipschitz and $L$-smooth with $G=5\sqrt{d}$ and $L=7$.
\end{restatable}
\begin{proof}
We recall that $\nabla_x f(x,y)=Dx+c+D^\top(x-y)$ and $\nabla_yf(x,y)=-(Dx+c)$.
We bound the Lipschitz parameter of $f$ as the supremum of the norm of the gradient. For all $(x,y)\in[0,1]^{2d}$ we have
\begin{align*}
    \|\nabla f(x,y)\|^2&=\|Dx+c+D^\top(x-y)\|^2+\|Dx+c\|^2\\
    &\le 6\|Dx\|^2+6\|c\|^2+2\|D^\top(x-y)\|^2  \tag{parallelogram inequality}\\
    &\le 6\|x\|^2+ 6d +2\|x-y\|^2  \tag{\Cref{lem:matrix_norms3}}\\
    &\le 20 d, \tag{since $\|x\|_2, \|y\|_2\le\sqrt{d}$}
\end{align*}
    where in the last inequality we exploit that $\|D\|\le\sqrt{\|D\|_1\|D\|_\infty}$ (\Cref{lem:matrix_norms3}) and that $\|D\|,\|D^\top\|\le 1$.
  Taking the square root on both sides, we conclude that the function $f$ is $G$-Lipschitz with $L\le 5\sqrt{d}$. 
  We now focus on the smoothness parameter of the function (\ie, the Lipschitz parameter of the function's gradient). 
  
  Then, it holds
\begin{align*}
    \|\nabla f(x,y)-\nabla f(x',y')\|^2&= \|D(x-x') + D^\top (x-x'-(y-y'))\|^2 + \|D(x-x')\|^2\\
    &\le 3\|D(x-x')\|^2+2\|D^\top(x-x'-(y-y'))\|^2\\
    &\le 3\|x-x'\|^2+4(\|x-x'\|^2+\|y-y'\|^2)\\
    &\le 7\|(x,y)-(x',y')\|^2,
\end{align*}
where we recall that $\|D\|\le 1$.
This proves that $L\le 7$ and concludes the proof.
\end{proof}

\paragraph{Constraints}
We define the constraint function $g$ by the induced set $K=\{(x,y)\in[0,1]^{2d}: \|x-y\|_\infty\le\Delta\}$, where $\Delta= \frac{\rho^*}{4}$. It is easy to see that there exists a function $g$ that implements such set.

In the following, we first prove the hardness result for \LRMINMAXJC. 
Then, we exploit the reduction from \LRMINMAXJC to \GDAFPJC in \Cref{lem:from_GDA_to_MM} to extend the hardness result to the problem of finding fixed points of gradient ascent-descent dynamics.
For both settings, restricting to single-dimensional deviations will be essential to provide tighter hardness results.

\paragraph{From \LRMINMAXJC to \LINVI} 

We provide a reduction for any $0<\gamma\le 1$, $\delta = \gamma \frac{\rho^*}{15}$, and $\epsilon=\gamma\frac{\rho^{*,2}}{60}$. It is easy to see that $\delta < \sqrt{\frac{2\epsilon}{L}}$ for any $\gamma\in(0,1]$ and thus we are in the local regime.
Now we want to prove that any solution to \LRMINMAXJC $(x^\star,y^\star)$ is a solution to \LINVI. 
To obtain a stronger hardness result, we focus on the specialized (but still \PPAD-Hard as proved in \Cref{th:linearvihard}) problem of finding an $x^\star$ that is robust to single-dimensional deviations, \ie
\begin{equation}\label{eq:VI_inside_MM}
 (Dx^\star+c)_i\cdot(x'_i-x^\star_i)\ge -\rho^*\quad\forall i \in [d], \, x'_i\in[0,1].
\end{equation}

For each $i\in [d]$ and $x'_i\in [0,1]$, we consider two cases depending on whether or not the deviation is \emph{inside} in the segment $(x_i^\star,y^\star_i)$, \ie, $x_i'\in[\min(x_i^\star,y^\star_i), \max(x_i^\star,y^\star_i)]$. \Cref{fig:cases} illustrates the different cases.

\minipara{Case $1$} We start from the case in which $x_i'\in[\min(x_i^\star,y^\star_i), \max(x_i^\star,y^\star_i)]$. Then, 
\[ 
(Dx^\star+c)_i\cdot(x'_i-x^\star_i)\ge - |(Dx^\star+c)_i| \cdot |x'_i-x^\star_i|  \ge - 2 \Delta \ge -\rho^*.
\]

\minipara{Case $2$} Conversely, we consider the case $x_i'\notin[\min(x_i^\star,y^\star_i), \max(x_i^\star,y^\star_i)]$. We denote by $Q_i=\mathbb{I}(|x_i'-x_i^\star|\ge|x'_i-y_i^*|)$ the indicator of whether $x_i^\star$ is closer to  $y_i^\star$ than $x_i'$ ($Q_i=1$) or not ($Q_i=0$). Then, we define the deviations
\[
\begin{cases}
    \bar x_i=\delta Q_i (x_i'-x_i^\star) + x^\star_i\\
    \bar x_j = x_j^\star&\forall j\neq i,
\end{cases}
\qquad\text{and}\qquad
\begin{cases}
    \bar y_i= \delta (1-Q_i) (x_i'-x_i^\star) + y^\star_i\\
    \bar y_j = y^\star_j&\forall j\neq i.
\end{cases}
\]
Intuitively, we move the point farther from $x'$ ($x^\star$ or $y^\star$) closer to it by an amount $\delta(x_i'-x_i^\star)$. 
Few considerations about $\bar x$ and $\bar y$ are in order. First, recall that $x_i'$ is outside of the interval $(x_i^\star, y_i^\star)$  by hypothesis. Hence, when $Q_i=1$, it must be the case that $x_i^\star-y_i^\star$ and $x'_i-x_i^\star$ have opposite sign. Similarly, when $Q_i=0$, it is the case that $y_i^\star-x_i^\star$ and $x'_i-x_i^\star$ have opposite sign.
Finally, it is also easy to see that $\|\bar x-x^\star\| \le \delta$ and $\|\bar y-y^\star\| \le \delta$. 

\begin{figure}[t!]
\centering
\begin{subfigure}{0.3\textwidth}
    \centering
    \scalebox{0.65}{\tikzset{every picture/.style={line width=0.75pt}} %
\tikzstyle{every node}=[font=\Large]

\begin{tikzpicture}[x=0.75pt,y=0.75pt,yscale=-1,xscale=.7]

\draw    (80,120) -- (417,120) ;
\draw [shift={(420,120)}, rotate = 180] [fill={rgb, 255:red, 0; green, 0; blue, 0 }  ][line width=0.08]  [draw opacity=0] (8.93,-4.29) -- (0,0) -- (8.93,4.29) -- cycle    ;
\draw    (80,115) -- (80,125) ;
\draw    (170,115) -- (170,125) ;
\draw    (260,115) -- (260,125) ;
\draw    (365,115) -- (365,125) ;
\draw    (220,115) -- (220,125) ;

\draw (78,128.4) node [anchor=north east] [inner sep=0.75pt]    {$0$};
\draw (367,128.4) node [anchor=north west][inner sep=0.75pt]    {$1$};
\draw (170,128.4) node [anchor=north] [inner sep=0.75pt]    {$x_{i}^{\star }$};
\draw (260,128.4) node [anchor=north] [inner sep=0.75pt]    {$y_{i}^{\star }$};
\draw (219,128.4) node [anchor=north] [inner sep=0.75pt]    {$x'_{i}$};

\end{tikzpicture}}
    \caption{Case 1}
    \label{fig:case1}
\end{subfigure}
\hfill
\begin{subfigure}{0.3\textwidth}
    \centering
    \scalebox{0.65}{\tikzset{every picture/.style={line width=0.75pt}} %
\tikzstyle{every node}=[font=\Large]

\begin{tikzpicture}[x=0.75pt,y=0.75pt,yscale=-1,xscale=.7]

\draw    (80,120) -- (417,120) ;
\draw [shift={(420,120)}, rotate = 180] [fill={rgb, 255:red, 0; green, 0; blue, 0 }  ][line width=0.08]  [draw opacity=0] (8.93,-4.29) -- (0,0) -- (8.93,4.29) -- cycle    ;
\draw    (80,115) -- (80,125) ;
\draw [line width=2.5]    (170,120) -- (217,120) ;
\draw [shift={(220,120)}, rotate = 180] [color={rgb, 255:red, 0; green, 0; blue, 0 }  ][line width=1.5]    (14.21,-4.28) .. controls (9.04,-1.82) and (4.3,-0.39) .. (0,0) .. controls (4.3,0.39) and (9.04,1.82) .. (14.21,4.28)   ;
\draw    (170,115) -- (170,125) ;
\draw    (260,115) -- (260,125) ;
\draw    (310,115) -- (310,125) ;
\draw    (365,115) -- (365,125) ;
\draw    (220,115) -- (220,125) ;

\draw (78,128.4) node [anchor=north east] [inner sep=0.75pt]    {$0$};
\draw (367,128.4) node [anchor=north west][inner sep=0.75pt]    {$1$};
\draw (170,128.4) node [anchor=north] [inner sep=0.75pt]    {$x_{i}^{\star }$};
\draw (260,128.4) node [anchor=north] [inner sep=0.75pt]    {$y_{i}^{\star }$};
\draw (310,128.4) node [anchor=north] [inner sep=0.75pt]    {$x'_{i}$};
\draw (220,128.4) node [anchor=north] [inner sep=0.75pt]    {$\bar x_{i}$};
\draw (195,116.6) node [anchor=south] [inner sep=0.75pt]    {$\delta $};

\end{tikzpicture}}
    \caption{Case 2. $Q_i=0$}
    \label{fig:case2}
\end{subfigure}
\hfill
\begin{subfigure}{0.3\textwidth}
    \centering
    \scalebox{0.65}{\tikzset{every picture/.style={line width=0.75pt}} %
\tikzstyle{every node}=[font=\Large]

\begin{tikzpicture}[x=0.75pt,y=0.75pt,yscale=-1,xscale=.7]

\draw    (80,120) -- (417,120) ;
\draw [shift={(420,120)}, rotate = 180] [fill={rgb, 255:red, 0; green, 0; blue, 0 }  ][line width=0.08]  [draw opacity=0] (8.93,-4.29) -- (0,0) -- (8.93,4.29) -- cycle    ;
\draw    (80,115) -- (80,125) ;
\draw [line width=2.5]    (260,120) -- (223,120) ;
\draw [shift={(220,120)}, rotate = 360] [color={rgb, 255:red, 0; green, 0; blue, 0 }  ][line width=1.5]    (14.21,-4.28) .. controls (9.04,-1.82) and (4.3,-0.39) .. (0,0) .. controls (4.3,0.39) and (9.04,1.82) .. (14.21,4.28)   ;
\draw    (170,115) -- (170,125) ;
\draw    (260,115) -- (260,125) ;
\draw    (118.5,115) -- (118.5,125) ;
\draw    (365,115) -- (365,125) ;
\draw    (219,115) -- (219,125) ;

\draw (78,128.4) node [anchor=north east] [inner sep=0.75pt]    {$0$};
\draw (367,128.4) node [anchor=north west][inner sep=0.75pt]    {$1$};
\draw (170,128.4) node [anchor=north] [inner sep=0.75pt]    {$x_{i}^{\star }$};
\draw (260,128.4) node [anchor=north] [inner sep=0.75pt]    {$y_{i}^{\star }$};
\draw (118.5,128.4) node [anchor=north] [inner sep=0.75pt]    {$x'_{i}$};
\draw (219,128.4) node [anchor=north] [inner sep=0.75pt]    {$\bar x_{i}$};
\draw (240,116.6) node [anchor=south] [inner sep=0.75pt]    {$\delta $};

\end{tikzpicture}}
    \caption{Case 2. $Q_i=1$}
    \label{fig:case3}
\end{subfigure}
\caption{Illustration of the case analysis. In \Cref{fig:case1} we have $x_i'\in[x_i^\star,y_i^\star]$. While in \Cref{fig:case2} and \Cref{fig:case3} we are in the setting in which $x_i'\notin[\min(x_i^\star,y^\star_i), \max(x_i^\star,y^\star_i)]$, where $Q_i=0$ and $Q_i=1$, respectively.}
\label{fig:cases}
\end{figure}

We now show that $\bar x$ and $\bar y$ are both in $[0,1]^d$ and $\|\bar y-x^\star\|_\infty,\|\bar x-y^\star\|_\infty\le \Delta$.
We break the analysis into two cases. When $Q_i=1$, then $\bar y=y^\star$ and thus $\bar y\in[0,1]^d$ and $\|\bar y-x^\star\|_\infty\le\Delta$ are trivially satisfied.
Moreover, $\bar x$ is a convex combination of $x^\star$ and $x'$, and thus it is guaranteed that $\bar x\in[0,1]^d$. Furthermore, since $\bar x$ is obtained by changing the $i$-th component of $x^\star$, we have
\[
\|\bar x-y^\star\|_\infty \le \max(\|x^\star-y^\star\|_\infty, |\bar x_i-y_i^\star|).
\]

We argue that $|\bar x_i-y_i^\star|$ is smaller then $\Delta$. Indeed we can rewrite $|\bar x_i-y^\star_i|$ as $|\delta(x_i'-x_i^\star)+x_i^\star-y_i^\star|$.
Then, we notice that $\delta(x_i'-x_i^\star)$ and $x_i^\star-y_i^\star$ have different sign and are both smaller than $\Delta$ since $\delta\le \Delta$.
Hence, we can use that $\|a+b\|\le\max(|a|,|b|)$ for any $a$ and $b$ with different signs. This shows that $|\bar x_i-y_i^\star|\le \Delta$. 

Conversely, when $Q_i=0$, then $\bar x=x^\star$ and thus $\bar x\in[0,1]^d$ and $\|\bar x-y^\star\|_\infty\le \Delta$ are trivially satisfied. Again we can write
\[
\|\bar y-x^\star\|_\infty \le \max(\|x^\star-y^\star\|_\infty, |\bar y_i-x_i^\star|),
\]
and by a similar argument as before $|\bar y_i-x_i^\star|\le\Delta$. Now we are left to check that $\bar y\in[0,1]^d$. If $y^\star_i \ge x^\star_i $ then $x^\star_i\ge x'_i$. Thus 
\[\bar y_i=\delta(x'_i-x^\star_i)+y^\star_i\ge (1-\delta) x^\star_i+ \delta x_i'\ge 0,\] and $\bar y_i\le y^\star_i\le 1$. 
Similarly, if $y^\star_i \le x^\star_i$ then $x^\star_i\le x'_i$. This clearly implies that $\bar y_i\ge y^\star_i\ge 0$ and $\delta(x_i'-x_i^\star)+y^\star_i\le (1-\delta)x^\star_i+\delta x_i'\le 1$. This shows that $\bar y\in[0,1]^d$. Thus we can conclude that both $\bar y$ and $\bar x$ are valid deviations.

Since $\bar y$ and $\bar x$ are valid local deviations and $(x^\star,y^\star)$ satisfies the  local optimality condition, it holds

\begin{equation}\label{eq:eq01}
(x^\star-y^\star)^\top(Dx^\star+c)\le (\bar x-y^\star)^\top(D\bar x+c)+\epsilon.
\end{equation}
and
\begin{equation}\label{eq:eq02}
(x^\star-y^\star)^\top (Dx^\star+c)\ge (x^\star-\bar y)^\top (Dx^\star+c) -\epsilon
\end{equation}

In the next step of the proof, we ``globalize'' such optimality condition, showing that the deviation to $x'$ only slightly increases the value of the VI.

We start from \Cref{eq:eq01}.
Using the fact that, by construction, $\bar x= x^\star + \kappa\cdot e_i$, where $\kappa = \delta Q_i(x_i'-x_i^\star)$, and using the fact that $|\kappa| \le \delta$, we can therefore write
\begin{align*}
    (\bar x-y^\star)^\top&(D\bar x+c)\\
    &=(x^\star+\kappa\cdot e_i-y^\star)^\top(D(x^\star+\kappa\cdot e_i)+c)\\
    &= (x^\star-y^\star)^\top(Dx^\star+c)+\kappa e_i^\top (Dx^\star+c)+\kappa (x^\star+\kappa e_i-y^\star)^\top (De_i)\\
    &\le (x^\star-y^\star)^\top(Dx^\star+c)+ \kappa e_i^\top (Dx^\star+c) +\delta \|x^\star-y^\star\|_\infty \|D\|_1+\delta^2 |D_{i,i}|\\
    &\le (x^\star-y^\star)^\top(Dx^\star+c) + \kappa e_i^\top (Dx^\star+c)+\delta\Delta +\delta^2\\
    &\le (x^\star-y^\star)^\top(Dx^\star+c) + \kappa e_i^\top (Dx^\star+c)+2\delta\Delta \tag{$\delta\le\Delta$}.
\end{align*}
Hence, combined with \Cref{eq:eq01} we obtain
\begin{align}\label{eq:eq11}
 \kappa e_i^\top(Dx^\star+c) = (Dx^\star+c)^\top(\bar x-x^\star) \ge -\epsilon-2\delta\Delta.
\end{align}

Then, we globalize the equilibrium condition of player $y$. \Cref{eq:eq02} directly implies that
\begin{align}\label{eq:eq22}
    (Dx^\star+c)^\top(\bar y-y^\star)\ge -\epsilon.
\end{align}

Summing \Cref{eq:eq11} and \Cref{eq:eq22} we obtain
\begin{equation}\label{eq:JCfinal}
(Dx^\star+c)^\top(\bar x+\bar y-x^\star-y^\star)\ge -2\epsilon-2\delta\Delta.
\end{equation}
Notice that either $\bar x_i\neq x_i^\star$ or $\bar y_i\neq y_i^\star$ and hence we could consider only one of the inequalities depending on $Q_i$. However, considering the two together simplifies the proof.

Finally, we observe that by combining $\bar x-x^\star$ and $\bar y-y^\star$ we implement the vector $\delta(x'_i-x_i^\star)e_i$. Formally,
\begin{equation}\label{eq:sum2}
\bar x+\bar y=x^\star+y^\star+\delta(x'_i-x_i^\star)e_i,
\end{equation}
where $e_i$ is the $i$-th standard basis vector. Hence, the joint deviation implements a deviation from $x^\star$ toward $x'$.

Thus, combining \Cref{eq:sum2} and \Cref{eq:JCfinal} we obtain that
\[
(Dx^\star+c)_i (x'-x_i^\star)\ge -2\left(\frac{\epsilon}{\delta}+\Delta\right)\ge -\rho^*,
\]
concluding the proof.

\paragraph{From \GDAFPJC to \LRMINMAXJC}
The relationship between fixed point of gradient descent-ascent dynamics and local min-max of \citet{daskalakis2021complexity} (\Cref{th:GDA_eq_minmax}) is not enough to obtain constant inapproximability of \GDAFPJC in $\ell_\infty$. This is because we could not design a function $f$ with constant Lipschitzness $G$. However, since we proved that \LRMINMAXJC is hard even with single component deviations (or equivalently ``norm 1'' deviations), we can use the specialized reduction that we show in \Cref{lem:from_GDA_to_MM}.

Now, with this result, we can prove hardness for \GDAFPJC by a reduction from \LRMINMAXJC.
Take the same instance of \LRMINMAXJC built at the beginning of \Cref{sec:formalproof}. 
It is easy to check that $K$ is defined such that $K_1(y)$ and $K_2(x)$ satisfy the conditions of \Cref{lem:from_GDA_to_MM} and that $G_\infty\le 3$.
Thus, we can use \Cref{lem:from_GDA_to_MM} with $\nu=0$, and $\alpha=\rho^{*,2}/120$. This implies that any $\alpha$ fixed point to \GDAFPJC, is a $(\delta,\epsilon)$-solution to \LRMINMAXJC, where 
$\delta= \rho^*/15$ and $\epsilon=\rho^{*,2}/60$.
Thus, \GDAFPJC is \PPAD-hard for any $\alpha<\rho^{*,2}/120$, where the function $f$ is $O(1)$-smooth, $G=O(\sqrt{d})$-Lipschitz, and nonconvex-concave (quadratic-linear). In \Cref{sec:constant_inapprox} we discuss how this result can be exploited to prove constant inapproximability of \GDAFPJC in $\ell_\infty$-norm.

\subsection{Proof of \Cref{th:qvihardness}}\label{sec:proofqvihard}

Take any \LINVI instance defined by the matrix $D\in[-1,1]^{d\times d}$, a vector $c\in [-1,1]^d$ and approximation $\alpha_\LINVI=\rho^*$, where $\rho^*$ is the constant in \Cref{th:linearvihard}. We showed that this problem is \PPAD-hard in \Cref{th:linearvihard}. 

   \paragraph{Objective function}
    We define the objective function $f:\Reals^d\times\Reals^d\to\Reals$ as $f(x,y)=x^\top (D y + c)$. 
    
    Similarly to \Cref{th:nonconvexconcavehardness}, we first prove some properties about the Lipschitzness and smoothness of the function. Recall that  we can restrict to $D$ such that  $\|D\|_1,\|D\|_\infty\le 1$.

\begin{restatable}{lemma}{lemLiptwo}
    The function $f$ is $G$-Lipschitz and $L$-smooth with $G=3\sqrt{d}$ and $L=1$.
\end{restatable}
\begin{proof}
We recall that $\nabla_x f(x,y)=Dy+c$ and $\nabla_yf(x,y)=D^\top x$.
We bound the Lipschitz parameter of $f$ as the supremum of the norm of the gradient. For all $(x,y)\in[0,1]^{2d}$ we have
\begin{align*}
    \|\nabla f(x,y)\|^2&=\|Dy+c+\|^2+\|D^\top x\|^2\\
    &\le 2\|Dx\|^2+2\|c\|^2+2\|D^\top x\|^2  \tag{Parallelogram inequality}\\
    &\le 2\|x\|^2+ 2d +2\|x\|^2\tag{\Cref{lem:matrix_norms3}} \\
    &\le 5 d.  \tag{Since $\|x\|_2, \|y\|_2\le\sqrt{d}$}
\end{align*}
  Taking a square root, we conclude that the function $f$ is $G$-Lipschitz with $L\le 3\sqrt{d}$. 
  We now focus on the smoothness parameter of the function (\ie, the Lipschitz parameter of the function's gradient). 
  We have
\begin{align*}
    \|\nabla f(x,y)-\nabla f(x',y')\|^2&= \|D(y-y')\|^2 + \|D^\top (x-x')\|^2\\
    &\le \|x-x'\|^2+\|y-y'\|^2\tag{\Cref{lem:matrix_norms3}}\\
    &\le \|(x,y)-(x',y')\|^2.
\end{align*}
This proves that $L\le 1$ and concludes the proof.
\end{proof}

\paragraph{Constraints}

      The constraint functions are given by  $g_2(x,y):=\|x-y\|_\infty\le 0$, while player $1$ is unconstrained. It is easy to see that $K_1(y)$ and $K_2(x)$ are non empty for each $(x,y)\in [0,1]^d$ and that the constraints have the required structure of independent constraints of \Cref{eq:constraints}.

   In the following, we follow the same blueprint of \Cref{th:nonconvexconcavehardness}
   First, prove the hardness result for \LRMINMAXCBC. 
Then, we exploit the reduction from \LRMINMAXCBC to \GDAFPCBC in \Cref{lem:from_GDA_to_MM} to extend the hardness result to the problem of finding fixed points of gradient ascent-descent dynamics.
Also, in this setting, restricting to single-dimensional deviations will be essential to provide tighter hardness results.
   
    \paragraph{From  \LRMINMAXCBC to \LINVI}

    We provide a reduction for any $0<\gamma\le 1$, $\delta = \gamma \frac{\rho^*}{15}$, $\epsilon=\gamma\frac{\rho^{*,2}}{60}$, and $\nu=\frac{\rho^*\delta}{4d}$. It is easy to check that $\delta < \sqrt{\frac{2\epsilon}{L}}$ for any $\gamma\in(0,1]$.
    
    Now we want to prove that any solution to \LRMINMAXJC $(x^\star,y^\star)$ is a solution to \LINVI. 
    To obtain a tighter hardness result, we focus on the specialized (but still \PPAD-Hard as proved in \Cref{th:linearvihard}) problem of finding a $z$ such that it is robust to single-dimensional deviations, \ie
    \begin{equation*}
 (Dx^\star+c)_i\cdot(x'_i-x^\star_i)\ge -\rho^*\quad\forall d \in [d], \, x'_i\in[0,1].
\end{equation*}

     Consider any $i\in [d]$, and  $x'_i\in [0,1]$. Then define $\bar x\in [0,1]^d$ as $\bar x_i=\delta x'_i+(1-\delta) x^\star_i$ and $\bar x'_j=x^\star_j$ for each $j \neq i$.
     
     We check that $\bar x_i$ is indeed a feasible deviation.  First, $\|\bar x-x^\star\|\le \delta$.
     Moreover, $\bar x_i$ is in $[0,1]$ as $\bar x_i$ is a convex combination of two points in $[0,1]$.
    
    Then by the local optimality condition of  $x^\star$ with respect to $\bar x$ we have:
    \begin{align*}
        x^{\star,\top} (D x^\star+c)&=x^{\star,\top} D(x^\star-y^\star)+x^{\star,\top} (Dy^\star+c)\\
        &\le \|D^\top x^\star\|_{1}\|x^\star-y^\star\|_{\infty}+x^{\star,\top} (Dy^\star+c)\\
        &\le d \nu+ x^{\star,\top} (Dy^\star+c)\\
        &\le d\nu + {\bar x}^{\top} (Dy^\star+c)+\epsilon\\
        &\le d\nu + {\bar x}^{\top} D(y^\star-x^\star)+{\bar x}^{\top} (Dx^\star+c)+\epsilon\\
        &\le 2d\nu + {\bar x}^{\top} (Dx^\star+c)+\epsilon\\
        &=2d\nu + \delta (x_i'-x_i^\star)\cdot (Dx^\star+c)_i+{x^\star}^{\top} (Dx^\star+c)+\epsilon,
    \end{align*}
    where the first inequality is the H{\"o}lder's inequality, the second holds since and $x\in[0,1]^d$ and $\|D\|_1\le 1$, and the third  one follows from $(x^\star,y^\star)$ being a solution to \LRMINMAXCBC with parameters $(\epsilon, \delta, \nu)$. The last two inequalities hold for similar reasons to the first two.
    Rearranging the above inequality and using the definition of $\nu$, $\epsilon$, and $\delta$ we obtain
    \[
        (Dx^\star+c)_i\cdot (x_i'-x_i^\star)\ge -\frac{\epsilon +2d\nu}{\delta}=-\rho,
    \]
    which holds for all $i\in [d]$ and $x_i'\in [0,1]$. This proves that $x^\star$ is a solution to the \LINVI instance.

    \paragraph{From  \GDAFPCBC  to \LRMINMAXCBC}

Similarly to what we have done in the proof of \Cref{th:nonconvexconcavehardness}, we reduce from \LRMINMAXCBC to \GDAFPCBC through \Cref{lem:from_GDA_to_MM}.

    Take the same instance of \LRMINMAXCBC built at the beginning of \Cref{sec:proofqvihard}. 
    It is easy to check that $K^\nu$ is defined such that $K^\nu_1(y)$ and $K^\nu_2(x)$ satisfy the conditions of \Cref{lem:from_GDA_to_MM}. Notice that working with the ``enlarged'' sets simply modifies the sets $K_1$ and $K_2$ without changing the properties of the constraints. 
    Moreover, it is easy to check that $G_\infty\le 1$.
    Thus, we can apply \Cref{lem:from_GDA_to_MM} with $\nu=\frac{\rho^*\delta}{4d}$, and $\alpha=\rho^{*,2}/120$. This implies that any $\alpha$ fixed point to \GDAFPCBC, is a $(\delta,\epsilon,\nu)$-solution to \LRMINMAXCBC, where $\delta= \rho^*/15$ and $\epsilon=\rho^{*,2}/60$.

    Thus, \GDAFPCBC is \PPAD-hard for any $\alpha<\rho^{*,2}/120$, where the function $f$  is $O(1)$-smooth, $G=O(\sqrt{d})$-Lipschitz, and convex-concave (linear-linear).

\section{Basic Mathematical Preliminaries}\label{app:basic}

\subsection{Matrix Norms}

\begin{definition}
For a matrix $M\in\Reals^{m\times n}$ we define the operator norm as
\[
\|M\|_{r,s} = \sup_{x\in\Reals^n:\|x\|_r=1} \|Mx\|_s.
\]
\end{definition}

When $r=s$, we simply write $\|M\|_r$. Some special cases that we use in this paper are $\|M\|_\infty=\max_{i\in[m]}\sum_{j\in[n]}|M_{i,j}|$ and $\|M\|_1=\max_{j\in[n]}\sum_{i\in[m]}|M_{i,j}|$. Clearly $\|M\|_1=\|M^\top\|_\infty$.

Finally, we use the following lemma.

\begin{lemma}\label{lem:matrix_norms3}
     Given a matrix $M\in\Reals^{m\times n}$ it holds that
     \[
     \|M\|_2\le\sqrt{\|M\|_1\|M\|_\infty}.
     \]
     In particular if $\|M\|_1,\|M\|_\infty\le 1$ then $\|Mx\|\le\|x\|$ and $\|M^\top x\|\le \|x\|$ for all $x\in\Reals^n$.
\end{lemma}
\begin{proof}
    Consider $\lambda$ the greatest eigenvalue of $M^\top M$ and $x$ an eigenvector. Now:
    \begin{align*}
        \|M\|_2^2&=\lambda
        =\frac{\|M^\top Mx\|_\infty}{\|x\|_\infty}
        \le\|M^\top M\|_\infty
        \le \|M^\top\|_\infty\|M\|_\infty
        \le \|M\|_1\|M\|_\infty.
    \end{align*}
    The proof is concluded by using Cauchy-Schwartz inequality for matrix norms on $\|Mx\|$.
\end{proof}

\subsection{Correspondences}
Here, we define some basic concepts about semicontinuity of correspondences.

\begin{definition}(Upper/Lower-semicontinuity)\label{def:semicont}
A correspondence $Q:X\rightrightarrows Y$ is:
\begin{itemize}
    \item \emph{upper-semicontinuous} at $x\in X$ if for any open set $V\subseteq Y$ such that $Q(x)\subseteq V$ there is a neighborhood $U$ of $x$ such that $\cup_{\tilde x\in U}Q(\tilde x)\subseteq V$.
    \item \emph{lower-semicontinuous} at $x\in X$ if for any open set $V\subseteq Q$ such that $Q(x)\cap V\neq \emptyset$ there is a neighborhood $U$ of $x$ such that $Q(\tilde x)\cap V\neq \emptyset$ for all $\tilde x\in U$.
    \item continuous at $x\in X$ if it is both upper- and lower-semicontinuous at $x$.
\end{itemize}
\end{definition}

We say that a correspondence $Q:X\rightrightarrows Y$ is continuous (resp., upper- or lower-semicontinuous) in a set $\tilde X\subseteq X$ if it is continuous (resp., upper- or lower-semicontinuous) at $x$ for all $x\in\tilde X$. 

The following well-known theorem shows the existence result of fixed points for semicontinuous correspondences. 

\begin{theorem}[Kakutani fixed point theorem]\label{thm:kakutani}
If $Q:X\rightrightarrows X$ is an upper-semicontinuous correspondence which is closed, convex, and non-empty valued, then there exists $x\in X$ such that $x\in Q(x)$.
\end{theorem}

\end{document}